\documentclass[oldversion]{aa}  %If author wishes unstructured abstract

\usepackage{graphicx}
\usepackage{multirow}
\usepackage{txfonts}
\usepackage{natbib}

\bibpunct{(}{)}{;}{a}{}{,} %to allow the A&A style
\usepackage{longtable,lscape}
\begin{document}
   \title{The 
influence of model parameters on the prediction of gravitational wave signals from stellar core collapse}
\titlerunning{Parameter dependent prediction of gravitational waves from stellar core collapse}
%
%\titlerunning{Parameter dependent prediction of GWs from stellar core collapse}

   \author{S. Scheidegger
          \and
           R. K\"appeli
           \and
          S. C. Whitehouse
          \and
          T. Fischer
          \and
          M. Liebend\"orfer}
\authorrunning{Scheidegger et al.}
   \offprints{S. Scheidegger}
   \institute{Department of Physics, University of Basel, Klingelbergstrasse
82, 4056 Basel, Switzerland\\
             \email{simon.scheidegger@unibas.ch }}
%
%   \date{Received 2009; *}
\date{Received August 31, 2009; Accepted January 18, 2010}
%
% \abstract{}{}{}{}{} 
% 5 {} token are mandatory
 %
  \abstract
{We present a gravitational wave (GW) analysis of an 
extensive series of three-dimensional
magnetohydrodynamical core-collapse simulations.
Our 25 models are based on a 15$M_{\odot}$ progenitor
stemming from (i) stellar evolution calculations, 
(ii) a spherically symmetric effective general
relativistic potential, either the Lattimer-Swesty (with three possible 
compressibilities) or the Shen equation of state for hot, dense 
matter, and (iii) a neutrino parametrisation
scheme that is accurate until about 5ms postbounce. 
For three representative models, we also included 
long-term neutrino physics by means of a
leakage scheme, which is based on partial implementation 
of the isotropic diffusion source approximation (IDSA).
We systematically investigated the effects of the equation of state,
the initial rotation rate, and both the toroidal 
and the poloidal magnetic fields  
on the GW signature. 
We stress the importance of including of postbounce neutrino 
physics, since it  quantitatively alters the GW signature.
Slowly rotating models, or those that do not rotate 
at all, show GW 
emission caused by prompt and proto-neutron star (PNS) convection.
Moreover, the signal stemming from prompt convection 
allows for the distinction between 
the two different nuclear equations of state
indirectly by different properties of the fluid instabilities.
For simulations with moderate or even fast rotation rates,
we only find the axisymmetric type I wave signature at core bounce.
In line with recent results, we could confirm that 
the maximum GW amplitude scales roughly linearly with the 
ratio of rotational to gravitational energy at core bounce below 
a threshold value of about $10\%$.
We point out that models set up with an initial
central angular velocity of $2\pi$ rad$s^{-1}$ or faster 
show nonaxisymmetric narrow-band GW radiation
during the 
postbounce phase. This emission process is 
caused by a low $T/|W|$ dynamical instability. 
Apart from these two points, we show that it is 
generally very difficult to discern 
the effects of the individual features of the input physics in 
a GW signal from a rotating core-collapse supernova
that can be attributed unambiguously to a specific model.
Weak magnetic fields do not notably influence the dynamical
evolution of the core and thus the GW emission.
However, for strong initial poloidal magnetic fields ($\gtrsim10^{12}$G),
the combined action of flux-freezing and field winding 
leads to conditions where the ratio of magnetic field pressure to 
matter pressure reaches about unity 
which leads to the onset of a jet-like supernova explosion.
The collimated bipolar out-stream of matter is then reflected
in the emission of a type IV GW signal.
In contradiction to axisymmetric simulations,
we find evidence that nonaxisymmetric fluid
modes can counteract or even suppress jet formation for models with strong
initial toroidal magnetic fields.
The results of models with continued neutrino emission 
show that including of the deleptonisation
during the postbounce phase is an indispensable issue for the
quantitative prediction of GWs 
from core-collapse supernovae,
because it can alter the GW amplitude up to a factor of 10 
compared to a pure hydrodynamical treatment.
Our collapse simulations 
indicate that corresponding events in our Galaxy would be detectable either
by LIGO, if the source is rotating, or at least by the advanced LIGO detector, 
if it is not or only slowly rotating.

}

   \maketitle
\section{Introduction}

Gravitational wave (GW) astronomy may soon become a reality
and will allow humankind to address questions about many different astrophysical 
objects that are
hidden from the electromagnetic detection. 
Within the past few years, the first generation of
the ground-based GW detectors LIGO (USA), VIRGO (Italy), GEO600 (Germany), 
and TAMA (Japan) have got very close to or
even reached design sensitivity
and collected partially coincident 
data, as discussed by \citet{abbott}.
Lately, the two 4 km LIGO detectors were 
upgraded to sensitivities increased by a factor of 2-3
(enhanced LIGO, see \citet{Adhikari})
and resumed observations in 2009.
Upgrades of the three 
LIGO interferometers and VIRGO are expected to 
be completed by 2014 and will increase the observable volume
by a factor of $\sim$ 1000.
Operating at such a high level of precision, GW 
detectors are sensitive to 
many different sources, such as compact binary coalescence from 
black-holes (BH) and neutron stars
up to distances of several 100 megaparsecs,
but might also very likely produce the
first detections of black-hole and neutron star mergers, 
core-collapse supernovae, and neutron star
normal mode oscillations, as recently reviewed in 
\citet{lrr-2009-2}.

The broad majority of GW sources can be subdivided into
two classes: (i) mathematically well-posed 
events such as BH-BH coalescence and (ii) scenarios that involve
matter.
The first category of sources can be modelled accurately, 
the waveform can be calculated with
high precision, and thus a matched-filter analysis can be applied.
On the other hand, the latter kind of GW sources 
can only be modelled imperfectly,
as matter effects carry large physics uncertainties and open up 
a huge 
parameter space for initial conditions. Moreover, even if there
were `perfect' models, there is still a turbulent, stochastic
element in some GW emission mechanisms that makes it strictly
impossible to compute templates. 
As a result, more general, un-modelled burst analysis techniques are used,
as summarised in \citet{abbott} and references therein.
One of the scenarios where burst data analysis must be applied
is the stellar core collapse, where in 
addition to the matter effects even the fundamental explosion mechanism
is not fully settled.
It is still unclear which processes may convert some of 
the released gravitational binding energy of order $\mathcal{O}(10^{53})$erg into
the typically observed kinetic and internal energy of the ejecta 
of $\mathcal{O}(10^{51})$erg = 1 Bethe [B] needed
to blow off the stellar envelope.
While most state-of-the art simulations investigate a mechanism based 
on neutrino heating in combination with hydrodynamical instabilities, 
such as \citet{2009ApJ...694..664M}, others explore the alternative magneto-rotational 
mechanism \citep{Leblanc1970,Kotake2004a,2007ApJ...664..416B,2008ApJ...683..357M,2009ApJ...691.1360T}, 
or the acoustic mechanism which is 
driven by strong core g-modes \citep{Burrows2006,2007ApJ...655..416B}. 
For a recent review see \citet{2007PhR...442...38J}. 
\citet{1988PThPh..80..861T}, \citet{1993ApJ...414..701G}, and lately also
\citet{2008PhRvD..77j3006N} and \citet{2009PhRvL.102h1101S}  
reported that a QCD phase transition may power a secondary shock wave
which triggers a successful hydrodynamical explosion.

Beside neutrinos, which have already been observed in the context of
stellar core collapse of SN1987A \citep{1988PhRvD..38..448H}, GWs could 
provide access to the electromagnetically hidden compact 
inner core of some such cataclysmic events. 
Since a core-collapse supernova is expected to 
show aspherical features \citep{2006Natur.440..505L},
there is  reasonable hope
that a tiny amount of the released binding energy will be 
emitted as GWs which could then provide us with valuable 
information about the angular momentum
distribution \citep{2008PhRvD..78f4056D} and 
the baryonic equation of state (EoS) \citep{2009A&A...496..475M}, 
both of which are
uncertain. Furthermore, they might help to constrain theoretically 
predicted SN mechanisms \citep{2009CQGra..26f3001O}.
Given the actual sensitivities of today's operating ground-based GW
observatories \citep{2008CQGra..25k4013W} combined with knowledge of 
waveforms from state-of-the art
modelling, GWs from a Galactic core-collapse supernova
may be considered to be within the detector limits. 
The collapsing iron core and the subsequently newly formed PNS 
can be subject to a whole variety of asymmetric hydrodynamical 
and nonaxisymmetric
instabilities which give rise to GW emission.

In this context probably most attention during the past three decades has been paid
to rotational core collapse and bounce dynamics. In this particular phase, 
the core spins up due to angular
momentum conservation, resulting in an oblate and 
time-dependent deformation that leads to 
strong GW emission \citep{1982A&A...114...53M,1991A&A...246..417M,Janka1996,1997A&A...320..209Z,
1998A&A...332..969R,Kotake2003a,Kotake2004,Obergaulinger2006,Ott2004,2007PhRvL..98z1101O,
2007CQGra..24..139O,2002A&A...393..523D,2007PhRvL..98y1101D,2008PhRvD..78f4056D,2008A&A...490..231S}.  
The steady improvement of the models
(e.g. the inclusion of GR, a micro-physical EoS, treatment of neutrino physics) recently led 
to a theoretically well understood 
single and generic, so-called type I wave form which is characterised by a large negative
peak at core bounce, followed by ring-down oscillations that damp quickly
(\citet{2007PhRvL..98z1101O,2007CQGra..24..139O,2007PhRvL..98y1101D,2008PhRvD..78f4056D}; with detailed references therein). 
Despite the reduction to a single wave form, 
the combined information of the GW amplitude and the location
of the narrow peak of the GW spectral energy density in frequency space
contains information that makes it still possible
to constrain progenitor- and postbounce rotation, but 
can barely distinguish between different finite-temperature EoS (see e.g. \citet{2008PhRvD..78f4056D},
who used the EoS of \citet{Lattimer1991} and \citet{Shen1998}).

Gravitational waves from magneto-rotational collapse were considered in detail by 
\citet{Kotake2004,2006PhRvD..74j4026S} and \citet{Obergaulinger2006}. 
As main differences compared to simulations which do not include magnetic fields it was found by 
these groups
that only in the very special case of precollapse fields
as strong as $\gtrsim 10^{12}$G the overall dynamics can be influenced. 
As \citet{Heger2005} have argued, such strong fields are unlikely to occur in 
standard core-collapse supernova progenitors.
The GW amplitude is then affected by (i) time-dependent magnetic fields, which contribute considerably 
to the overall energy density, and (ii) by bipolar magnetohydrodynamic (MHD) jet outflows which give rise to a so-called type IV signal 
with memory \citep{Obergaulinger2006}.
Physically, such a memory effect in the GW signal arises
from the temporal history of asymmetric matter outflow, leaving behind a constant
offset in the amplitude  \citep{Thorne1989}.

Recently it has been argued through numerical simulations 
of equilibrium neutron star models or full core-collapse simulations 
that PNSs with a high degree 
of differential rotation can be subject to nonaxiymmetric rotational instabilities at low $\beta$ values (= $T/|W|$, 
ratio of rotational to gravitational energy), 
leading to strong narrow-band GW emission \citep{2003ApJ...595..352S,2005ApJ...618L..37W,
2005ApJ...625L.119O,2006AIPC..861..728S,2006ApJ...651.1068O,2007CQGra..24..139O,2007CoPhC.177..288C,2007PhRvL..98z1101O,
2008A&A...490..231S}. However little is known about the true nature of the instability at present.
Previous work has failed to establish an analytical instability criterion, and 
the dependence of the instability 
on PNS rotation rate and degree of differential rotation
is still unclear, as it was pointed out by \citet{2009CQGra..26f3001O}.

Current state-of-the-art stellar evolution calculations \citep{Heger2005} tell us that 
iron cores of stars
generally lose most of their angular momentum during their evolution due 
to magnetic torques. Therefore, they
cannot become subject to strong rotationally-induced aspherities. However, anisotropic neutrino
emission, convection and standing accretion shock instability 
(SASI, see e.g. \citet{Blondin2003}) - driven deviations from spherical symmetry which can lead to the emission of 
GWs of sizable amplitudes are likely 
to occur inside the PNS and the post-shock gain region and last for probably hundreds of ms  \citep{1997A&A...317..140M,M2004,2007ApJ...655..406K,2009A&A...496..475M,
2009ApJ...707.1173M,2009ApJ...704..951K,2009ApJ...697L.133K}.
Although there is qualitative consensus among the different core-collapse supernova
groups about the aforementioned features being 
emitters of stochastic broad-band signals, 
their detailed quantitative character still remains quite uncertain, since self-consistent 3D simulations 
with proper long-term neutrino transport were not carried out so far, but would in principle be 
required.
The most recent, very elaborate 2D simulations in that context were performed 
by \citet{2009A&A...496..475M}. Their numerical setup includes 
long-term multi-flavor neutrino transport, an effective relativistic potential and two
different EoS.
The first 100ms of after bounce they observed GWs from early prompt postbounce convection, 
peaking somewhat around or below 100Hz, depending on the employed EoS. 
After this early episode, the GW emission in their models is dominated by 
a growing negative amplitude related to 
anisotropic neutrino emission at frequencies $\lesssim 200$Hz, while
the GW signal associated with non-radial mass 
motions stems from density 
regimes $10^{11}-10^{13}$gcm$^{-3}$ and 
peaks in the frequency range of $300-800$Hz, highly sensitive 
to the nuclear EoS.

In the core-collapse scenario, PNS pulsations can provide 
another mechanism for GW emission. \citet{2006PhRvL..96t1102O} pointed out that
in the context of the acoustic mechanism \citep{Burrows2006}
excited core g-mode oscillations might emit very strong GWs. 
The GW signal is due to the nonlinear quadrupole components of the pulsations that,
at least initially, are of $l=1$ g-mode character \citep{2006PhRvL..96t1102O}.

For recent reviews on GWs from core-collapse supernovae 
with complete lists of references see \citet{2009CQGra..26f3001O} and \citet{2006RPPh...69..971K}.

In this paper, we present the gravitational wave analysis of a
comprehensive set of three-dimensional MHD  
core-collapse supernova simulations in order to investigate 
the dependencies of the resulting GW signal on the 
progenitor's initial conditions. Our calculations encompass presupernova models
from stellar evolution calculations, a finite-temperature nuclear EoS 
and a computationally efficient treatment of the deleptonisation and 
neutrino emission during the collapse. General relativistic corrections
to the spherically symmetric Newtonian gravitational potential 
are taken into account. Moreover, several models incorporate 
long-term neutrino physics by means of a leakage scheme.
As for the progenitor configuration, we systematically consider not 
only variations in the precollapse rotation rate, but also
in the nuclear EoS and the magnetic field topology.
This study extends the work of 
\citet{2008A&A...490..231S}, who investigated
only two models with similar input physics.
In this way, we carried out the so far largest parameter 
study of 3D MHD stellar collapse 
with respect to the prediction of GWs.

The structure of the paper is as follows.
In section 2 we briefly describe the numerical methods, input physics
and progenitor configurations applied in our core-collapse simulations.
Furthermore, we describe how we extract GWs from our 
model set. In sec. 3 we discuss results of 25 three-dimensional MHD simulations
with respect to the GW signal from convection, rotational core bounce,
nonaxisymmetric rotational instabilities and very strong magnetic fields.
In sec. 4, we summarise and present conclusions.

\section{Numerical Methods}

\subsection{The 3D MHD code and its input physics}
The algorithm used to solve the time-dependent, Newtonian MHD 
equations in our 3D simulations is based on a simple and fast cosmological MHD code of
\citet{2003ApJS..149..447P,Liebendorfer2006}, which has been 
parallelised with a hybrid combination of 
MPI and openMP and improved and adapted to the requirements of core-collapse supernova 
simulations \citep{2009arXiv0910.2854K}.
The 3D computational domain consists of a central cube of $600^3$km$^3$ volume,
treated in equidistant Cartesian coordinates with a grid spacing of 1km.
It is, as explained in detail in \citet{2008A&A...490..231S}, embedded in a 
larger spherically symmetric computational domain  that is
treated by the time-implicit hydrodynamics code 'Agile' \citep{Liebend2002}.
With this setup, the code scales nicely to at least 8000 parallel processes
\citep{2009arXiv0910.2854K}.
The ideal MHD equations read 

\begin{eqnarray}
\frac{\partial\rho}{\partial t}+\nabla\cdot(\rho\textbf{v})& = & 0 \\
\label{equ:mass}
\frac{\partial\rho\textbf{v}}{\partial t} \nabla\cdot(\textbf{v}\rho\textbf{v}-\textbf{b}\textbf{b}) + \nabla P & = & -\rho\nabla\Phi 
\label{equ:momentum} \\
\frac{\partial E}{\partial t} + \nabla\cdot\left[(E+P)\textbf{v}-\textbf{b}\cdot(\textbf{v}\cdot\textbf{b})\right] & = & -\rho\textbf{v}\cdot\nabla\Phi
\label{equ:energy}\\
\frac{\partial\textbf{b}}{\partial t} -\nabla\times (\textbf{v}\times\textbf{b}) & = & 0
\label{equ:MHD}
\end{eqnarray}
expressing the conservation of mass, momentum, energy and magnetic flux, respectively. 
$\rho$ stands for mass density, $\textbf{v}$ is the velocity vector and $E=\rho e + \frac{\rho}{2}v^2 +\frac{b^2}{2}$ 
the total energy (the sum of internal, kinetic and magnetic energy).
The magnetic field is given by $\vec{B} = \sqrt{4\pi}\vec{b}$ and the
total pressure by $P=p+\frac{b^2}{2}$ (the sum of gas and magnetic pressure).
The MHD equations are evolved from a condition 
\begin{equation}
 \nabla\cdot\textbf{b}=0 \; .
\label{equ:divergence}
\end{equation}
We employ the constrained transport method \citep{1988ApJ...332..659E} 
to guarantee the divergence-free time evolution
of the magnetic field.
The right hand side of Eq. \ref{equ:momentum} and Eq. \ref{equ:energy} take into account
the effect of gravitational forces on the magnetohydrodynamical variables. The gravitational
potential $\Phi$ obeys the Poisson equation 
\begin{equation}
 \Delta\Phi=4\pi G\rho \; .
 \label{equ:Poisson}
\end{equation}
We only implement the monopole term of the
gravitational potential by 
a spherically symmetric mass integration that includes 
general relativistic corrections \citep{Marek2006}. 
In this approach, the Newtonian gravitational potential $\Phi$ is replaced 
by an effective potential $\Phi_{eff}$ (`case A' potential, see \citet{Marek2006})
\begin{equation}
\Phi_{eff}(r) = \int_{r}^{\infty}\frac{dr'}{r'^{2}} 
\left[ \frac{m_{eff}}{4\pi}+r'^3\cdot(p +p_{\nu} )  
\right]\cdot\frac{1}{\Gamma^2}\left(\frac{\rho + e + p}{\rho}\right)
\label{equ:effpot}
\end{equation}
with $p$ being the gas pressure, $p_{\nu}$ the 
neutrino pressure and $e$ the internal energy density. The effective 
mass is given by 
\begin{equation}
m_{eff}(r) = 4\pi \int_{0}^{r}dr'r'^2 \left(\rho + e + E \right) \; ,
\label{equ:effmass} 
\end{equation}
where $E$ is the neutrino energy. 
The metric function $\Gamma$ is given by
\begin{equation}
 \Gamma= \sqrt{1 + v_{r}^2 - \frac{2m_{eff}}{r}} \; .
\end{equation}
where $v_{r}$ is the radial fluid velocity.
Note that \citet{2008A&A...489..301M}
recently proposed an effective relativistic 
gravitational potential for rapidly rotating
configurations. However, since it would have added
another degree of freedom to our parameter set, 
we did not consider it in this study.

The system of the MHD equations must be closed by a finite-temperature
EoS.
In core-collapse supernova simulations, an EoS has to handle several
different regimes.
For temperatures below 0.5MeV,
the presence of nuclei and time dependent nuclear processes
dominate the internal energy evolution.
For simplicity, an ideal gas of Si-nuclei is assumed,
which, for our GW study, 
is sufficiently accurate to describe the low baryonic pressure
contribution at low densities in the outer core.
At higher temperatures, where matter is in nuclear statistical 
equilibrium, we employ two alternative EoS: The Lattimer \& Swesty EoS
(\citet{Lattimer1991}, LS EoS) and the one by \citet{Shen1998a}.
The first EoS is based on a phenomenological compressible liquid drop model;
it also includes surface effects as well 
as electron-positron and photon contributions.
The LS EoS assumes a nuclear symmetry energy of 29.3MeV 
and we will perform simulations with three choices of the 
nuclear compressibility modulus K (180,220,375MeV)
as provided by \citet{Lattimer1991}.
Since variations in K affect the stiffness of the nuclear component 
of the EoS, this enables us to probe the effects of variations in stiffness
while keeping the general EoS model fixed.
The pure baryon EoS from \citet{Shen1998a} has a compressibility
of K=281MeV and a symmetry energy of 36.9MeV. It is based on
relativistic mean field theory and the Thomas-Fermi approximation.
For matter in non-NSE (T\(<0.44\)MeV), the Shen EoS is coupled
to the baryonic and electron-positron EoS given in \citet{1999ApJS..125..277T} and
\citet{2000ApJS..126..501T}. It employs an ideal gas for the nuclei
and additionally includes contributions from ion-ion-correlations and photons.

\subsection{The treatment of neutrino physics}

The treatment of neutrino physics is an essential ingredient of core-collapse supernova
simulations \citep{Mezzacappa2005}.
Multidimensional core-collapse supernova simulations 
therefore must rely on more or less severe approximations of the neutrino physics. 
In state-of-the-art 2D simulations, one way in which Boltzmann transport is approximated 
is the so-called `ray-by-ray plus scheme' \citep{2006A&A...457..281B}. It 
solves the full transport in separate 1D angular segments, 
where the neighbouring rays are coupled. 
Other groups rely on multi-group flux limited diffusion (MGFLD,
\citet{2008ApJ...685.1069O,2009ApJS..181....1S}).
MGFLD treats all neutrinos in seperate energy groups, 
drops the momentum space angular dependence of the radiation
field and evolves the zeroth moment of the specific intensity instead
of the specific intensity itself.
\citet{2008ApJ...685.1069O} also 
compared MGFLD with angle-dependent (i.e. partly Boltzmann) 
transport in 2D.
In three dimensions, simulations have been performed using 
`grey' flux-limited diffusion \citep{Fryer2004}, which oversimplyfies 
the important 
neutrino spectrum. 
It is important to resolve the neutrino spectrum,
since the charge-current interaction rates go with the square of the
neutrino energy.

\begin{figure}
   \centering
    \includegraphics[width=8.8cm]{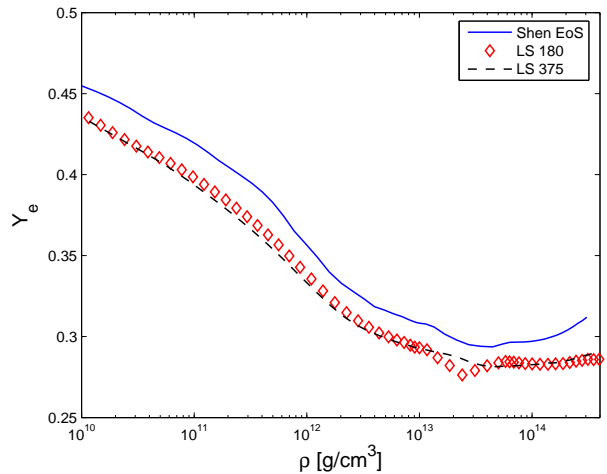}
        \caption{Electron fraction $Y_{e}$ as 
a function of density, obtained from detailed general relativistic,
spherically symmetric three-flavour Boltzmann neutrino transport.
The calculations were carried out using a 
$15M_{\odot}$ progenitor from \citet{Woosley1995}
and a finite temperature EoS of either 
\citet{Lattimer1991} or \citet{Shen1998a}.
In the figure, we denote the LS EoS version 
with a compressibility of 180MeV as LS180,
the one carried out with 375MeV as LS375. 
The Shen EoS is named as such.}
             \label{fig1.eps}
\end{figure}

In our 3D MHD simulations, we apply a
parametrised deleptonisation scheme \citep{Liebendorfer2005}. 
Detailed spherically-symmetric collapse calculations with Boltzmann 
neutrino transport show that the electron fraction $Y_{e}$ in different 
layers in the homologously collapsing core follow a similar deleptonisation 
trajectory with density. The local electron fraction of a fluid 
element can be parametrised as a function of density $Y_{e}(\rho)$.
For our 3D simulations, we apply tabulated $Y_{e}$ profiles which were obtained
from detailed general relativistic, spherically symmetric 
three-flavour Boltzmann neutrino transport.
Exemplary profiles are shown in Fig. \ref{fig1.eps}.
The offset between the $Y_{e}$ profiles from the  
LS- and the Shen EoS is consistent with the different
asymmetry energies of the two EoS.
This in turn is reflected in different neutrino reaction
rates and thus a different $Y_{e}$ at a given density.
These effects have been discussed in \citet{2008ApJ...688.1176S}
and \citet{2009A&A...499....1F} for massive
progenitor stars in the range of 40-50$M_{\odot}$.
However, this parametrisation scheme is only valid until a few milliseconds
after bounce, since it cannot account for the neutronisation burst,
as explained in \citet{Liebendorfer2005} and \citet{2008A&A...490..231S}.

%%%%%%%%%%%%%%%%%%%%%%%%%%%%%%%%%%%%%%%%%%%%%%%%%%%%%%%%%%%%%%%%%%%%%%%

%%%%%%%%%%%%%%%%%%%%%%%%%%%%%%%%%%%%%%%%%%%%%%%%%%%%%%%%%%%%%%%%%%%%%%%

\begin{figure}
   \centering
    \includegraphics[width=8.8cm]{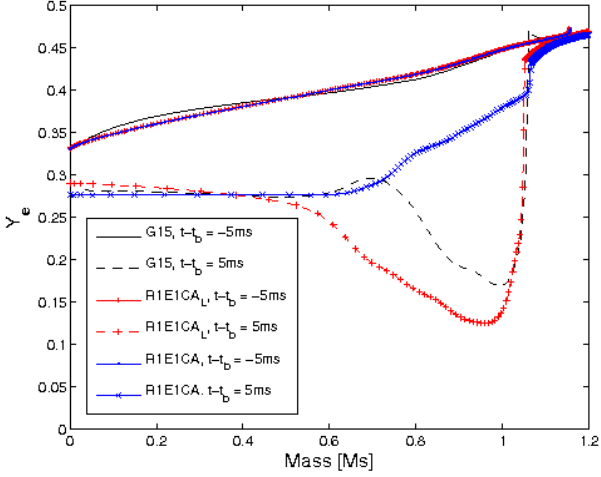}
        \caption{Comparison of the $Y_{e}$ 
profiles of the almost non-rotating 3D models 
R1E1CA and R1E1CA$_{L}$ with 
the spherically symmetric model G15, 
which is based on general relativistic
three-flavour neutrino 
Boltzmann transport \citep{Liebend2005} 
as function of the enclosed mass at 5ms before and after bounce.
Note that while the simulation R1E1CA, which includes only 
the neutrino parametrisation scheme for the collapse phase,
cannot model the neutrino burst, the 
`leakage'-IDSA model R1E1CA$_{L}$ can.}
             \label{fig2.eps}
   \end{figure}

\noindent
In order to give an estimate of how long the above approximation
is able to predict quantitatively reliable postbounce (pb) GW signals for times
$t \gtrsim5$ms after bounce, we carried out 
three representative simulations 
(R1E1CA$_{L}$, R3E1AC$_{L}$ and R4E1FC$_{L}$, see Tab. \ref{table:1})
that incorporate a postbounce treatment for neutrino transport and 
deleptonisation.
The scheme we apply is based on a partial implementation of the
isotropic diffusion source approximation (IDSA; \citet{2009ApJ...698.1174L}). 
The IDSA splits the distribution function $f$ of the neutrinos into two
components, a trapped component $f^{t}$ and a streaming component $f^{s}$,
representing neutrinos of a given species and energy which find the
local zone opaque or transparent, respectively. The total distribution
function is the sum of the two components, $f =  f^{t} + f^{s}$. The two
components are evolved using separate numerical techniques, coupled by a
diffusion source term. The trapped component transports neutrinos by
diffusion to adjacent fluid elements, while in this paper the streaming
component is discarded ($f^s = 0$), implying that these neutrinos are lost from the
simulation immediately. 
The part of the IDSA for trapped neutrinos is implemented in three
dimensions, including both
electron neutrinos and electron anti-neutrinos 
(Whitehouse \& Liebend\"orfer 2010, in preparation).
The use of this partial
implementation of the IDSA enables us to capture the neutrino burst and 
to continue our simulations into
the postbounce regime, as shown in Fig. \ref{fig2.eps}.

%%%%%%%%%%%%%%%%%%%%%%%%%%%%%%%%%%%%%%%%%%%%%%%%%%%%%%%%%%%%%%%%%%%%%%%

%%%%%%%%%%%%%%%%%%%%%%%%%%%%%%%%%%%%%%%%%%%%%%%%%%%%%%%%%%%%%%%%%%%%%%%
\noindent
The leakage scheme is switched on at core bounce, 
when the central density 
of the core reaches its first maximum value.
At present, we neglect the emission 
of $\mu$ and $\tau$ neutrinos and their antineutrinos, 
which in principle play an important role in the cooling 
of the PNS.
However, since our 
leakage schemes 
neglects any absorption of transported neutrinos, it 
already overestimates the cooling of the PNS, 
even without the treatment of the $\mu$ and $\tau$
neutrinos.
This is shown in \ref{fig3.eps}.

\subsection{Initial model configurations}

We construct the initial conditions of our simulations by 
a parametric approach.
All our models are launched from a $15M_{\odot}$ progenitor star 
from the stellar evolutionary calculations of \citet{Woosley1995}. 
Angular momentum was added to the presupernova model according 
to a shell-type rotation law \citep{1985A&A...146..260E}
\begin{equation}
\Omega(r) = \Omega_{i,c}\cdot\frac{A^2}{A^2+r^2} \; , 
\end{equation}
where we define $r = \sqrt{x^2+y^2+z^2}$ as spherical radius, and (x,y,z)
as Cartesian coordinates.
The constant A is the degree of differential rotation that controls the steepness
of the angular velocity profile, $\Omega_{i,c}$ the
initial central rotation rate, and $r$ is the distance from the  origin.
In our whole model set we choose $A=500$km as degree
of differential rotation.

In order to guarantee a divergence-free initial state, 
the initial magnetic field configuration was set up employing
its definition via the vector potential $\vec{A}$.
The components of $\vec{A}$ we chose are

\begin{equation}
\vec{A} = \left(-\frac{B_{pol}}{2}y, \frac{B_{pol}}{2}x , \sqrt{x^2 + y^2}B_{tor} \right)
\label{equ:5}
\end{equation}
In order to mimic a dipole-like field, we scale the vector potential
with density according to
\begin{equation}
\vec{\tilde{A}}=\sqrt{\frac{\rho}{\rho_{ref}}}\vec{A} \; .
\end{equation}
Finally, the magnetic field is derived from this vector potential 
\begin{equation}
\vec{B} = \nabla \times \vec{\tilde{A}} \; .
\end{equation}
The initial toroidal- and poloidal components
of the magnetic field are specified at a reference density of 
$\rho_{ref}=5 \times 10^7$gcm$^{-3}$ according to \citet{Heger2005}.

We compute a total of 25 models, changing the combination of
total angular momentum, the EoS, and toroidal- and poloidal magnetic 
fields. The model parameters are summarised in table \ref{table:1}.
The models are named after the combinations of 
initial central rotation rate, the EoS, and toroidal- and poloidal
magnetic fields.
The first two letters of the model name represent the initial
central rotation rate $\Omega_{c,i}$ [rads$^{-1}$] according to

\vspace{.1cm}
{\small
\begin{tabular}{c|c|c|c|c|c}
0  &  0.3  & $\pi$ & $2\pi$  & $3\pi$ & 4$\pi$\\
\hline
R0 &  R1 & R2 & R3 & R4 & R5
\end{tabular}
},
\vspace{.1cm}

\noindent
the second two letters stand for the applied EoS

\vspace{.1cm}
{\small
\begin{tabular}{c|c|c|c}
LS  &  LS  & LS &  Shen \\
(K = 180MeV) & (K = 220MeV) & (K =325 MeV) \\
\hline
E1 & E2 & E3 & ST
\end{tabular},
}
\vspace{.1cm}
\noindent
while the last two letters assign the order 
of magnitude of the toroidal-, and poloidal
field strength [G] according to 

\vspace{.1cm}

{\small
\begin{tabular}{c|c|c|c|c|c}
$10^6$  & $10^7$  & $10^9$ &  $10^{10}$ & $10^{11}$& $10^{12}$\\
\hline
A & B & C & D & E & F
\end{tabular}
}.

\vspace{.1cm}
\noindent
The final subscript $L$ signs simulations which were 
carried out with the leakage scheme.
Some of the values which we adopt
as rotation rate and magnetic fields correspond 
to the values suggested in \citet{Heger2005}.
However, we point out that some of the initial rotation rates 
($\Omega_{c,i}\gtrsim \pi$rads$^{-1}$)
and magnetic fields ($B\gtrsim 10^{10}$G)
are larger and stronger compared to current predictions from stellar evolution 
calculations.
However, we computed these models 
in order to cover a wide  parameter space for our three-dimensional models 
with neutrino transport approximations.

\begin{table*}
\begin{minipage}[t]{180mm}
\caption{Summary of initial conditions}
\footnote{
The subscript $i$ stands for $initial$, $b$ for $bounce$, 
while $f$ stands for $final$. $\rho_{b}$ is the 
maximum central density at the time of 
core bounce.
$B_{pol,i}$ and $B_{tor,i}$ abbreviate the 
initially imposed toroidal and poloidal magnetic fields, whereas
$E_{m}/|W|$ stands for the ratio of magnetic 
to gravitational energy. The subscript \textit{L} denotes 
models which were carried out with the leakage scheme.}
\renewcommand{\footnoterule}{}  % to avoid a line before footnotes
\label{table:1}
\centering
\begin{tabular}{cccccccccc}
\hline\hline\\
Model & $\Omega_{c,i}$ [rads$^{-1}$] & $\beta_{i}$ & $\beta_{b}$ & EoS & $\rho_{b}[\frac{g}{cm^3}]$& $B_{tor,i}$[G] & $B_{pol,i}$[G] &  ${E_{m}/|W|}_{b}  $ &  ${E_{m}/|W|}_{f}  $  \\
\hline\\
 R0E1CA & $0$ & $0$ & $0$ & E1 &$4.39\times10^{14}$ & $5\times10^9$ & $1\times10^6$ &$1.8\times10^{-9}$& $3.8\times10^{-8}$\\

\hline \\
 R0E3CA & $0$ & $ 0$& $0$ & E3 &$4.17\times10^{14}$ & $5\times10^9$ & $1\times10^6$ & $1.8\times10^{-9}$& $3.1\times10^{-8}$\\

\hline \\
 R0STCA &  $0$ & $ 0$& $0$ & ST & $3.38\times10^{14}$  & $5\times10^9$ & $1\times10^6$ &$1.7\times10^{-9}$& $3.3\times10^{-8}$  \\

\hline \\
 R1E1CA & $0.3$ & $0.59\times10^{-5}$  & $1.7 \times10^{-4}$ & E1 &$4.53\times10^{14}$ & $5\times10^9$ & $1\times10^6$ &$1.7\times10^{-9}$& $3.4\times10^{-8}$\\
   
\hline \\
 R1E3CA & $0.3$ &$0.59\times10^{-5}$  & $1.7 \times10^{-4}$ & E3 &$4.17\times10^{14}$ &$5\times10^9$ & $1\times10^6$&$1.7\times10^{-9}$&$3.6\times10^{-8}$\\
     
\hline \\
 R1E1DB & $0.3$ & $0.59\times10^{-5}$  & $1.7 \times10^{-4}$ & E1 &$4.53\times10^{14}$ & $5\times10^{10}$ & $1\times10^7$  &$1.7\times10^{-7}$& $3.0\times10^{-6}$ \\
    
\hline \\
 R1STCA & $0.3$ & $0.59\times10^{-5}$  & $1.7 \times10^{-4}$ & ST &$3.38\times10^{14}$ & $5\times10^{9}$ & $1\times10^6$&$1.7\times10^{-9}$&$3.2\times10^{-8}$ \\

\hline \\
 R1E1CA$_{L}$ & $0.3$ & $0.59\times10^{-5}$  & $1.8 \times10^{-4}$ & E1 & $4.38\times10^{14}$ &
$5\times10^{9}$ & $1\times10^6$&$6.1\times10^{-9}$&$1.2\times10^{-7}$\\

\hline \\
 R2E1AC & $\pi$ & $ 0.64\times10^{-3}$& $1.6\times10^{-2}$ & E1 &$4.27\times10^{14}$ & $1\times10^6$ & $1\times10^9$ & $6.5\times10^{-9}$ & $1.27\times10^{-5}$\\

\hline \\
 R2E3AC & $\pi$ & $ 0.64\times10^{-3}$& $1.6\times10^{-2}$& E3 &$4.00\times10^{14}$ & $1\times10^6$ & $5\times10^9$&$6.4\times10^{-9}$ &$5.1\times10^{-6}$ \\
  
\hline \\
 R2STAC & $\pi$ &  $ 0.64\times10^{-3}$ &$1.6\times10^{-2}$  & ST &$3.30\times10^{14}$ & $1\times10^6$ & $5\times10^9$ &$6.4\times10^{-9}$&$3.6\times10^{-6}$ \\

\hline \\
 R3E1AC & $2\pi $ & $0.26\times10^{-2}$ & $5.2\times10^{-2}$ & E1 & $3.80\times10^{14}$& $1\times10^6$ & $5\times10^9$ &$8.0\times10^{-9}$&$6.7\times10^{-6}$ \\
     
\hline \\
 R3E2AC & $2\pi$ & $0.26\times10^{-2}$ & $5.1\times10^{-2}$ & E2 & $3.65\times10^{14}$& $1\times10^6$ & $5\times10^9$& $8.9\times10^{-9}$&$7.2\times10^{-6}$  \\
      
\hline \\
 R3E3AC & $2\pi$ & $0.26\times10^{-2}$ & $5.1\times10^{-2}$ & E3 & $3.64\times10^{14}$ & $1\times10^6$ & $5\times10^9$& $9.0\times10^{-9}$&$5.8\times10^{-6}$\\

\hline \\
 R3STAC & $2\pi$ & $0.26\times10^{-2}$ & $5.1\times10^{-2}$ & ST & $3.01\times10^{14}$ & $1\times10^6$ & $5\times10^9$ &$8.9\times10^{-9}$ & $4.7\times10^{-6}$\\

\hline \\
 R3E1CA & $2\pi $ & $0.26\times10^{-2}$ & $5.2\times10^{-2}$ & E1 & $3.82\times10^{14}$& $5\times10^9$ & $1\times10^6$ &$3.1\times10^{-9}$ & $1.4\times10^{-8}$\\
      
\hline \\
 R3E1DB & $2\pi $ & $0.26\times10^{-2}$ & $5.2\times10^{-2}$ & E1 & $3.81\times10^{14}$ &$5\times10^{10}$ & $1\times10^7$&$3.1\times10^{-7}$ & $1.6\times10^{-6}$\\
   
\hline \\
 R3E1AC$_{L}$ & $2\pi $ & $0.26\times10^{-2}$ & $5.1\times10^{-2}$   & E1 & $3.65\times10^{14}$ &$1\times10^{6}$ & $5\times10^9$&$8.7\times10^{-9}$ &$8.6\times10^{-5}$\\

\hline \\
 R4E1AC & $3\pi $ & $0.57\times10^{-2}$ & $8.6\times10^{-2}$ & E1 & $3.22\times10^{14}$ & $1\times10^6$ & $5\times10^9$ &$1.2\times10^{-8}$&$1.3\times10^{-5}$ \\

\hline \\
 R4STAC & $3\pi $ & $0.57\times10^{-2}$ & $9.0\times10^{-2}$ & ST &$2.69\times10^{14}$ & $1\times10^6$ & $5\times10^9$ &$1.5\times10^{-8}$ &$1.4\times10^{-5}$ \\

\hline \\
 R4E1EC & $3\pi $ & $0.57\times10^{-2}$ & $8.6\times10^{-2}$ & E1 & $3.22\times10^{14}$  & $1\times10^{11}$ & $5\times10^9$ &$2.4\times10^{-8}$&$1.3\times10^{-5}$\\
     
\hline \\
 R4E1FC & $3\pi $ & $0.57\times10^{-2}$ &  $8.6\times10^{-2}$ & E1 &  $3.19\times10^{14}$ &$1\times10^{12}$ & $5\times10^9$& $1.3\times10^{-4}$& $1.3\times10^{-4}$ \\
    
\hline \\
 R4E1FC$_{L}$ & $3\pi $ & $0.57\times10^{-2}$ &  $8.7\times10^{-2}$  & E1 &  $3.14\times10^{14}$  & $1\times10^{12}$ & $5\times10^9$&$1.3\times10^{-4}$ & $2.0\times10^{-4}$  \\

\hline \\
 R4E1CF & $3\pi $ & $0.57\times10^{-2}$ &  $8.2\times10^{-2}$  & E1  & $3.22\times10^{14}$   
& $5\times10^{9}$ & $1\times10^{12}$ &$5.8\times10^{-4}$ & $5.9\times10^{-3}$ \\

\hline \\
 R5E1AC & $4\pi $ & $ 1.02\times10^{-2}$ & $10.2\times10^{-2}$ & E1 & $2.47\times10^{14}$ &  $1\times10^{6}$ & $5\times10^9$&$1.3\times10^{-8}$&$1.2\times10^{-5}$\\

\hline
\end{tabular}
\end{minipage}
\end{table*}
%%%%%%%%%%%%%%%%%%%%%%%%%%%%%%%%%%%%%%%%%%%%%%%%%%%%%%%%%%%%%%%%%%%%%%%

\subsection{Gravitational Wave extraction}

The two independent polarisations of the 
dimensionless gravitational wave field $h_{\mu\nu}=g_{\mu\nu}-\eta_{\mu\nu}$
in the transverse traceless gauge are given by 
\begin{equation}
 h_{ij}^{TT}(\textbf{X},t)=\frac{1}{R}(A_{+}e_{+}+A_{\times}e_{\times}) \; .
\label{equ:1}
\end{equation}
The spatial indices $i,j$ run from 1 to 3 \citep{1973grav.book.....M}.
$R$ is the distance from the source to the observer and the unit polarisation
tensors $e_{+}$ and $e_{\times}$
in spherical coordinates are represented as
\begin{eqnarray}
 e_{+} & = &e_{\theta}\otimes e_{\theta}-e_{\phi}\otimes e_{\phi} \;  \\
\label{equ:2}
 e_{\times} & = & e_{\theta}\otimes e_{\phi}+e_{\phi}\otimes e_{\theta} \; .
\label{equ:3}
\end{eqnarray}
In the slow-motion limit \citep{1973grav.book.....M,1990ApJ...351..588F}
the amplitudes $A_{+}$ and $A_{\times}$ are
linear combinations of the second time derivative of the transverse
traceless mass quadrupole tensor $t_{ij}^{TT}$.
In Cartesian coordinates, the quadrupole tensor is expressed as
\begin{equation}
t_{ij}^{TT}=\frac{G}{c^4}\int dV \rho\left[x_{i}x_{j}-\frac{1}{3}\delta_{ij}(x_{1}^2+x_{2}^2+x_{3}^2)\right]^{TT} \; .
\label{equ:4}
\end{equation} 
For convenience we evaluate below the GW amplitudes 
along the polar axis ($\theta=\phi=0$, denoted as subscript I)
\begin{eqnarray}
A_{\rm{+I}} & = & \ddot{t}^{TT}_{xx}-\ddot{t}^{TT}_{yy} \; , \\
A_{\rm{\times I}} & = & 2\ddot{t}^{TT}_{xy} \; ,
\label{equ:5a}
\end{eqnarray}
and in the equatorial plane ($\theta=\frac{\pi}{2}$, $\phi=0$, denoted as II)
\begin{eqnarray}
A_{\rm{+II}} & = & \ddot{t}^{TT}_{zz}-\ddot{t}^{TT}_{yy} \; , \\
A_{\rm{\times II}} & = & -2\ddot{t}^{TT}_{yz} \; .
\label{equ:6}
\end{eqnarray}
Since a direct evaluation of Eq. (\ref{equ:4}) is numerically problematic, as discussed
in \citet{1990ApJ...351..588F}, 
we apply alternative reformulations of the standard quadrupole formula,
in which one or both time derivatives are replaced by hydrodynamic variables using the 
continuity and momentum equations.
In the \textit{first moment of momentum density formula} (see \citet{1990ApJ...351..588F}),
the first time derivative of the 
quadrupole moment yields 
\begin{equation}
\dot{t}_{ij}^{\;TT}=\frac{G}{c^4}\int dV \rho\left[v_{i}x_{j} + v_{j}x_{i} -\frac{2}{3}\delta_{ij}\left(v_{1}x_{1}+v_{2}x_{2}+v_{3}x_{3}\right)\right]^{TT} \; .
\label{equ:7}
\end{equation}
In the \textit{stress formulation} \citet{1990MNRAS.242..289B}, $\ddot{t}^{TT}_{ij}$ is given by  
\begin{equation}
\ddot{t}_{ij}^{\;TT}=\frac{G}{c^4}\int dV \rho\left(2v_{i}v_{j}-x_{i}\partial_{j}\Phi_{eff}-x_{j}\partial_{i}\Phi_{eff}\right)^{TT} \; ,
\label{equ:8}
\end{equation}
where $\Phi_{eff}$ is the gravitational potential.
We apply both ways of computing the GW signal and compare them
by computing the overlap \citep{2009MNRAS.399L.164G} of the resulting waveforms via
\begin{eqnarray}
O(h_{1},h_{2}) & = & \frac{\langle h_{1}|h_{2}\rangle}{\sqrt{\langle h_{1}|h_{1}\rangle \langle h_{2}|h_{2}\rangle}} \; , \\
\langle h_{1}|h_{2}\rangle & = & 4\Re\int_{0}^{\infty}df \frac{\tilde{h_{1}}(f)\tilde{h_{2}}(f)}{S_{h}(f)} \; .
\label{equ:overlap}
\end{eqnarray}
where $S_{h}(\nu)$ [Hz$^{-1}$] is the power spectral density of the strain
noise from a given detector, e.g. LIGO, which were kindly provided
by Shoemaker (2007, private communication) 
The results show
good agreement with numerical deviations of a few percent, as displayed in
Table \ref{table:2}.
There, we computed the overlap of GW trains from the
representative models R1E1CA$_{L}$, R3E1CA$_{L}$ and 
R4E1FC$_{L}$, extracted by using
Eq. \ref{equ:7} and Eq. \ref{equ:8}. We assume 
optimal orientation of detector and source as well
as the GW to be emitted along the polar axis. 
\begin{table}
\begin{minipage}[t]{\columnwidth}
\caption{Overlap}
\centering
\renewcommand{\footnoterule}{}  % to avoid a line before footnotes
\label{table:2}
\begin{tabular}{ccr}
\hline\hline
\\
Model & $\Omega_{c,i}$ [rads$^{-1}$] & Overlap \\

\hline
\\
R1E1AC$_{L}$ & 0.3 &  96\%      \\ 
R3E1AC$_{L}$ & 2$\pi$& 92\%     \\ 
R4E1AC$_{L}$ & 3$\pi$ & 97\%     \\ 
\hline
\end{tabular}
\end{minipage}
\end{table}
Below, we generally apply Eq. \ref{equ:7} if not stated 
otherwise.
Given our spherically-symmetric effective GR approach to the
solution of the Poisson equation, this expression is
physically best motivated, since it does not depend on 
spatial derivatives of the gravitational potential.

We also point out that the Newtonian quadrupole 
formalism to extract the gravitational radiation
is not gauge invariant and only
valid in the Newtonian slow-motion limit. 
However, it was shown by \citet{2003PhRvD..68j4020S} 
that the above method seems to be sufficiently accurate compared to more elaborate
techniques, as it preserves phase while being 
off in amplitude by $\sim10\%$ in neutron star pulsations. 
The energy carried away by gravitational radiation can be calculated by the following expression: 
\begin{eqnarray}
E_{GW} & = & \frac{c^3}{5G}\int \left[\frac{d}{dt}\left(I_{ij}-\frac{1}{3}\delta_{ij}I_{ll}\right)\right]^2 dt \\
&=&\frac{2c^3}{15G}\int dt\left[\dot{I}_{xx}^2+\dot{I}_{yy}^2 + \dot{I}_{zz}^2\right.\nonumber\\
&-& \dot{I}_{xx}\dot{I}_{yy}- \dot{I}_{xx}\dot{I}_{zz}-\dot{I}_{yy}\dot{I}_{zz}\nonumber\\
&+&\left. 3(\dot{I}_{xy}^2+\dot{I}_{xz}^2+\dot{I}_{yz}^2)\right] \; , \nonumber 
%\label{equ:energygw}
\end{eqnarray}
where $I_{ij}=\ddot{t}^{\;TT}_{ij}$. 
The equivalent frequency integral yields
\begin{eqnarray}
E_{GW} & = & \frac{4c^3}{15G}\int_{0}^{\infty}\nu^2d\nu  \left[{\hat{I}}_{xx}^2+{\hat{I}}_{yy}^2 + {\hat{I}}_{zz}^2\right.\nonumber\\
&-& {\hat{I}}_{xx}{\hat{I}}_{yy}- {\hat{I}}_{xx}{\hat{I}}_{zz}-{\hat{I}}_{yy}{\hat{I}}_{zz}\nonumber\\
&+&3\left.\left({\hat{I}}_{xy}^2+{\hat{I}}_{xz}^2+{\hat{I}}_{yz}^2\right)\right] \; , 
\label{equ:9}
\end{eqnarray}
where $\hat{I}_{ij}(\nu)$ is the Fourier transform of the quadrupole amplitude $I_{ij}(t)$.

In order to calculate the contribution to the GW signal due to magnetic stresses,
we generalised Eq. (\ref{equ:8}), taking into account contributions from the magnetic
field. Following the derivations of \citet{Kotake2004a} and \citet{Obergaulinger2006}, the 
Cartesian quadrupole gravitational wave amplitude in the MHD case yields
\begin{equation}
\ddot{t}_{ij}^{\;TT}=\frac{G}{c^4}\int dV \left[2f_{ij}-\rho\left(x_{i}\partial_{j}\Phi_{eff}+x_{j}\partial_{i}\Phi_{eff}\right)\right]^{TT} \; ,
\label{equ:bfieldgw}
\end{equation} 
where $f_{ij}=\rho v_{i} v_{j}-b_{i}b_{j}$, and $b_{i}=B_{i}/\sqrt{4\pi}$.
Below, the GW source is assumed to be 
located at the Galactic centre at $R=10$kpc. 
We estimate the signal-to-noise ratios (SNR) for optimal filtering searches 
according to \citet{1998PhRvD..57.4535F}:
\begin{equation}
 \left(\frac{S}{N}\right)^{2}= 4\int \frac{|\hat{h}(\nu)|^{2}}{S_{h}(\nu)}d\nu=4\int \frac{|\hat{h}_{+}(\nu)|^{2} + |\hat{h}_{\times}(\nu)|^{2}}
{S_{h}(\nu)}d\nu  \; ,
\label{equ:SNR}
\end{equation}
where $\hat{h}(\nu)$ is the Fourier transform of the gravitational wave amplitude,
and $S_{h}(\nu)$ is the power
spectral density of strain noise in the detector.
We assume optimal orientation of
detector and source. and the Fourier spectra were normalised according 
to Parseval's theorem.
%%%%%%%%%%%%%%%%%%%%%%%%%%%%%%%%%%%%%%%%%%%%%%%%%%%%%%%%%%%%%%%%%%%%%%%

\section{Results}

In the subsequent four subsections we will discuss
the GW signature of our 25 models.
While presenting the resulting GW patterns,
we will pay special attention 
to possible imprints of different finite temperature EoS, rotation rates, 
nonaxisymmetric instabilities, magnetic fields and a
postbounce leakage scheme on the predicted
3D GW signals.
The models' initial conditions and similar relevant quantities are
summarised in Table \ref{table:1}, whilst the 
GW data is listed in Table \ref{table:3}.

%%%%%%%%%%%%%%%%%%%%%%%%%%%%%%%%%%%%%%%%%%%%%%%%%%%%%%%%%%%%%%%%%%%%%%%%

\begin{table*}
\begin{minipage}[t]{180mm}
\caption{Summary of GW related quantities} 
\footnote{
$t_{f}$ is the time after core bounce when the 
simulation was stopped. $E_{GW}$ is the total energy released in gravitational radiation. 
We present the maximum amplitudes at different stages of 
their time-evolution in polar (I) and equatorial (II) direction. 
The subscripts $b$ and $pb$ stand for bounce and
postbounce. $f_{b}$ denotes the peak frequency of the GW burst at bounce, 
while $f_{TW}$ stands for the spectral peak from the 
narrow band emission caused by a low $T/|W|$ instability.
The electronic wave forms of all models are available upon request.}
\renewcommand{\footnoterule}{}  % to avoid a line before footnotes
\centering
\begin{tabular}{ccrcrrrrrr}
\hline\hline\\
Model & $t_{f}$[ms]
&$E_{GW} [M_{\odot}c^2]$ & dir. & $|A_{+,b,max}|$ & $|A_{\times,b,max}|$ & 
$|A_{+,pb,max}|$ & $|A_{\times,pb,max}|$& $f_{b} [Hz]$ & $f_{TW}$[Hz]  \\

\hline

\multirow{2}{*}{R0E1CA}& \multirow{2}{*}{130.8} & \multirow{2}{*}{$2.15\times10^{-11}$}  & I  &  3  &  $<1$   & $ 3 $  &  $<1$  & - & -  \\
          &                 &    & II &  2  &  $<1$    & 5    & $1$ & -& - \\

\hline

\multirow{2}{*}{R0E3CA}& \multirow{2}{*}{103.9} & \multirow{2}{*}{$5.72\times10^{-11}$}   & I  & 2   &   $<1$    & 10  & $<1$ & -  & -  \\
          &                &     & II & 2   &   $<1$    & 8 & $<1$ & -& - \\

\hline

\multirow{2}{*}{R0STCA} & \multirow{2}{*}{70.3} & \multirow{2}{*}{$1.35\times10^{-11}$} & I  & 3 &$<1$ & 4 &$<1$ & -&-   \\
         &                  &         & II  & 1 & $<1$    & 3 & $<1$ &-&-\\

\hline 

\multirow{2}{*}{R1E1CA}& \multirow{2}{*}{112.8} & \multirow{2}{*}{$1.36\times10^{-10}$}  & I  & 3   & $<1$    & 12  & 6 & - & - \\
         & &                     & II &  3   & $<$1 & 15  & 1 & -& -  \\

\hline

\multirow{2}{*}{R1E3CA}& \multirow{2}{*}{130.4}  &  \multirow{2}{*}{$1.43\times10^{-10}$} & I  & 2   &  $<1$    & 10  & 5 &- &-\\
          &             &        & II & 3   & $<1$   & 17 & $<1$& - & -\\

\hline

\multirow{2}{*}{R1E1DB}& \multirow{2}{*}{112.8}  & \multirow{2}{*}{$1.24\times10^{-10}$}    & I  &  3  & $< 1$    &  12 & 6 & -  & -\\
          &             &        & II &  3  & $<1$ & 15 & 2 & - & -  \\

\hline

\multirow{2}{*}{R1STCA}& \multirow{2}{*}{45.8}    &\multirow{2}{*}{$2.01\times10^{-11}$}    & I  &  2  &  $ <$1   &  3 &1  & -  & -\\
          &             &        & II &  2  & $<$1 & 6 &$<$1  &- &- \\

\hline

\multirow{2}{*}{R1E1CA$_{L}$} & \multirow{2}{*}{92.9}  & \multirow{2}{*}{$1.04\times10^{-10}$}  & I  & 3 & $<1$  &4 & 2 & -  & -   \\
          &            &         & II & 2 &$<$1 &  10 & 1 & -  & -
                                                          \\
\hline

\multirow{2}{*}{R2E1AC} & \multirow{2}{*}{127.3}     & \multirow{2}{*}{$5.52\times10^{-9}$}      & I  &  2  &  2    & 6 &5 & -& -\\
          &             &         & II & 105 & $<$1  & 1  &3 & 841 & - \\

\hline
\multirow{2}{*}{R2E3AC} & \multirow{2}{*}{106.4}    &  \multirow{2}{*}{$5.31\times10^{-9}$}      & I  &  2  &  2    & 4  & 4 & - & - \\
          &             &         & II & 104 & $<$1  &  1 &$<1$ & 803 & - \\

\hline
\multirow{2}{*}{R2STAC} &  \multirow{2}{*}{64.0}     &  \multirow{2}{*}{$7.62\times10^{-9}$}     & I  &  3  &  3   & 6 &6 & - & -  \\
          &             &         & II & 133  & $<$1  & 2   & 1& 680 & - \\

\hline

 \multirow{2}{*}{R3E1AC} & \multirow{2}{*}{62.5}    & \multirow{2}{*}{$5.58\times10^{-8}$}        & I  &  2   &  1   &  15 & 16 & - &725\\ 
           &            &         & II &  393 & $<1$ &  10 & 2 & 880 & - \\

\hline

\multirow{2}{*}{R3E2AC} & \multirow{2}{*}{45.5}    &  \multirow{2}{*}{$6.53\times10^{-8}$}        & I &   2  &    1 &  11  & 13 & -  &     \\
           &            &         & II&  409 &   $<1$&  5  &   3  & 882 & -                    \\
\hline

\multirow{2}{*}{R3E3AC} & \multirow{2}{*}{57.5}    &  \multirow{2}{*}{$6.44\times10^{-8}$}       & I &  2   &    2  &   6    & 9  &- &  \\
           &            &         & II&  409 &   $<1$&   6    & 3 & 891 & - \\   

\hline

\multirow{2}{*}{R3STAC}  & \multirow{2}{*}{50.2} & \multirow{2}{*}{$1.05\times10^{-7}$} & I   &  5    &   4   &  9  & 11  &- & 935  \\
           &  &                   & II  &  526  &  $<1$ &   5       &  3 & 854 & - \\

\hline

\multirow{2}{*}{R3E1CA}  & \multirow{2}{*}{69.4} & \multirow{2}{*}{$8.05\times10^{-8}$} & I  &  3  &  3   & 9  & 9 &-&-\\
          &     &                & II &  426  & $<1$ &5 &3 &897 & -\\
\hline

\multirow{2}{*}{R3E1DB}  & \multirow{2}{*}{62.7}  &  \multirow{2}{*}{$7.76\times10^{-8}$} & I  &  8  &   8   & 13 & 11 & -  &- \\
          &       &              & II &  425  & $<1$&8 & 3 & 886& -\\
\hline

\multirow{2}{*}{R3E1AC$_{L}$} & \multirow{2}{*}{196.7}  & \multirow{2}{*}{$2.14\times10^{-7}$}  & I  & 1 & 1  & 136  & 136  & -  & 909 \\
          &       &              & II & 437   &$<1$  & 70 & 9 & 909& -  \\

\hline

\multirow{2}{*}{R4E1AC}& \multirow{2}{*}{98.7}    & \multirow{2}{*}{$7.74\times10^{-8}$}   & I  &  1   &  $<1$   & 37  & 37 & -& 662 \\
          &         &                           & II & 512  &  $<1$  &20 & 6 &385 & - \\

\hline

\multirow{2}{*}{R4STAC} &  \multirow{2}{*}{67.2}   & \multirow{2}{*}{$1.91\times10^{-7}$}   & I  & 2    &  1  & 102 & 109 & -&  902\\
          &        &            & II &  536   & $<1$    & 53 & 3 & 396 & - \\

\hline

\multirow{2}{*}{R4E1EC}& \multirow{2}{*}{100.8}    &  \multirow{2}{*}{$7.51\times10^{-8}$}  & I  & 1   &  1    & 25  & 20 & -  &611\\
          &      &               & II &  492  & $<1$ & 15 & 7 & 866 &  \\

\hline

\multirow{2}{*}{R4E1FC}& \multirow{2}{*}{80.1}    & \multirow{2}{*}{$7.29\times10^{-8}$} & I  &  $<1$    &  $<1$   & 16  & 19 &  - & 828\\
          &        &             & II &  516  & $<1$ & 9 & 3 & 859 & - \\

\hline

\multirow{2}{*}{R4E1FC$_{L}$}    & \multirow{2}{*}{97.6} & \multirow{2}{*}{$3.42\times10^{-7}$}   & I  &  2  &  1   & 316  & 298 &  - &673 \\
          &        &             & II & 536 & $<1$ & 157 & 6 & 485 & - \\
\hline

\multirow{2}{*}{R4E1CF}    & \multirow{2}{*}{19.1} & \multirow{2}{*}{$6.50\times10^{-8}$}   & I  &    1&   1  & 4 &4 & - & \\
          &        &             & II & 518 & $<1$ & ring-down & 2 & 370 & - \\

\hline

\multirow{2}{*}{R5E1AC} & \multirow{2}{*}{93.2}   & \multirow{2}{*}{$1.20\times10^{-8}$}   & I  &  1 &  1  &  18 & 18 &-& 727 \\
          &          &           & II &  238   & $<1$ & 10 & 11 & 317 &- \\

\hline

\end{tabular}
\label{table:3}
\end{minipage}
\end{table*}

%%%%%%%%%%%%%%%%%%%%%%%%%%%%%%%%%%%%%%%%%%%%%%%%%%%%%%%%%%%%%%%%%%%%%%%

\subsection{Non- or slowly rotating core collapse}

%%%%%%%%%%%%%%%%%%%%%%%%%%%%%%%%%%%%%%%%%%%%%%%%%%%%%%%%%%%%%%%%%%%%%%%

\subsubsection*{General remarks}

\begin{figure}
   \centering
    \includegraphics[width=8.8cm]{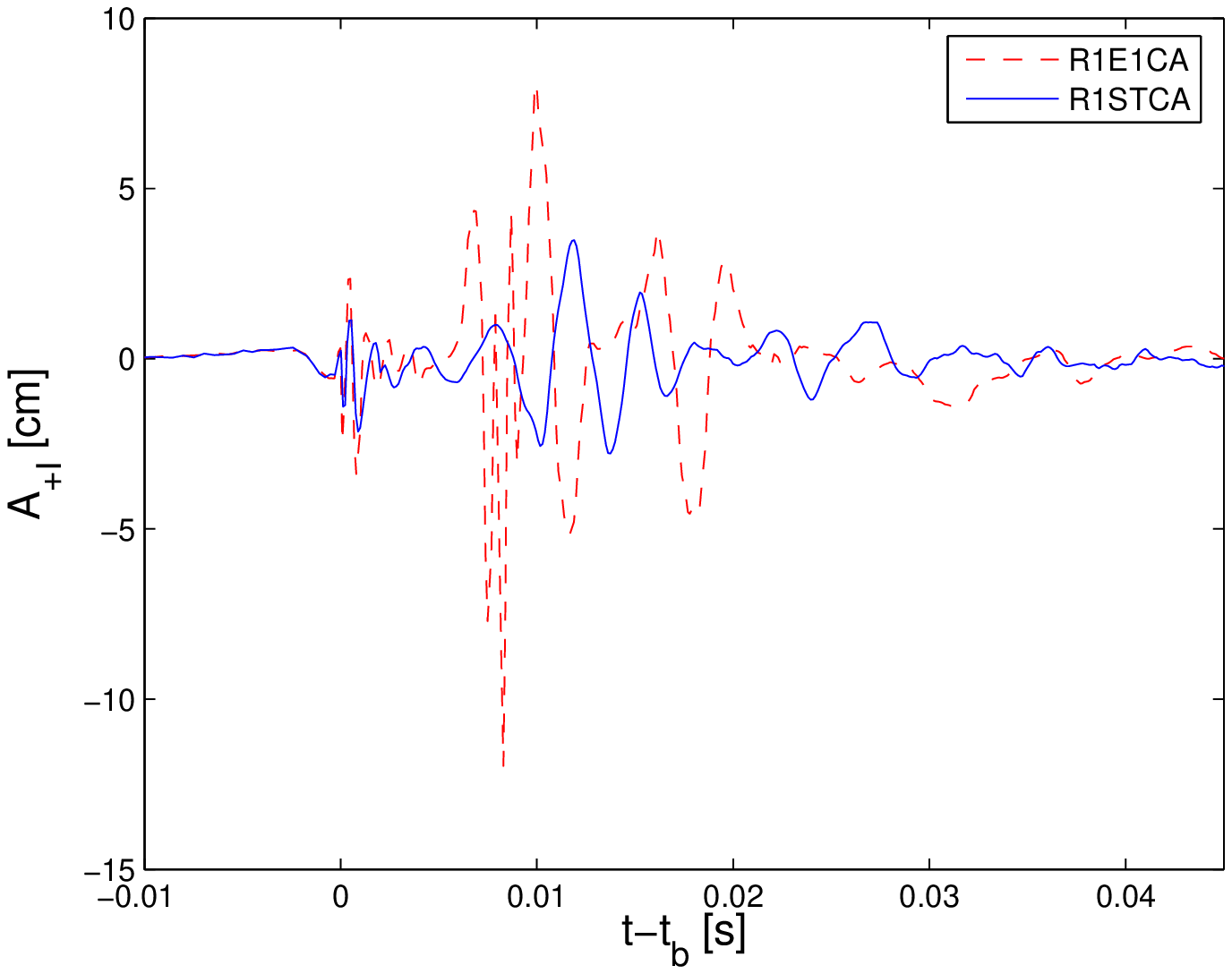}
        \caption{
Time evolution of the GW amplitude A$_{\rm{+I}}$ from the 
slowly rotating models R1STCA (full line) and 
R1E1CA (dashed line). Since convection is a stochastic 
process, the GW amplitudes are rather insensitive 
to the location of the observer. Hence, we display only
one representative polarisation.}
             \label{fig3.eps}
   \end{figure}

Non- and slowly rotating progenitors 
($\Omega_{c,i}=0\ldots0.3$rads$^{-1}$
in our model set)
all undergo quasi-spherically symmetric core collapse.
As the emission of GWs intrinsically depends on 
dynamical processes
that deviate from spherical symmetry, 
the collapse phase in our models (that 
neglect inhomogenities in the progenitor star)
therefore does not provide any kind of signal, as shown 
in Fig. \ref{fig3.eps}
for $t-t_{b}< 0$.
However, subsequent pressure-dominated core 
bounce, where
the collapse is halted due to the stiffening 
of the EoS 
at nuclear density $\rho_{nuc}\approx 2\times 
10^{14}$gcm$^{-3}$, 
launches a shock wave that plows through the 
infalling material, 
leaving behind a negative entropy gradient that 
induces so-called `prompt'
convective activity 
(e.g. \citet{M2004,Buras2006,2008PhRvD..78f4056D,
2009A&A...496..475M,2009CQGra..26f3001O}).
The GW burst which is accompanied by such aspherities 
starts several ms after bounce when convective 
overturn starts to be
effective.
The criterion for convective instability 
(the `Ledoux condition') is generally 
expressed as \citep{1959flme.book.....L,1988PhR...163...63W}
\begin{equation}
 \left(\frac{\partial\rho}{\partial Y_{e}}\right)_{P,s} 
\left(\frac{\partial Y_{e}}{\partial r}\right) +  
\left(\frac{\partial\rho}{\partial s}\right)_{P,Y_{e}} 
\left(\frac{\partial s}{\partial r}\right) >0 \; , 
 \label{equ:Ledoux}
\end{equation}

\noindent
where $Y_{e}$, $\rho$, $r$ and $s$ are electron fraction, density, stellar 
radius and entropy per baryon respectively.
Thermodynamic consistency requires $(\partial\rho/\partial s)_{Y_{e},P}<0$, which implies that a 
negative entropy gradient always acts in a destabilising manner.
Additionally, the neutronisation burst, occurring some $\sim5$ms after bounce, 
causes a negative lepton gradient at the edge of the PNS which further drives convection 
(cf. Fig. 13 of \citet{Sumiyoshi2004}, \citet{Buras2006,2006ApJ...645..534D}).
Note that convection is be weakened in the rotational plane by positive specific angular
momentum gradients in rotating cores \citep{1978ApJ...220..279E}.

%%%%%%%%%%%%%%%%%%%%%%%%%%%%%%%%%%%%%%%%%%%%%%%%%%%%%%%%%%%%%%%%%%%%%%%

%%%%%%%%%%%%%%%%%%%%%%%%%%%%%%%%%%%%%%%%%%%%%%%%%%%%%%%%%%%%%%%%%%%%%%%

\subsubsection*{Models without deleptonisation in the postbounce phase:
Effects of the EoS and magnetic fields on the GW signature}

%%%%%%%%%%%%%%%%%%%%%%%%%%%%%%%%%%%%%%%%%%%%%%%%%%%%%%%%%%%%%%%%%%%%%%%

  \begin{figure}
   \centering
    \includegraphics[width=8.8cm]{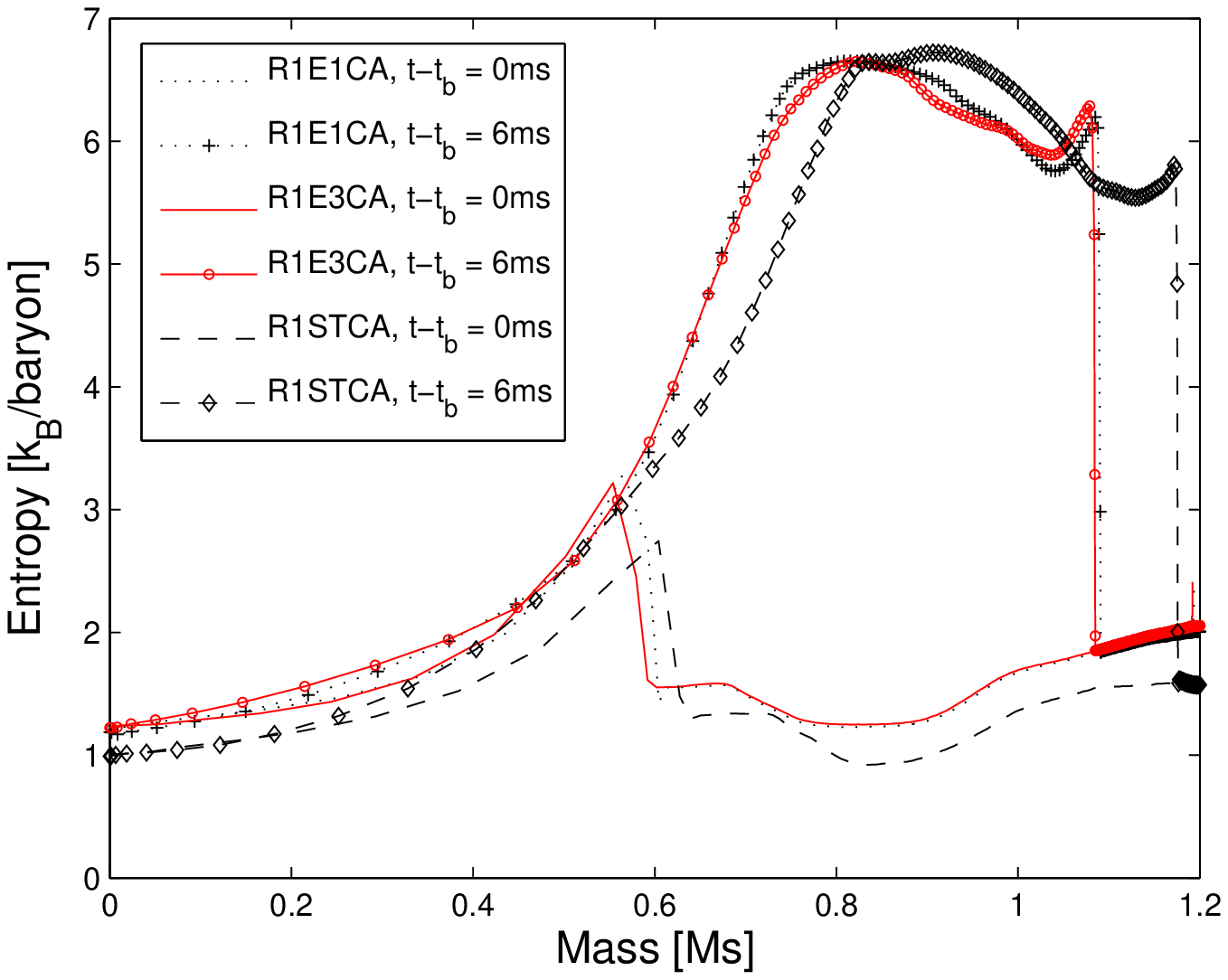}
        \caption{Spherically averaged density profiles from the 
slowly rotating models R1E1CA (red full line), R1E3CA (black dotted 
line) and R1STCA (black dashed line) at bounce.
The second entropy time slice is chosen to be approximately 
at the onset of 
the GW signal from prompt convection.}
             \label{fig4.eps}
   \end{figure}

%%%%%%%%%%%%%%%%%%%%%%%%%%%%%%%%%%%%%%%%%%%%%%%%%%%%%%%%%%%%%%%%%%%%%%%
  \begin{figure}
   \centering
    \includegraphics[width=8.8cm]{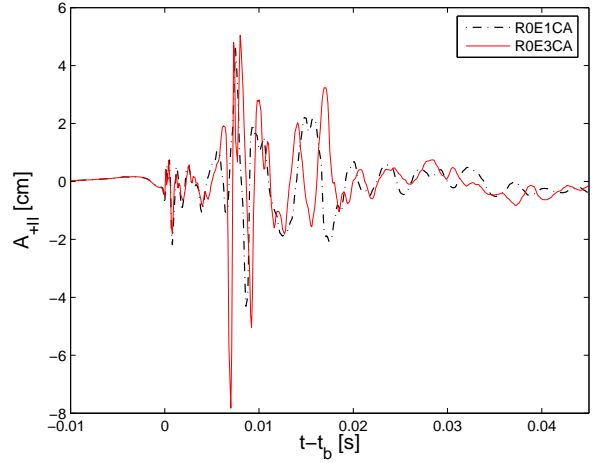}
        \caption{Time evolution of the 
GW amplitude A$_{\rm{+II}}$ from the 
non-rotating models R0E1CA (dashed) and 
R0E3CA (full line).}
             \label{fig5.eps}
   \end{figure}

%%%%%%%%%%%%%%%%%%%%%%%%%%%%%%%%%%%%%%%%%%%%%%%%%%%%%%%%%%%%%%%%%%%%%%%

During the early postbounce stage ($t-t_{b}\lesssim20 -30$ms), 
prompt convective motion is predominantly driven by a 
negative entropy gradient, 
as pointed out e.g. by 
\citet{2009arXiv0912.1455S} 
(see also \citet{2008PhRvD..78f4056D,2009A&A...496..475M,2009CQGra..26f3001O}).
The GW burst to be associated with 
prompt convection sets in $\sim6$ms 
after bounce in models based on the 
LS EoS, generally
$\sim2-3$ms before the same feature 
occurs in the corresponding simulations
using the Shen EoS, as indicated in
Fig. \ref{fig3.eps}. The reason for this 
behaviour is that the shock wave in the 
`Shen'-models carries more energy compared 
to those in models using a LS EoS.
Therefore, the shock wave stalls at slightly 
later times at larger radii (see Fig. \ref{fig4.eps}), 
and the conditions for convective activity are 
delayed compared to the LS runs.
Note that the convective overturn causes a smoothing of
the negative entropy gradient.
As a result, the GW amplitude quickly decays
($t\gtrsim30$ms after bounce)
and is not revived during the later evolution
of the models without 
deleptonisation in the postbounce phase, as 
displayed in Figs. \ref{fig3.eps} and \ref{fig5.eps}.
Simulations that incorporate the Shen 
EoS return up to a factor of 2 smaller maximum 
amplitudes compared to their counterparts, 
as can be deduced from Fig. \ref{fig3.eps} and 
Table \ref{table:3}. 

\begin{figure}
   \centering
    \includegraphics[width=8.8cm]{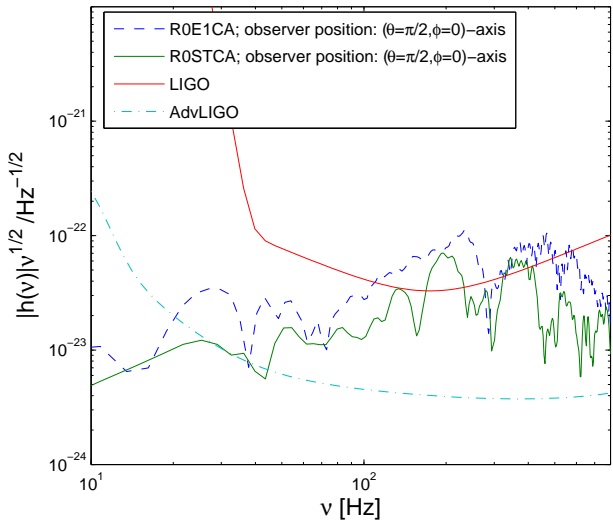}
        \caption{Spectral energy distribution from the 
models R0E1CA (dashed line) and R0STCA (full line) 
for a spectator in the equatorial plane at a distance of 10kpc
compared with the LIGO strain sensitivity 
(Shoemaker 2007, private communication) 
and the planned performance of Advanced LIGO.
Optimal orientation between source and detector is assumed.}
             \label{fig6.eps}
   \end{figure}

We also find that 
the GWs from simulations which were 
carried out with the stiff E3 show no significant
deviations from those computed with 
E1 (see Table \ref{table:3} and Fig. \ref{fig5.eps}). 
The minor deviations of the GWs are entirely
due to the stochastic nature of convection.
The waveform spectra from the LS models 
cover a broad frequency
band ranging from $\sim150 - 500$Hz, as 
displayed in Fig. \ref{fig6.eps}.
The spectral peak of the Shen models is shifted
somewhat to lower frequencies, covering a 
range from $\sim150-350$Hz (see Fig. \ref{fig6.eps}).
When comparing the energy $E_{GW}$ which is
emitted during the first 30ms after bounce 
from the LS models to that of the corresponding
Shen models, we find the latter models emit
less energy (see Table \ref{table:3}). 
This discrepancy is due to 
the lower emission at higher frequency $\nu \gtrsim 350$Hz
in the Shen models and the fact that the emitted
energy $dE_{GW}/d\nu$ is proportional to $\nu^2$.
Moreover we find that already slow rotation (rotation 
rate R1) leads to a deformation of the PNS. 
This can be quantified by considering
e.g. a density cut at 
$\rho=10^{11}$gcm$^{-3}$, 
which is just 
inside the convectively 
unstable region.
For the slowly rotating model R1E1CA, 
this point is located at a radial 
distance of 80km along the polar axis at
20ms after bounce. 
However, in the equatorial plane, 
the same position 
is reached one radial grid zone 
further out relative to the origin
due to the action of centrifugal forces.
The short time variation in the quadrupole 
due to rotation combines with that 
of prompt convection and together they lead
to somewhat stronger GW emission 
in the slowly rotating case compared to 
the non-rotating model set.
This effect is strongest for the GW amplitude 
A$_{\bf{+II}}$, which is the `axisymmetric' 
(l=2, m=0) component of the wave field.
Despite this feature, the frequency content
of models which only differ in rotation rate, 
stays practically unaltered.

\begin{figure}
   \centering
    \includegraphics[width=8.8cm]{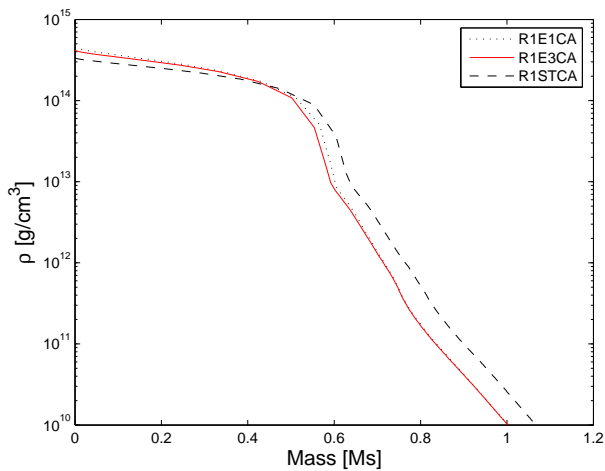}
        \caption{Spherically averaged density profiles from models
R1E1CA (dotted line), R1E3CA (full line) and R1STCA (dashed line) 
at core bounce ($t-t_{b}=0$).}
             \label{fig7.eps}
   \end{figure}

We found the key controlling factors that govern
the GW emission from prompt convection
to be the i) radial location
of the convectively unstable zones and 
ii) the related characteristic dynamical 
timescales involved, 
for which we use as rough estimate 
$t_{dyn} \sim \Delta_{r}/\overline{c_{s}}$
\footnote{$\overline{c_{s}}=1/\Delta_{r}\int_{r}c_{s}(r)dr$
is the radially averaged sound speed of a convectively unstable
layer with a radial extension of $\Delta_{r}$.}
(Ott 2009, private communication).
Note that both i) and ii) are implicitly determined 
by the applied EoS, being responsible e.g. for the
local speed of sound and the radial PNS density profile.
Our core-collapse simulations that use a 
version of the LS EoS (E1 or E3) 
show at core bounce maximum central 
densities up to $\sim25\%$ higher
than the corresponding models that apply 
the Shen EoS (see Tab.\ref{table:1}). 
Moreover, the LS-models possess a PNS which is more strongly 
condensed in central regions and has a steeper density gradient
further out.
The densities of e.g. 
R1E1CA and R1STCA intersect at $\sim0.5M_{\odot}$ 
($\sim 8$km), as displayed 
in Fig. \ref{fig7.eps}.
In addition, models using the E1 show somewhat 
higher central densities compared to the ones 
applying the stiffer
E3 variant, and densities of e.g. R0E1CA and R0E3CA 
cross at $\sim7$km.
Nevertheless these two variants of the LS 
EoS differ only in their
compressibility. The radial structures
resemble each other strongly, unlike
to the models carried 
out using the Shen EoS, as shown in Fig. \ref{fig7.eps}.
Hence, the particular similarity in the GW characteristics 
of LS runs (see Fig. \ref{fig5.eps}) is not unexpected
since the region which is convectively unstable is 
restricted to roughly the same radial- and density regimes 
($R\approx30$-$70$km, $\rho\approx 10^{11}-1.1\times10^{12}$ gcm$^{-3}$)
and therefore bound to the same dynamical 
timescale $t_{dyn}$
and its corresponding frequency band.
For the Shen models on the other side 
the region of negative entropy gradient, 
whose boundaries are radially constrained 
by $R\approx60$-$90$km, 
contains considerably less
matter than its LS analogues. 
This explains the smaller GW amplitudes.
A narrower density spread 
($\rho\approx 7\cdot10^{10}-1.2\times10^{11}$ gcm$^{-3}$) 
in the unstable region 
leads to a more restricted peak frequency band at lower values 
(see Fig. \ref{fig4.eps}).

%%%%%%%%%%%%%%%%%%%%%%%%%%%%%%%%%%%%%%%%%%%%%%%%%%%%%%%%%%%%%%%%%%%%%%%%

%%%%%%%%%%%%%%%%%%%%%%%%%%%%%%%%%%%%%%%%%%%%%%%%%%%%%%%%%%%%%%%%%%%%%%%%

Computing the SNRs
we find that all the simulations
discussed above lie just below the detector 
limits of LIGO if we assume them to be located
at a Galactic distance of 10kpc. Their 
single-detector optimal-orientation
SNR is just a little above unity.
However, for a successful detection, 
at least a SNR of 7 to 8 
is necessary.
Note that the SNRs of models with different
EoS do not differ much at this stage since all
share a similar spectral energy distribution 
within the window of LIGO's maximum sensitivity.
The current detector sensitivity does not allow
for the detection of the high frequency tail 
of the LS models. 
Note, however, that for planned future detectors
such as the Advanced LIGO facility, things change
dramatically. As a direct consequence, these new
detectors would permit the distinction between 
the prompt convection GW signal from the LS and
Shen EoS, since the full spectral information would
be available.
However, we find it impossible to 
discriminate between 
the different LS EoS variants.
Hence, our simulations indicate that the 
GW signature depends more strongly 
on the asymmetry energy than the 
compressibility parameter of the EoS.

These results are partly different from those 
previously published.
Recently, \citet{2009A&A...496..475M} 
as well as \citet{2009CQGra..26f3001O}
reported to observe GW from
prompt convection in state-of-the-art 2D 
simulations which were launched 
from similar initial conditions as ours, 
namely the same
progenitor star and the soft variant of 
the LS EoS (K=180MeV).
Whilst the extracted 
GW amplitudes from early prompt convection 
of \citet{2009A&A...496..475M} 
(cf. their model M15LS-2D)   
are in rough agreement with our results,
the spectrum of their wave train peaks at considerably 
lower frequencies, namely around about $\sim100$Hz.
We suppose that this discrepancy is due 
mainly to different radial locations of the 
unstable regions
and the consequently encompassed amount of overturning 
matter.
\citet{2009CQGra..26f3001O} 
computed two models with different resolution.
While one of their models (s15WW95) in particular fits 
our results well in all characteristic GW 
features, namely the size of amplitudes, 
band of emission and the amount of emitted
energy, the better resolved model (s15WW95HR)
showed that convection is much weaker 
due to less seed perturbations and hence the GW signal and the 
total amount of emitted 
energy considerably lower. 
We recently also tested this issue with a better resolved 
model (cf. model R1$_{HR}$ of \citet{2009arXiv0912.1455S}, 
with a grid spacing of 0.6km). 
This better resolved model showed considerably smaller seed
perturbations around $t-t_{b} \sim 0$, as grid alignment 
effects are better suppressed at core bounce; 
hence prompt convection then is much weaker and a
smaller GW amplitude ($\sim 50$\%) is emitted.
However, better numerical resolution also leads 
to less numerical dissipation in the system, which eases 
the dynamical effects that follow. Thus, 
\citet{2009arXiv0912.1455S} found for 
$\sim 10 \lesssim t \lesssim 20$ms 
considerably stronger GW emission from early 
prompt convection compared to 
the 1km resolved models.

Finally, we find that imposing initial 
magnetic fields ten times as strong as 
the values suggested in
\citet{Heger2005} does not influence the model dynamics
and therefore the GW signal at all, 
as can be deduced from Table \ref{table:3} 
by a comparison of models R1E1CA and R1E1DB.

%%%%%%%%%%%%%%%%%%%%%%%%%%%%%%%%%%%%%%%%%%%%%%%%%%%%%%%%%%%%%%%%%%%%%%%

\subsubsection*{Model with deleptonisation in the postbounce phase}

%%%%%%%%%%%%%%%%%%%%%%%%%%%%%%%%%%%%%%%%%%%%%%%%%%%%%%%%%%%%%%%%%%%%%%%

  \begin{figure*}
   \centering
    \includegraphics[width=12cm]{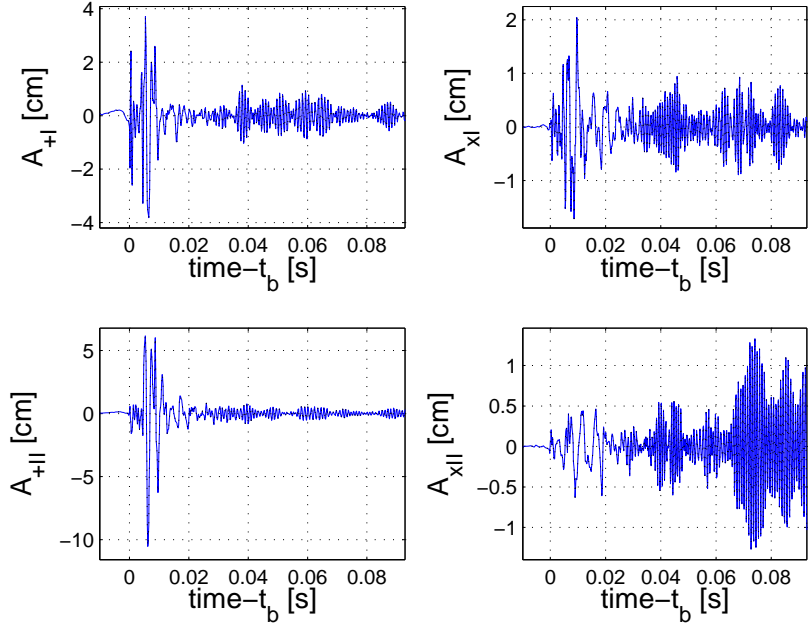}

        \caption{Model R1E1CA$_{L}$'s 
time evolution of the quadrupole 
amplitudes A$_{\rm{+I}}$, A$_{\rm{xI}}$, A$_{\rm{+II}}$, 
and A$_{\rm{xII}}$.}
             \label{fig8.eps}
   \end{figure*}
%%%%%%%%%%%%%%%%%%%%%%%%%%%%%%%%%%%%%%%%%%%%%%%%%%%%%%%%%%%%%%%%%%%%%%%

  \begin{figure}
   \centering
    \includegraphics[width=8.8cm]{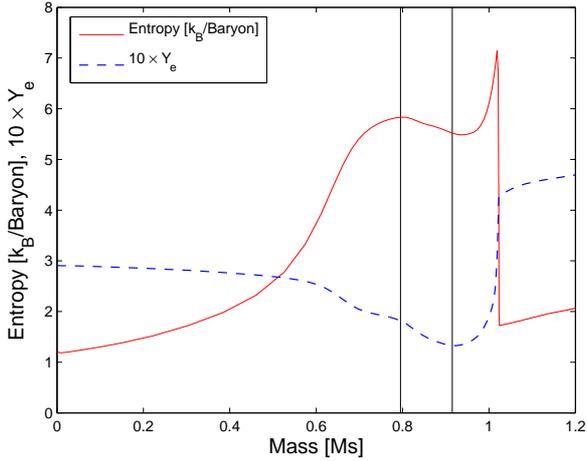}
        \caption{Model R1E1CA$_{L}$'s spherically 
averaged specific entropy (full line)
and 10 $\times$ electron fraction Y$_{e}$ (dashed line) profiles
are plotted versus the enclosed mass $\sim 5$ms after bounce.
The radial position of simultaneous negativ entropy- and lepton 
gradients are marked by
the vertical lines. }
             \label{fig9.eps}
   \end{figure}
%%%%%%%%%%%%%%%%%%%%%%%%%%%%%%%%%%%%%%%%%%%%%%%%%%%%%%%%%%%%%%%%%%%%%%%

  \begin{figure}
   \centering
    \includegraphics[width=8.8cm]{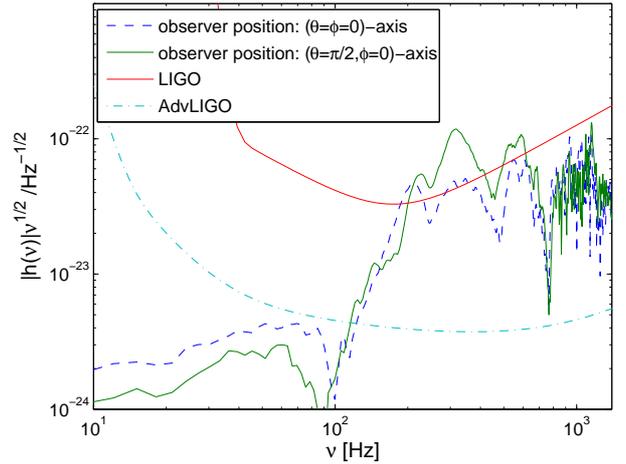}
        \caption{Model R1E1CA$_{L}$'s spectral 
energy distribution of the GW signals 
at a distance of 10kpc.
Note that the spectrum interval from 
$\sim 150-500$Hz is caused by prompt 
convective activity, while the higher modes
result from lepton gradient driven postbounce PNS convection.}
             \label{fig10.eps}
   \end{figure}

%%%%%%%%%%%%%%%%%%%%%%%%%%%%%%%%%%%%%%%%%%%%%%%%%%%%%%%%%%%%%%%%%%%%%%%

Due to the absence of accurate postbounce neutrino transport,
it is unclear how reliably the models discussed in 
the previous subsection
predict the GW signals from the early 
postbounce period.
In order to investigate this question, we carried out 
one computationally expensive 
simulation, model R1E1CA$_{L}$, 
that includes the emission of neutrinos
after bounce but neglects the neutrino heating, which
becomes relevant at $t-t_{b}\gtrsim50$ms.
The absence of the latter makes it impossible 
for this model to track the long-term postbounce
neutrino-driven convection.
However, in comparison with purely MHD models, 
the inclusion of neutrino leakage makes is possible to distinguish 
effects on the GW signal due to entropy
and lepton-gradient driven convection.

A detailed comparison of model R1E1CA$_{L}$ with its
purely magnetohydrodynamical counterpart R1E1CA 
shows that both follow a similar dynamical
behaviour until about 20ms after bounce.
Asphericities leading to GW emission 
are predominantly driven by entropy- and not lepton-induced
convection in 
this supernova stage.
Consequently, the wave trains emitted within 
this early period
fit each other qualitatively 
(cf. Fig. \ref{fig3.eps} and Fig. \ref{fig8.eps}).
However, we find some quantitative deviations:
The GWs of R1E1CA$_{L}$ reach lower 
maximum values 
(see Table \ref{table:3}), as 
the `Ledoux'
unstable region encompasses less mass 
(Fig. \ref{fig9.eps}), and
the presence of neutrino cooling 
leads to a more rapid smoothing of the 
entropy gradient compared to the models
discussed in the previous subsection.
However, since the overturning 
matter in the top layers of the PNS has 
the same radial position, densities 
and dynamical timescales 
as in model R1E1CA, these models have  
similar GW spectra, peaking between $\sim150 - 500$Hz
(cf. Fig. \ref{fig6.eps} and Fig. \ref{fig10.eps}).
Therefore the physically simpler models still
provide reasonably accurate GW predictions in
frequency space
until about 20ms after bounce, 
although the amplitudes are overestimated 
a few $\times 10$\% 
(cf. Figs. \ref{fig3.eps} and \ref{fig8.eps}).
Model R1E1CA$_{L}$'s later postbounce evolution 
($t\gtrsim 20$ms) differs strongly
compared to its purely hydrodynamical 
counterpart R1E1CA.
A negative radial lepton gradient, caused by
the neutronisation burst and 
subsequent deleptonisation, drives convection
inside the lower layers of the PNS 
\citep{2006ApJ...645..534D}
at a radial position
of $\sim$ 10-30km and a density range of
$\sim 10^{12}-10^{14}$gcm$^{-3}$ and therefore
causes now the entire GW emission.
The cooling PNS contracts with time, which causes 
the convective zones to migrate towards smaller
radii and shrink. However, we point out that 
our leakage scheme overestimates the neutrino cooling 
processes, as shown in Fig. \ref{fig2.eps}.
Hence, this mechanism proceeds too quickly 
for model R1E1CA$_{L}$ (Ott 2009, private communication).
The PNS convection exhibits GW emission of roughly 
$\sim 0.5-1$cm amplitude,
as can be seen in Fig. \ref{fig8.eps} ($t \gtrsim 20$ms).
The corresponding spectral distribution is shown 
in Fig. \ref{fig10.eps}.
A broad peak rises between $\sim700-1200$Hz 
and reflects the dynamical timescale $t_{dyn}$ of the 
violent overturn activity of a millisecond scale 
inside the PNS.
The model's SNR for LIGO at 10kpc is again 
around unity.
The high frequency tail of the spectrum 
($\gtrsim 700$Hz),
which is present due to PNS convection, 
cannot contribute to the SNR as 
it lies below the current 
detector sensitivity.
Our computed GW strains for 
PNS convection agree roughly 
in amplitude with the ones found 
in \citet{2009CQGra..26f3001O} 
for axisymmetric MGFLD models. 
However, the amount of released 
energy emitted is found to be about one order of 
magnitude higher compared to his simulations.
This discrepancy is most likely due to 
the lower average frequency content of the GWs 
in Ott's model ($\sim 350$Hz, see his Fig. 7), 
as $dE_{GW}/df\propto \nu^2$.
We suppose the reason for this mismatch to 
be the different radial 
location of the convectively unstable region
and thus the different 
related dynamical timescale $t_{dyn}$.

\subsection{Rapidly rotating core collapse}

%%%%%%%%%%%%%%%%%%%%%%%%%%%%%%%%%%%%%%%%%%%%%%%%%%%%%%%%%%%%%%%%%%%%%%%%
\subsubsection{Core bounce}

%%%%%%%%%%%%%%%%%%%%%%%%%%%%%%%%%%%%%%%%%%%%%%%%%%%%%%%%%%%%%%%%%%%%%%%%

 \begin{figure*}
   \centering
 \includegraphics[width=8.8cm]{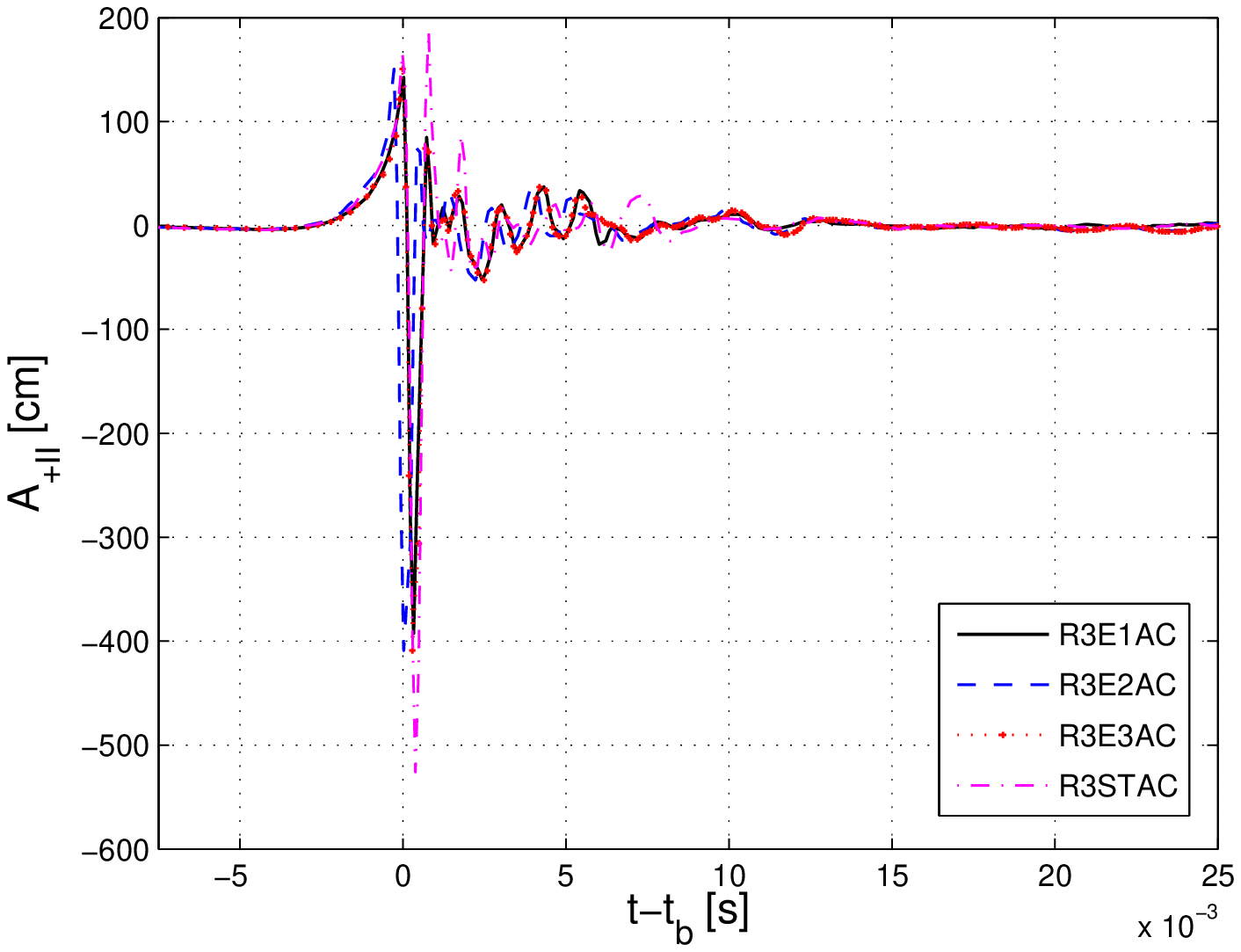}
  \includegraphics[width=8.8cm]{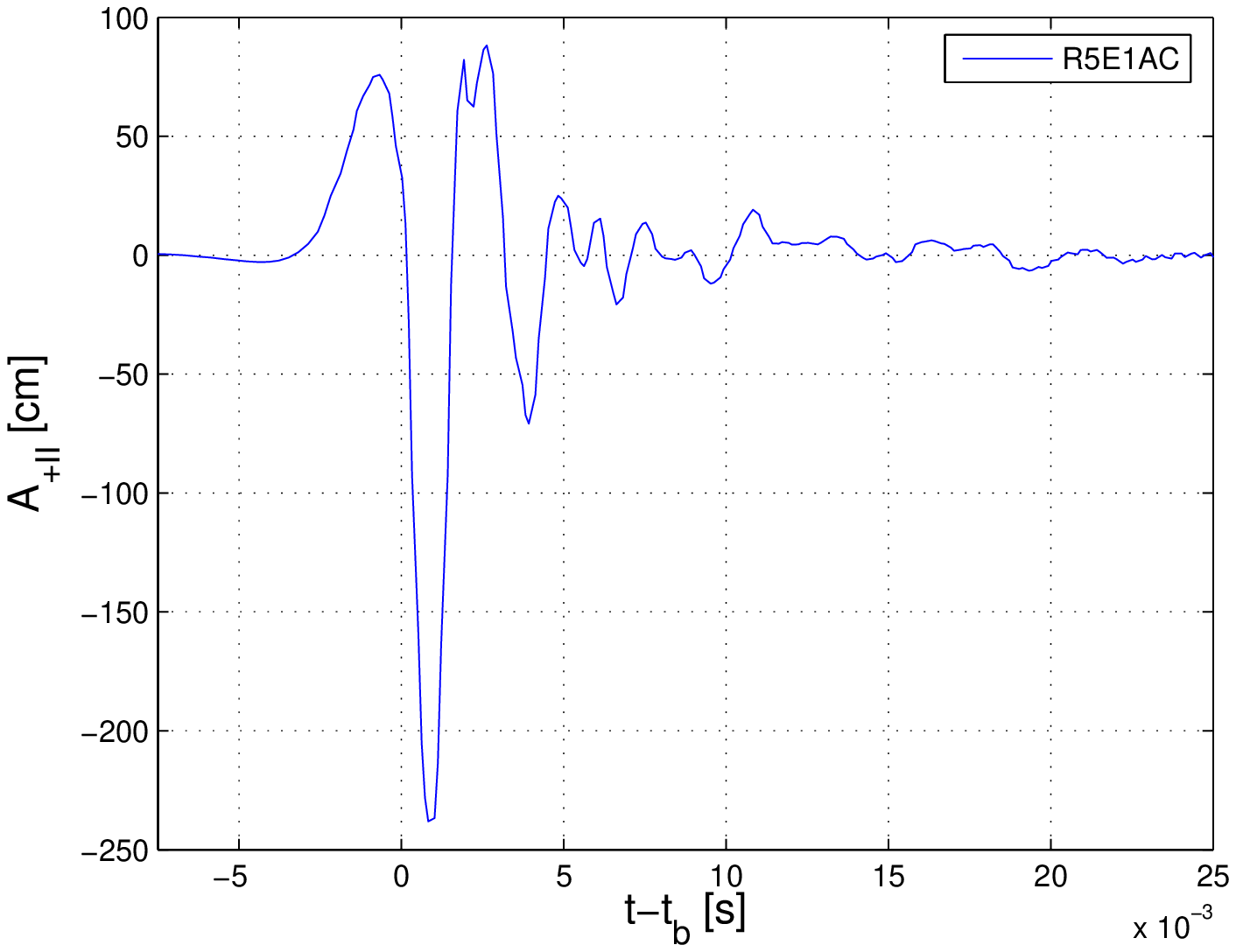}
\caption{\textbf{Left:} Time evolution of the 
GW amplitude A$_{\rm{+II}}$ of the 
models R3E1AC (solid line), R3E2AC (dashed line)
R3E3AC (dotted line) and R3STAC (dashed-dotted line). 
\textbf{Right:} Model R5E1AC's GW amplitude A$_{\rm{+II}}$.}
             \label{fig11.eps}
   \end{figure*}
%%%%%%%%%%%%%%%%%%%%%%%%%%%%%%%%%%%%%%%%%%%%%%%%%%%%%%%%%%%%%%%%%%%%%%%%
\begin{figure}
   \centering
    \includegraphics[width=8.8cm]{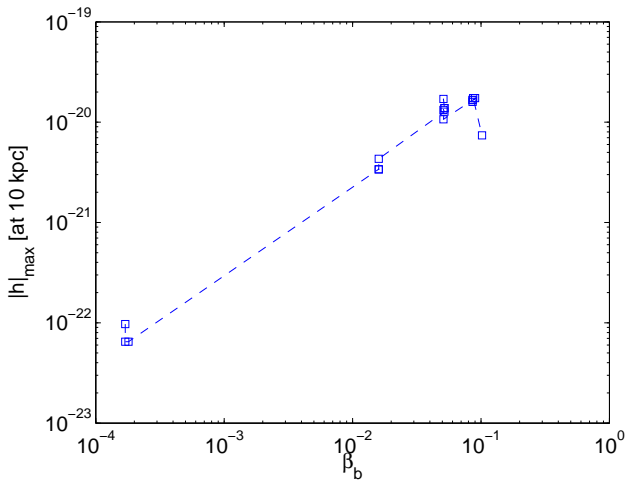}
        \caption{Summary of all model's 
dimensionless peak  gravitational wave amplitude 
$|h|_{max,b}$ at a distance of 10kpc and at 
core bounce versus the rotation rate $\beta_{b}$. 
While $|h|_{max,b}$ scales roughly 
linearly with $\beta_{b}$ for rotation rates 
$\beta_{b} \lesssim 10\%$, 
the growing centrifugal force reduces $|h|_{max,b}$
for $\beta_{b}\gtrsim10 \%$.}
             \label{fig12.eps}
\end{figure}
%%%%%%%%%%%%%%%%%%%%%%%%%%%%%%%%%%%%%%%%%%%%%%%%%%%%%%%%%%%%%%%%%%%%%%%%
\begin{figure}
   \centering
    \includegraphics[width=8.8cm]{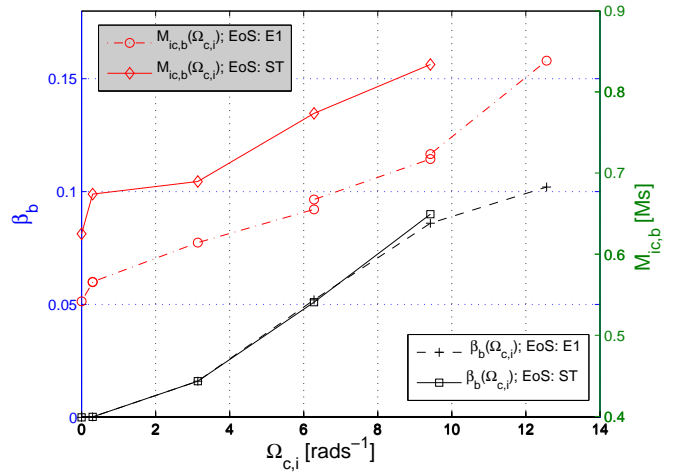}
        \caption{Precollapse central 
angular velocity $\Omega_{c,i}$ versus
the rotation rate rate $\beta_{b}$ or 
the mass of the inner core $M_{ic,b}$
at core bounce for models run with the E1 or ST EoS.
We define the mass of the inner core as the mass 
enclosed by the entropy maximum at core-bounce, i.e.
the unshocked region.}
             \label{fig13.eps}
\end{figure}

%%%%%%%%%%%%%%%%%%%%%%%%%%%%%%%%%%%%%%%%%%%%%%%%%%%%%%%%%%%%%%%%%%%%%%%%

%

\subsubsection*{Effects of the rotation rate 
on the GW signature}

Rapidly rotating progenitors 
($\Omega_{c,i}=\pi\ldots4\pi$rads$^{-1}$ 
in our model set) undergo different core-collapse dynamics 
compared to the previously discussed non- and slowly rotating
models.
Conservation of angular momentum in combination with 
contraction leads to a 
massive spin-up and hence oblate deformation of the core. 
The collapse is halted either by pure stiffening of 
the EoS above nuclear saturation
density, or, if rotation is sufficiently strong, 
by a combination of the 
centrifugal and the nuclear forces. 
The abrupt slowdown of axisymmetrically 
axisymmetrically-arranged and quickly rotating bulk matter gives rise
to rapid temporal variations in the quadrupole tensor, 
resulting in the emission of GWs.
Note that the core remains essentially axisymmetric
during the collapse and the 
early postbounce times ($t-t_{b}\lesssim 10$ms), 
as already pointed out in 
\citet{2007PhRvL..98z1101O,2007CQGra..24..139O}.
\noindent
%%%%%%%%%%%%%%%%%%%%%%%%%%%%%%%%%%%%%%%%%%%%%%%%%%%%%%%%%%%%%%%%%%%%%%%
Within the chosen parameter space of the rotation rate, 
all our models exhibit
a so-called type I GW burst (\citet{1997A&A...320..209Z}, 
see Fig. \ref{fig11.eps}) around core bounce, no matter 
what the initial choice of the EoS or the magnetic 
field configuration. This was previously 
also found by \citet{2008PhRvD..78f4056D}
in 2D GR simulations without magnetic fields.

%%%%%%%%%%%%%%%%%%%%%%%%%%%%%%%%%%%%%%%%%%%%%%%%%%%%%%%%%%%%%%%%%%%%%%%%

The question now arises what kind of information could possibly be
delivered from a quasi-axisymmetric type I GW burst, since it
is a priori unclear how degenerate it is with respect to the model 
parameters such as the EoS, rotation rate and so forth. This
was already investigated in great detail
by \citet{2008PhRvD..78f4056D} who performed
an extensive set of 2D GR core-collapse simulations.
Our 3D results show the same systematics:
the peak amplitude 
$|h|_{max,b} \equiv$ A$_{\bf{+II}}/$R 
scales about linearly with $\beta_{b}$ for 
models up to a moderately rapid rotation 
($ \beta_{b} \lesssim 10\%$, see Fig. \ref{fig12.eps}).
The Fourier-transforms 
of the bounce wave trains ($\pm$ 5ms 
relative to core bounce) 
show for most models in the indicated parameter 
a spectrum with a narrow bandwidth, 
peaking around $\sim800-900$Hz
(see Tab. \ref{table:3}). 
Moreover, with growing rotation rate, 
prompt convective overturn in the rotational 
plane is suppressed by the influence of positive angular momentum
gradients. This effect was 
first pointed out by \citet{1978ApJ...220..279E} and
is known as the Solberg-H\o iland instability 
criterion.
%%%%%%%%%%%%%%%%%%%%%%%%%%%%%%%%%%%%%%%%%%%%%%%%%%%%%%%%%%%%%%%%%%%%%%%%

The outcome of our 3D models confirms the statement of 
\citet{2008PhRvD..78f4056D} that
two parameters are essential for the behaviour of the 
GW amplitude around $t-t_{b} = 0$, 
namely the mass of the inner core at bounce,
which we denote as $M_{ic,b}$, and the initial central 
rotation rate 
$\Omega_{c,i}$.
Over the parameter range covered by our models, 
$\beta_{b}$ is 
a strictly monotonic function of the initial 
central angular velocity $\Omega_{c,i}$, as 
displayed in Fig. \ref{fig13.eps}.
The mass of the 
inner core $M_{ic,b}$ is linked to
$\beta_{b}$ via 
its dependence on $\Omega_{c,i}$ 
(see Fig. \ref{fig13.eps}). 
The positive mass offset of the inner core 
(which is approximately constant for all 
rotational configurations)
that 
occurs when switching from E1
to ST is interpreted as follows.
In a static initial configuration, the mass of 
the inner core is proportional to the square of
the electron fraction $Y_{e}$ and the entropy per 
baryon \citep{Goldreich1980}.
As the minimum of $Y_{e} (\rho)$ for E1 appears
at $\sim 0.276$ compared to $\sim 0.293$ 
for the Shen EoS 
(see Fig. \ref{fig1.eps}), we 
attribute the mass difference primarily 
to this relative difference, and
secondarily to changes in the specific entropy which occurs 
as the LS EoS permits more efficient electron capture.
However, note that we may overestimate the
spread of the inner core mass $M_{ic,b}$ 
in dependence with rotation
compared to simulations which are carried out with full
neutrino transport (Janka 2009, private communication).
The stronger the rotation becomes 
(at $\Omega_{c,i}\gtrsim3\pi$), 
the increased centrifugal forces start 
to play a dominant role, slowing 
down the entire dynamics of the collapse 
and causing the core to rebound 
at sub- or just above supra-nuclear matter densities.
The imprint of such behaviour is found in the GW signature
by a 
smaller maximum amplitude and lower
peak frequency compared to slower rotating models, 
as shown in Fig. \ref{fig11.eps}
and Tab. \ref{table:3}. 
Similar to \citet{2008PhRvD..78f4056D}, 
we also find that $h_{max}$ depends sensitively
on the competition of both the amount 
of imposed quadrupolar 
deformation due to rotation,
and on the other hand on the average 
density level in the for GW emission
dynamically relevant region of the inner core.
The density in the central region of the PNS 
is lowered considerably 
by centrifugal forces,
however, no longer compensated by a prominent
quadrupolar deformation which results from rapid rotation.
The `optimal' configuration for strong GW 
emission now is overshot,
which causes a 
smaller maximum amplitude and a lower
peak frequency.

%%%%%%%%%%%%%%%%%%%%%%%%%%%%%%%%%%%%%%%%%%%%%%%%%%%%%%%%%%%%%%%%%%%%%%%%
Our findings stand in very good qualitative agreement with 
\citet{2008PhRvD..78f4056D}, who recently performed a large
set of 2D core-collapse simulations in GR, nearly identical
micro-physical input, but without magnetic fields.
However, there are some quantitative differences.
For our models that undergo a pressure-dominated bounce, 
we find
spectra which peak 
in average some 100-150Hz higher than the models
in \citet{2008PhRvD..78f4056D}. It has been shown by 
Dimmelmeier (2007, private communication)
that the difference stems from the fact that 
fully relativistic
calculations shift the GW bounce 
spectrum to lower frequencies in
comparison to the ones using an 
effective, spherically symmetric
gravitational potential.
Furthermore, for comparable 
precollapse rotational configurations, 
our models return higher peak GW values.
We suspect the size of the inner core is the major cause
of this difference. The mass of the inner core for all 
our simulations is roughly $\sim 0.1 M_{\odot}$ bigger
than the ones of \citet{2008PhRvD..78f4056D}.
If we take into account 
that we are using different 
electron capture rates
(\citet{1985ApJS...58..771B} 
versus \citet{Langanke2003b} and \citet{2003PhRvL..91t1102H}),
we consider the mismatch to be understood, as
updated rates
cause the mass of the inner core in our models to shrink.

%%%%%%%%%%%%%%%%%%%%%%%%%%%%%%%%%%%%%%%%%%%%%%%%%%%%%%%%%%%%%%%%%%%%%%%%
\subsubsection*{Effects of magnetic fields  
on the GW signature}
%%%%%%%%%%%%%%%%%%%%%%%%%%%%%%%%%%%%%%%%%%%%%%%%%%%%%%%%%%%%%%%%%%%%%%%%

The presence of magnetic fields in our models
slows down the accretion of angular momentum 
onto the PNS
via field winding.
For example, the poloidal field's stress 
acts on fluid particles moving in 
the x-y plane in a direction opposing the motion,
leading to a deceleration.
Thus, while the GW signal from the initially 
`weakly' magnetised model R4E1AC 
is already strongly affected by 
centrifugal forces at bounce 
(the frequency peak at bounce is at $\sim$ 385Hz,
while the central angular velocity is 
$\Omega_{c,b}\sim$ 6200rads$^{-1}$), 
the initially more 
strongly magnetised but otherwise 
comparable model R4E1FC 
still undergoes a pressure 
dominated bounce with a frequency peak
at $\sim $860Hz and a central angular velocity of 
$\Omega_{c,b}\sim$ 5600rads$^{-1}$.
However, note that this effect only gets
prominent for initial magnetic fields 
that are by two orders of magnitude stronger than
suggested by \cite{Heger2005}.
We will discuss the issue of strong magnetic 
fields in more detail 
in sec. \ref{section:Bfield}.

%%%%%%%%%%%%%%%%%%%%%%%%%%%%%%%%%%%%%%%%%%%%%%%%%%%%%%%%%%%%%%%%%%%%%%%%
\subsubsection*{Effects of the EoS on the GW signature}
%%%%%%%%%%%%%%%%%%%%%%%%%%%%%%%%%%%%%%%%%%%%%%%%%%%%%%%%%%%%%%%%%%%%%%%%

\begin{figure}
   \centering
    \includegraphics[width=8.8cm]{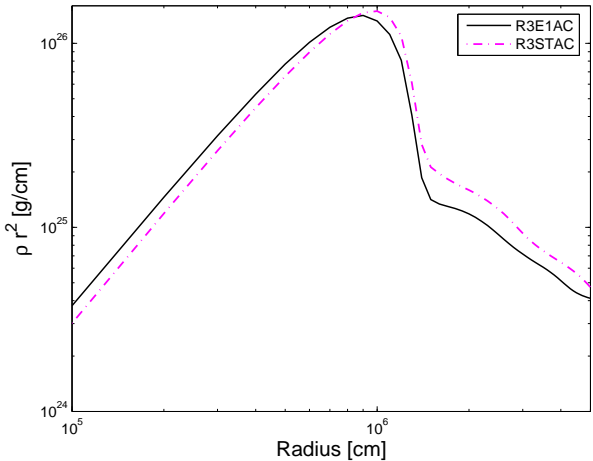}
        \caption{Radial profiles of 
the weighted density $\rho r^2$ of the spherically averaged data at 
time $t=0$ relative to core bounce for model R3E1AC and
its opposing model R3STAC.}
             \label{fig14.eps}
\end{figure}

%%%%%%%%%%%%%%%%%%%%%%%%%%%%%%%%%%%%%%%%%%%%%%%%%%%%%%%%%%%%%%%%%%%%%%%

In order to understand the dependence of the GW 
burst at bounce on the EoS, 
we repeated several simulations changing only
the EoS while keeping the other parameters 
fixed. 
The most prominent change occurs 
when switching from E1 to
ST.
Applying the latter EoS in our 
models leads to systematically 
larger absolute GW amplitudes at lower 
frequencies compared to its
counterparts, as shown in Tab. 
\ref{table:3} and the left panel of Fig. \ref{fig11.eps}.
This was also observed e.g. 
by \citet{Kotake2004},
where the two EoS were compared.
The shift to lower frequencies can be 
explained by the fact that the typical 
timescale of 
the GW burst at bounce is 
given by the free-fall timescale 
$\tau_{dyn}\sim1/\sqrt{G\overline{\rho}_{ic}}$, 
where $\overline{\rho}_{ic}$ is the mean density of the
inner core.
Since E1 leads to 
substantially higher central 
densities at core bounce
compared to ST in simulations 
that are not dominated 
by centrifugal forces, the spectral peak
of the GW signal
is shifted to higher frequencies.
The maximum GW burst amplitude depends 
on the dynamical timescale, the mass 
of the inner core as well as on 
the global density 
distribution inside the core, 
as pointed out e.g. 
in \citet{Kotake2004} and \citet{2008PhRvD..78f4056D}.
The dimensionless GW amplitude 
is roughly proportional to $M_{ic}$, divided
by the square of the dynamical timescale
($h\propto M_{ic}/{\tau^2_{dyn}}$). 
Hence, it scales approximately linearly with
density. This implies that one could 
expect higher GW peak amplitudes from the 
more compact cores in the case of 
E1. 
However, since models using E1 
are slightly more compact in central
regions, they exhibit lower densities
in the outer layers.
This is displayed in Fig. 
\ref{fig7.eps}.
Following the reasoning of 
\citet{2008PhRvD..78f4056D},
we display the quantity 
$\rho r^2 $ in Fig. \ref{fig14.eps}. 
It is the 
essential quantity in the integrand 
of the quadrupole GW formula 
(see Eq. \ref{equ:4}).
From this plot it is apparent 
that models run with 
E1 
have higher 
$\rho r^2$ at small radii, ST
yields higher values at intermediate 
and large 
radii.
Due to their larger volume, these regions
contribute more to the total 
quadrupole integral.
For fast rotators ($\Omega_{c,i}\gtrsim 3\pi$), 
centrifugal forces start to play a dominant role, 
as already previously 
observed by \citet{2008PhRvD..78f4056D}. 
Here, the relative difference between the
GW signatures of models run with 
two EoS decreases, since in 
the regime of lower densities, the EoS
do not differ significantly. 
When comparing, e.g., 
models R3E1AC with R3STAC, 
the absolute size of the burst 
amplitudes vary roughly 25\%,
whereas R4E1AC and R4STAC are 
only discriminated by $\sim 4$\% 
(see Tab. \ref{table:3}).
In summary we state that it 
seems very difficult to reveal 
information about 
the different two EoS 
by considering the GW signature 
from core bounce alone. 
For models run with either the LS or the
Shen EoS and all other parameters being 
indentical, the differences brought
about by the EoS are clearly distinguishable.
However, since a small variation in one 
of the other parameters can easily have
a similar effect as the EoS change, it 
will be nearly impossible to constrain
the nuclear EoS in the general case.

When changing the LS compressibility 
from K=180MeV to 
K=220MeV or K=375MeV, 
the features of the collapse dynamics
and the corresponding  
GW emission remain practically unaltered, as 
displayed in the left panel of Fig. \ref{fig11.eps}.
The only notable difference occurs in 
the vicinity of core bounce
at the center of the PNS: 
Models run with the softest  
version of the LS EoS (E1)
allow the core to 
bounce at slightly higher central density
compared to E2 and E3 (see Tab. \ref{table:1}).
At the same time, models run with E1 exhibit
lower densities at larger radii compared to 
cases where E2 or E3 was applied
(see Fig. \ref{fig7.eps}).
However, 
since the differences in 
the radial density profiles are relatively 
low, these effects cancel each other if we
consider again the quantity $\rho r^2$. This
leads to GW amplitudes 
of similar size and frequencies. 

%%%%%%%%%%%%%%%%%%%%%%%%%%%%%%%%%%%%%%%%%%%%%%%%%%%%%%%%%%%%%%%%%%%%%%%

\subsubsection{Gravitational waves 
from the nonaxisymmetric rotational instability}

\subsubsection*{General remarks}

%%%%%%%%%%%%%%%%%%%%%%%%%%%%%%%%%%%%%%%%%%%%%%%%%%%%%%%%%%%%%%%%%%%%%%%
\begin{figure}
   \centering
    \includegraphics[width=8.8cm]{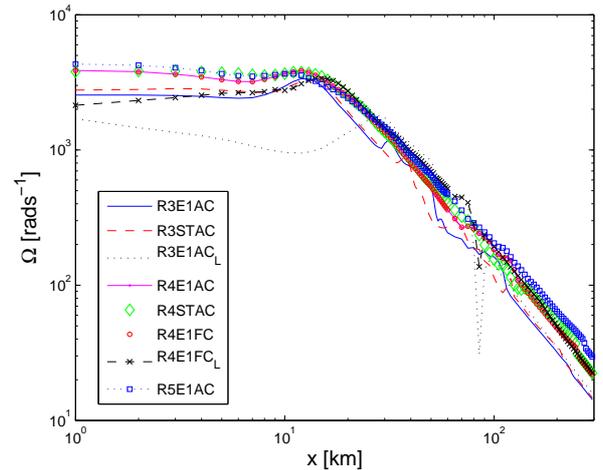}
        \caption{Angular velocity profile 
along the positive x-axis for different models at 10ms 
after bounce.
Note that the hump in the angular velocity profile  
at about $\sim11$km seems to be a generic feature, caused by
accreting material with rather high specific angular 
momentum which accumulates
on the nuclear density region of the PNS, 
as discussed in \citet{2006ApJS..164..130O}.
It is most pronounced in models with an initial  
rotation rate $\lesssim$ R3
and gets flatter for high initial rotation rates.}
             \label{fig15.eps}
\end{figure}
%%%%%%%%%%%%%%%%%%%%%%%%%%%%%%%%%%%%%%%%%%%%%%%%%%%%%%%%%%%%%%%%%%%%%%%

  \begin{figure*}
   \centering
   \includegraphics[width=\textwidth]{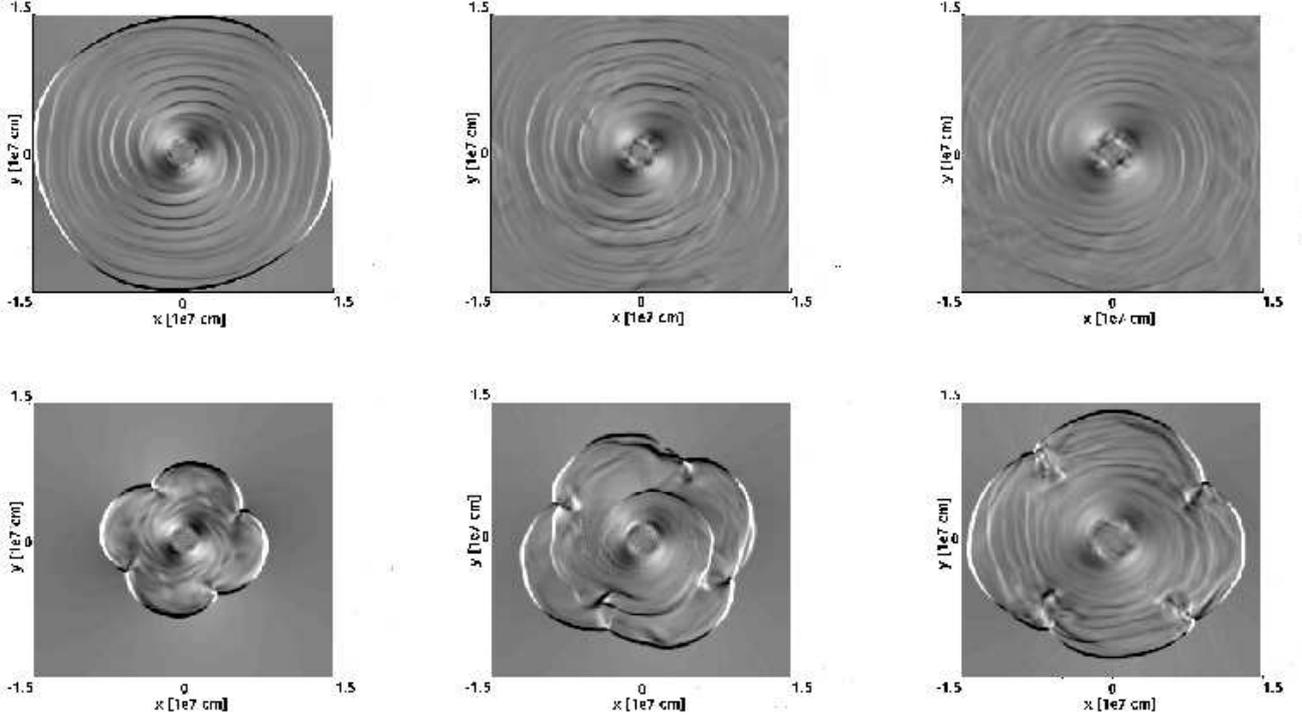}
        \caption{Snapshots of the vorticity's 
z-component $\vec{w_{z}}=(\nabla\times \vec{v})_{z}$ 
in the equatorial plane for models R4STAC 
(upper panels, $t-t_{b}=10,29,63$ms) 
and R4E1FC$_{L}$ (lower panels, $t-t_{b}=10,29,54$ms) 
at three representative instants of their evolution. 
The innermost 300$^2$km$^2$ are displayed, 
and the colour is encoded in units of [s$^{-1}$], ranging
from -5000 (white) to 5000 (black).
The upper panels of vorticity plots 
show nicely a two-armed $m = 2$ spiral pattern, 
while the middle plot of the 
lower panel mainly shows the same for a $m = 1$ 
mode.}
             \label{fig16.eps}
   \end{figure*}

%%%%%%%%%%%%%%%%%%%%%%%%%%%%%%%%%%%%%%%%%%%%%%%%%%%%%%%%%%%%%%%%%%%%%%%

\begin{figure*}
    \centering
    \includegraphics[width=8.cm]{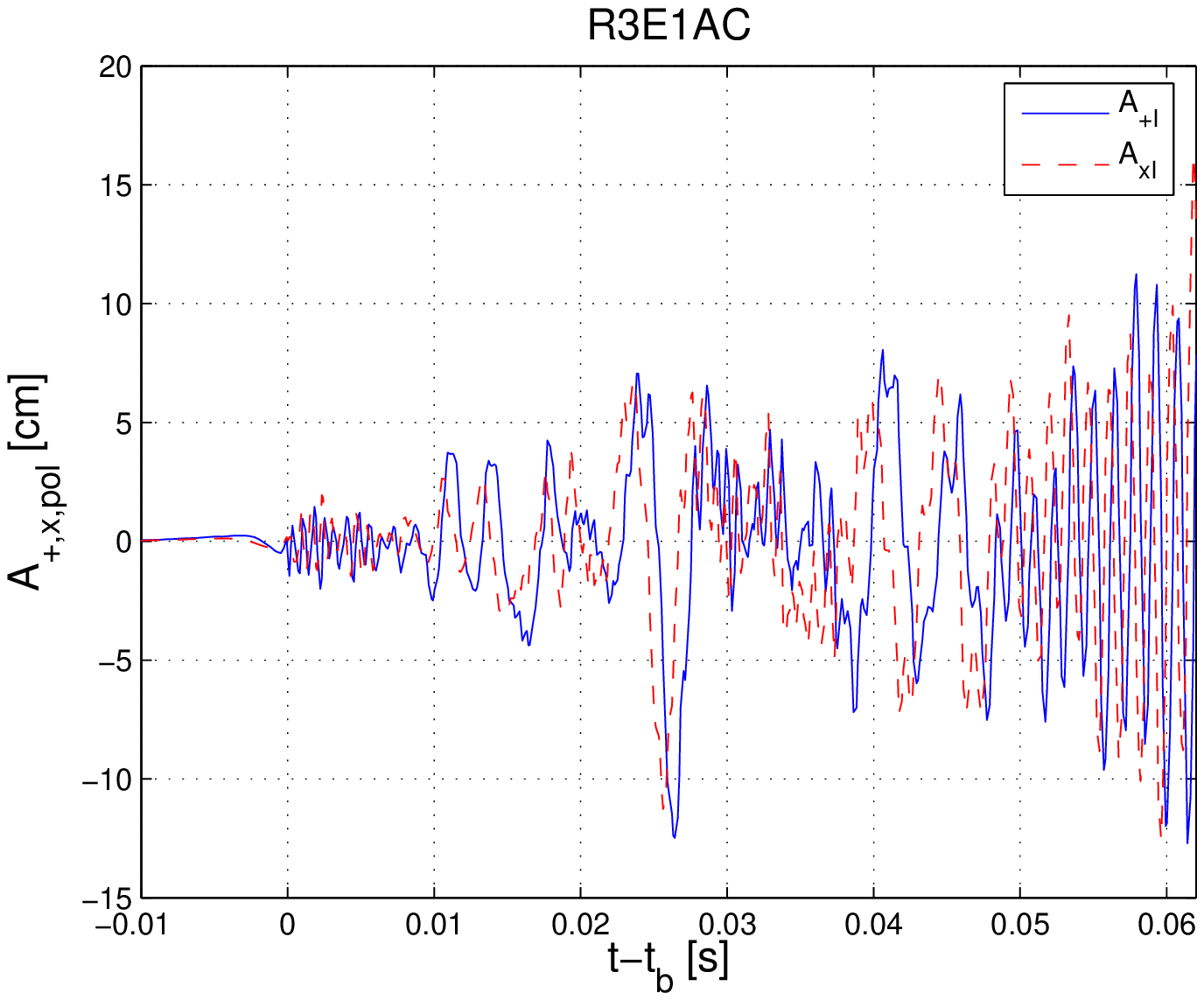}
    \includegraphics[width=8.cm]{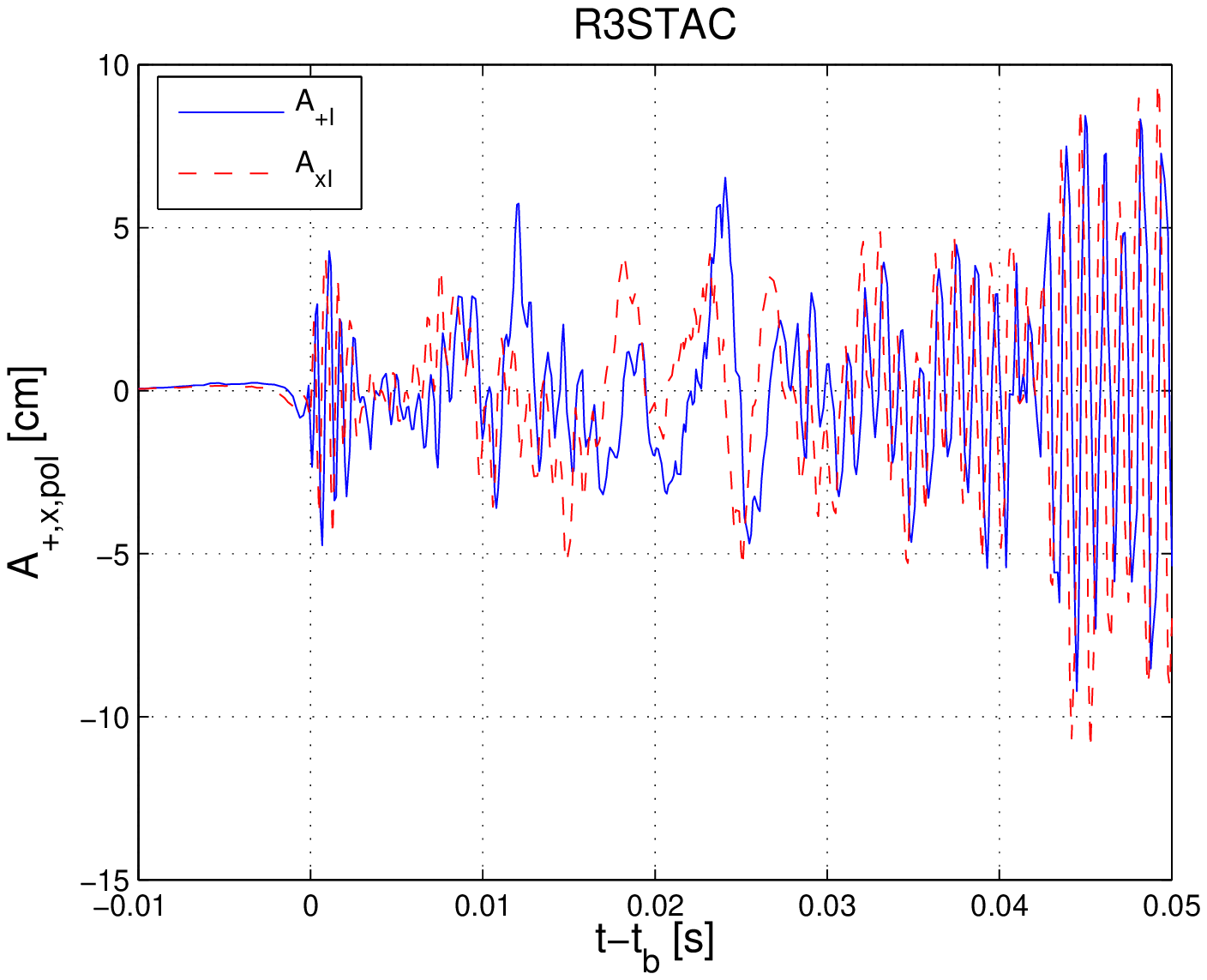}
    \includegraphics[width=8.cm]{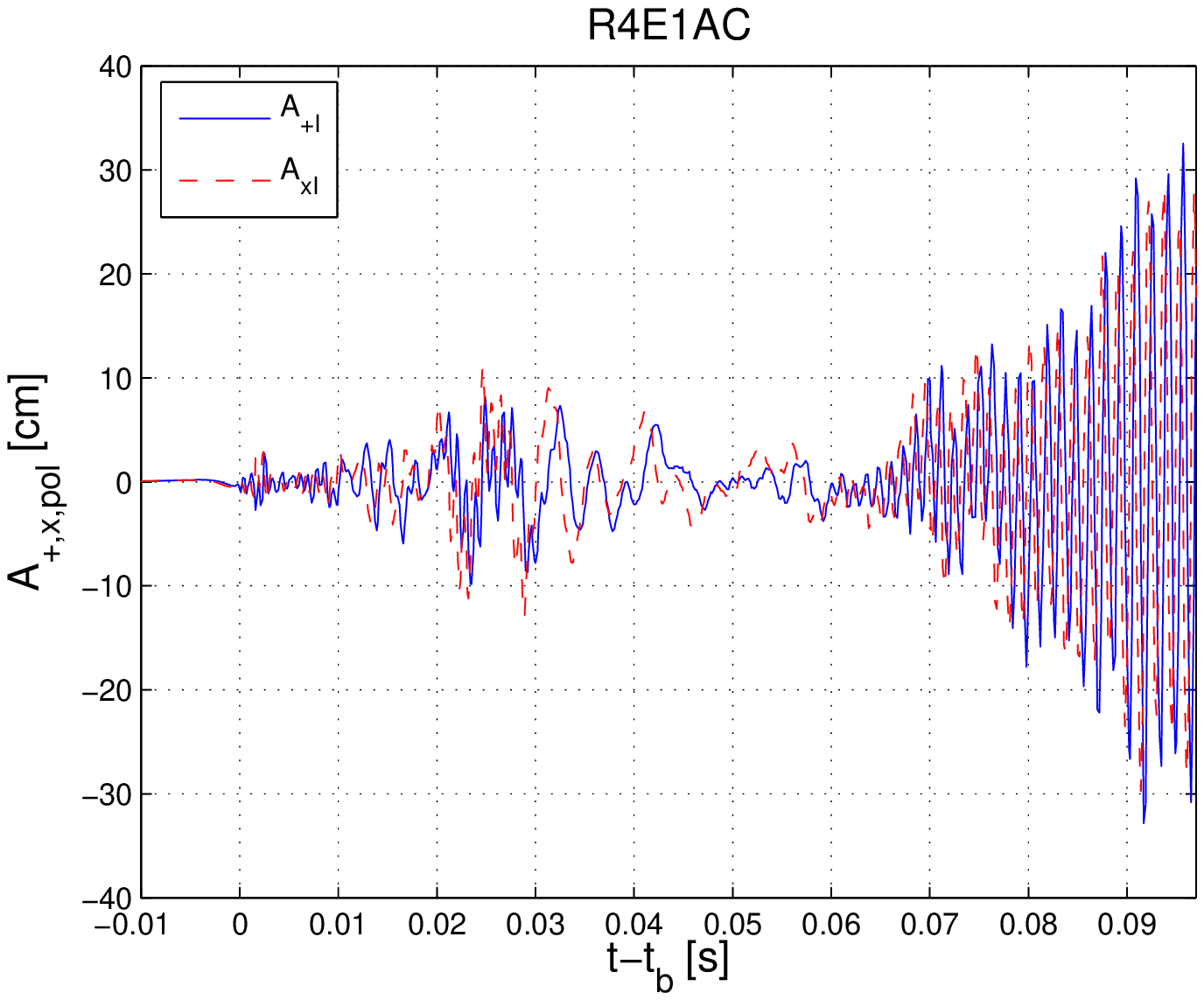}
    \includegraphics[width=8.cm]{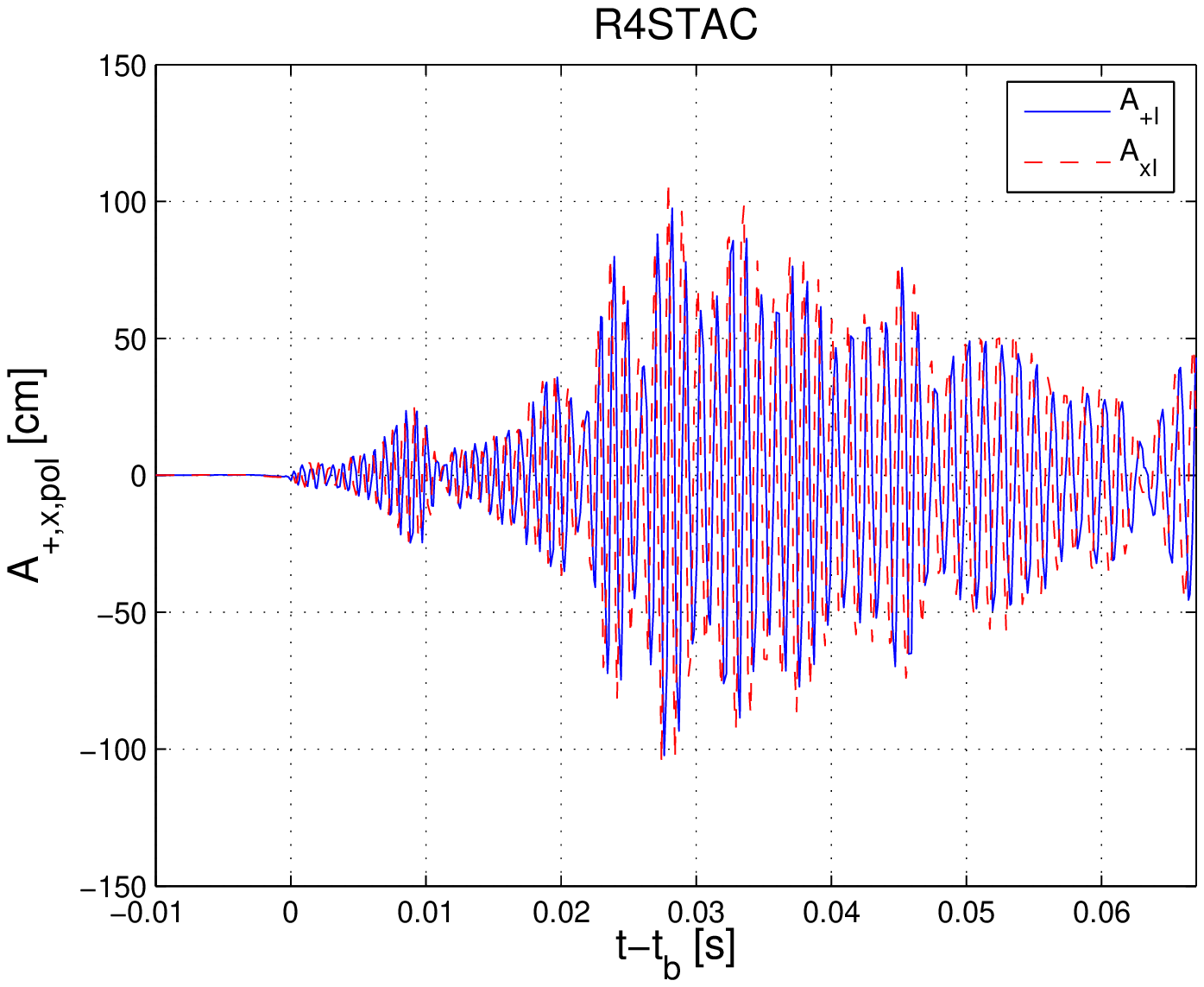}
    \caption{The upper left panel shows the time evolution 
of the GW amplitudes A$_{+}$ and A$_{\times}$ of 
model R3E1AC emitted along the 
polar axis. The upper right
displays the same for model R3STAC. The lower two 
panels contain the same information
for R4E1AC and R4STAC from left to right. 
Note that A$_{+}$ and A$_{\times}$
oscillate at the same frequency, phase shifted by $\pi/2$.}
             \label{fig17.eps}
\end{figure*}
%%%%%%%%%%%%%%%%%%%%%%%%%%%%%%%%%%%%%%%%%%%%%%%%%%%%%%%%%%%%%%%%%%%%%%%
\begin{figure}
   \centering
    \includegraphics[width=8.8cm]{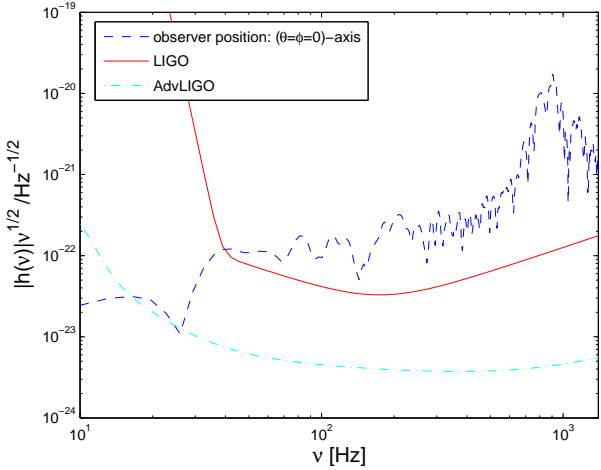}
        \caption{Spectral energy distribution of 
the GW signal emitted along the polar 
axis from model R4STCA at a distance of 10kpc.}
             \label{fig18.eps}
\end{figure}
%%%%%%%%%%%%%%%%%%%%%%%%%%%%%%%%%%%%%%%%%%%%%%%%%%%%%%%%%%%%%%%%%%%%%%%

Rotating proto-neutron stars can be subject 
to non-axisymmetric rotational instabilities in situations
when $T/|W|_{dyn}=\beta$ exceeds a certain critical value.
Since the growing instabilities carry the object's 
spheroidal- into a triaxial configuration with a time-dependent 
quadrupole moment, strong GW emission is to be expected.
The best understood type of instability is the classical
dynamical bar mode rotational instability with a threshold value of
$\beta_{dyn}\sim27$\%.
However, strong evidence was found by \citet{2008PhRvD..78f4056D} 
that it is unlikely that the PNS reaches 
rotation rates required for it to become
unstable during the core-collapse of 
and the early postbounce phase 
the iron core.
Another possibility is the 
secular instability, triggered at 
moderately high $\beta_{sec} \sim 14$\%
if a dissipative mechanism is present.
It grows on the relatively slow dissipative timescale of the
order of a second (see, \citet{1978trs..book.....T}).
Since none of our models reach such high $\beta$-values,
both instabilities cannot play 
any role in our simulations.

However, recent work, some of which has
been carried out in idealised setups and 
assumptions
\citep{2002MNRAS.334L..27S,2003ApJ...595..352S,
2006AIPC..861..728S,2005ApJ...618L..37W,
2006ApJ...651.1068O,2007CoPhC.177..288C} 
and later also in more self-consistent
core-collapse simulations 
\citep{2007CQGra..24..139O,2007PhRvL..98z1101O,2008A&A...490..231S}, 
suggests that a differentially rotating PNS 
can become dynamically unstable at 
$T/|W|$-values as low as $\sim1$\%.
Today this so-called low `$T/|W|$' 
instability is interpreted as being 
a resonance phenomenon \citep{2005ApJ...618L..37W}.
The underlying mechanism 
is suspected to be the amplification of azimuthal 
(non-axisymmetric) modes at 
\textit{co-rotation points}, where 
the pattern speed $\sigma_{p}=\sigma/m$ of the unstable mode
matches the local angular velocity,
($\propto \exp[i(\sigma t - m\Phi)]$) 
where $\sigma$ is the mode's 
eigenfrequency. 
The PNS is differentially rotating outside a radius of $\sim 10$km, 
as displayed in Fig. \ref{fig15.eps}.
This differential rotation provides a reservoir of shear energy 
may be tapped 
by the instability. The latter leads to spiral waves, 
as displayed in 
Fig. \ref{fig16.eps}, which transfer
angular momentum outwards 
(see, e.g., \citep{1999ApJ...513..805L}).
Note that this entire phenomenon appears 
to be closely related to the 
Papaloizou-Pringle instability which 
occurs in accretion discs
around a central gravitating body \citep{1985MNRAS.213..799P}.

In order to investigate the growth of the non-axisymmetric structures, 
we monitor the PNS by decomposing the density at a given 
radius $R$ in the equatorial plane ($z=0$) 
into its azimuthal Fourier components: 

\begin{equation}
\rho(R,z,\phi)  =  \sum\limits_{m=-\infty}^{\infty}C_{m}(R,z)e^{im\phi} \;  \\
\label{equ:rho} 
\end{equation}

\begin{equation}
C_{m}  = \frac{1}{2\pi}\int_{0}^{2\pi}\rho(R,z,\phi)e^{-im\phi}\rm d\phi \; .
\label{equ:rho2} 
\end{equation}

The gravitational wave morphology resulting 
from the nonaxisymmetric process generally shows
narrow-band and highly periodic signals which 
persist until the end of our simulations. This is
shown in Fig. 
\ref{fig17.eps} and the upper panels
of Fig. \ref{fig21.eps}.
Bearing in mind that the effectively measured GW
amplitude scales with the
number of GW cycles \textit{N} as  
$h_{\rm{eff}}\propto h\sqrt{N}$ (see \citet{Thorne1989})
and that it could last for several hundreds of 
ms as our simulations suggest,
the chances of being able to detect such
kind of signal are enhanced 
compared e.g. to the short duration burst signal 
at bounce, as one can see in Fig. \ref{fig18.eps}.
In the lower panels of Fig. \ref{fig21.eps} the 
normalised mode amplitudes of the models R3E1AC$_{L}$ and R4E1FC$_{L}$, with
\begin{equation} 
A_{m}=|{C_{m}}|/C_{0}=\frac{1}{C_{0}}\sqrt{\Re(C_{m})^2+\Im(C_{m})^2},
\end{equation}

\noindent
are plotted in order to follow the behaviour of unstable modes.
We generally find modes with density wave 
numbers $m = {1,2,3}$ being triggered,
with the $m = 1$ or $m = 2$ as the overall dominant ones, depending 
on the individual model.
Furthermore, note that all modes have the 
same pattern speed,
as previously observed by 
\citet{2007CQGra..24..139O,2007PhRvL..98z1101O} 
and \citet{2008A&A...490..231S}.
In Fig. \ref{fig16.eps}, the upper panels of vorticity plots 
show nicely a two-armed $m = 2$ spiral pattern, 
while the middle plot of the 
lower panel mainly shows the same for a $m = 1$ 
mode, as we expect from a mode analysis of 
the corresponding model R4E1FC$_{L}$
presented in 
Fig. \ref{fig21.eps}.
After the early linear growth phase, the modes saturate due to 
Kelvin-Helmholtz shear instabilities, 
which break the spirals apart in the outer layers,
as displayed in 
Fig. \ref{fig16.eps} and previously observed and 
discussed in e.g. \citet{2007CoPhC.177..288C}.
Note the close relation between the $m = 2$ bar mode
and the emission of GWs. The growth and saturation of this mode
is imprinted on the GWs emitted. As soon as it exceeds
the Cartesian $m = 4$ noise background, strong 
GW emission at a frequency
corresponding to twice the $m = 2$ 
pattern speed along the pole emerges, with the $+$ and
$\times$ polarisations shifted by a quarter cycle, 
as one could expect from
GWs emitted by a spinning bar.
%%%%%%%%%%%%%%%%%%%%%%%%%%%%%%%%%%%%%%%%%%%%%%%%%%%%%%%%%%%%%%%%%%%%%%%

\subsubsection*{Models without deleptonisation in the postbounce phase:
Effects of the EoS and magnetic fields on the GW signature}
%%%%%%%%%%%%%%%%%%%%%%%%%%%%%%%%%%%%%%%%%%%%%%%%%%%%%%%%%%%%%%%%%%%%%%%

\begin{figure}
   \centering
    \includegraphics[width=8.8cm]{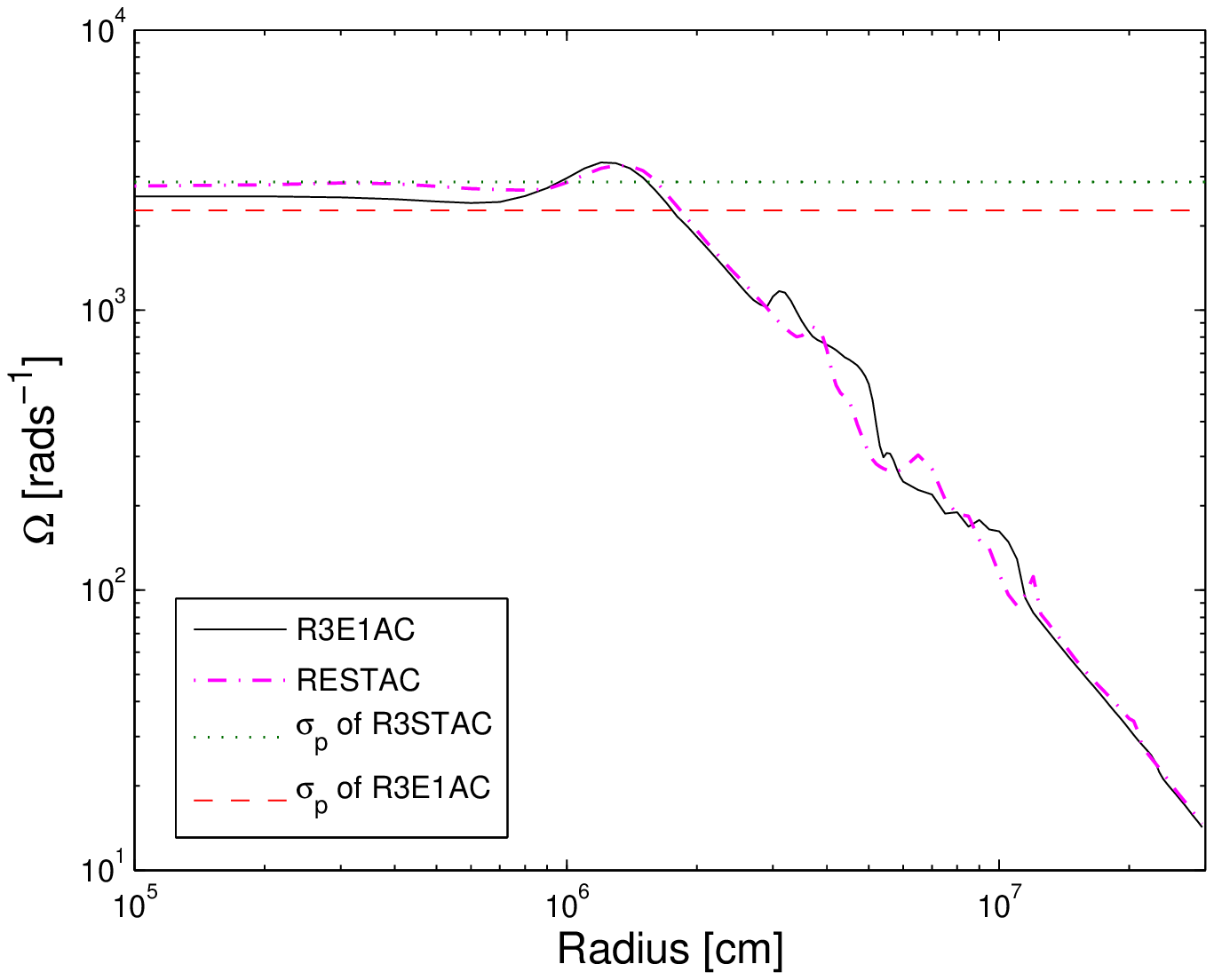}
    \includegraphics[width=8.8cm]{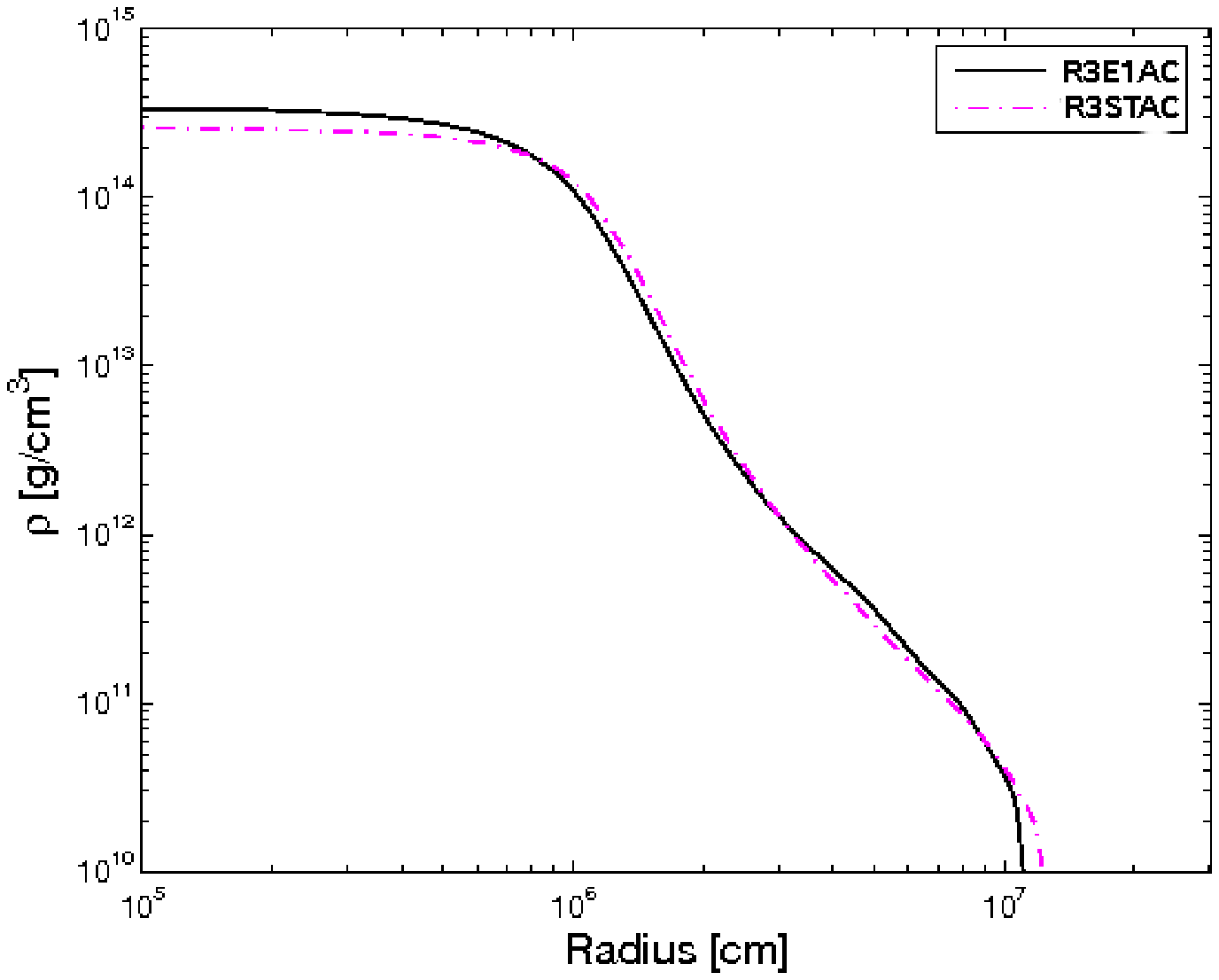}
        \caption{The upper panel shows the angular 
velocity profile at $t-t_{b}\approx$ 10 ms 
along the positive x-axis for the models R3E1AC 
and R3STAC. 
The pattern speeds of the corresponding 
simulations are indicated by horizontal lines. 
The lower panel displays the spherically 
averaged radial density profiles at the same
particular time.}
             \label{fig19.eps}
\end{figure}
%%%%%%%%%%%%%%%%%%%%%%%%%%%%%%%%%%%%%%%%%%%%%%%%%%%%%%%%%%%%%%%%%%%%%%%

We now turn the discussion in more detail to individual simulations. 
Models with an initial rotation rate of 
at least R3 ($\beta_{b}\sim 5.2\%$)
sooner or later become low $\beta$ dynamically unstable 
in our parameter set
(see Table \ref{table:3}).
Note, however, that \citet{2009arXiv0912.1455S} recently 
found models to become dynamically unstable at even slower 
initial rotation rates 
($\beta_{b}\sim 2.3\%$).

When comparing models that were carried out with the Shen- and the 
LS (E1) EoS, 
we find that the `Shen'-models emit GWs 
at significantly higher frequencies than 
their LS counterparts.
This result is a consequence of the fact 
that the specific angular momentum, which scales
as $j_{z}\propto\rho\omega_{z}$, 
is roughly preserved on a mass shell \citep{Keil1996}:
The innermost part of the
PNS in `Shen'-simulations, which rotates 
nearly in perfect solid body rotation 
at the pattern speed (see Fig. \ref{fig19.eps}),
has lower density in this radial region.  
Hence the central rotation rate must in turn
be higher to fulfill the conservation law. 
As an example, we compare the models R3E1AC 
and R3STAC in detail.
While R3E1AC rotates at $t-t_{b} = 10$ms 
with a central rotation 
rate of $\omega_{R3E1AC} \sim 2300$rads$^{-1}$ 
and has a central density
$\rho_{R3STAC}$ $\sim$ 3.38$\times10^{14}$gcm$^{-3}$,
R3STAC revolves with the values of $\omega_{R3STAC}\sim 2800$ 
rads$^{-1}$ and $\rho_{R3E1AC}\sim2.63\times10^{14}$gcm$^{-3}$,
as displayed in Fig. \ref{fig19.eps}.
Doing the maths, the ratios of $\omega_{R3E1AC}/\omega_{R3STAC}$
to $\rho_{R3STAC}/\rho_{R3STAC}$ are about the same.
We also state that the dynamical instability in the `Shen'cases
grows faster generically than in the LS simulations, as shown
in Fig. \ref{fig17.eps}.
%%%%%%%%%%%%%%%%%%%%%%%%%%%%%%%%%%%%%%%%%%%%%%%%%%%%%%%%%%%%%%%%%%%

We generally observe slower growth 
of the T/$|$W$|$ unstable modes 
in situations where we applied
stronger initial poloidal- than toroidal magnetic fields.
Although such magnetic fields ($B_{pol,i} > 
B_{tor,i}$) may not be motivated
by stellar evolution calculations \citep{Heger2005},
it is still important to study their effects.
Model R3E1AC for example starts to emit strong 
GWs due to the low $T/|W|$
instability around $\sim 50$ms after bounce, while
model R3E1DB does not within the duration 
of the simulation, which we followed until
 $\sim 65$ms after bounce 
(note, however, that the latter model shows strong
growth of the m=1,2,3 modes although GW emission 
due to the low $T/|W|$ did not set in yet).
The poloidal fields are able to suppress the dynamical
instability for some time as they 
slow down the spin-up of the PNS.
The detailed discussion of this issue is postponed to sec. 
\ref{section:Bfield}.

%%%%%%%%%%%%%%%%%%%%%%%%%%%%%%%%%%%%%%%%%%%%%%%%%%%%%%%%%%%%%%%%%%%
Centrifugal forces set a limit to the maximum frequency of the
GW signal in similar fashion to the situation 
at core bounce. As discussed
in the previous subsection, the limit is 
somewhere around $\sim 935$Hz (which is twice the pattern speed!).
The faster the initial rotation rate, the stronger the influence of
centrifugal forces, which slow down in the postbounce phase 
the advection 
of angular momentum onto the PNS. 
The result is a
slower central rotation rate, a lower
pattern speed and thus GW emission at lower frequencies.
Beside these semi-quantitative statements which allow
for the distinction of the simulations' input physics 
by its GW signature
on a model-to-model basis, where only one parameter is 
varied while keeping the others fixed, 
it is in general very difficult to 
discern effects of individual features of the input physics 
in a GW signal that cannot unambiguously be attributed 
to a specific model.
The degeneracy in the simulation results is large with respect 
to the rotation rate, 
the magnetic fields and the underlying EoS.

%%%%%%%%%%%%%%%%%%%%%%%%%%%%%%%%%%%%%%%%%%%%%%%%%%%%%%%%%%%%%%%%%%%%%%%

\subsubsection*{Models with deleptonisation 
during the postbounce phase}

%%%%%%%%%%%%%%%%%%%%%%%%%%%%%%%%%%%%%%%%%%%%%%%%%%%%%%%%%%%%%%%%%%%%%%%

\begin{figure}
   \centering
    \includegraphics[width=8.8cm]{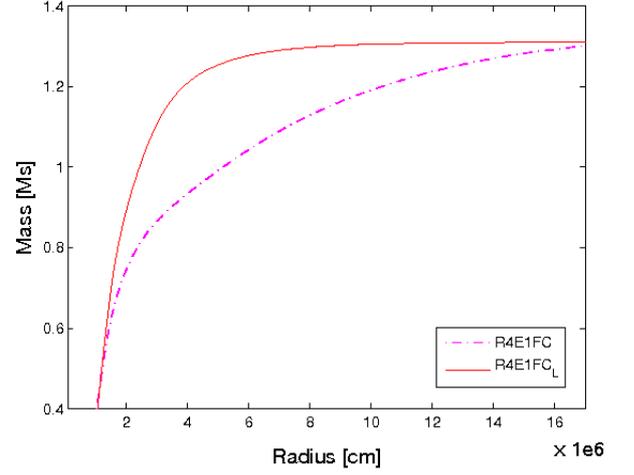}
        \caption{The figure shows 
the enclosed mass [$M_{\odot}$] as a 
function of radius for models R4E1FC and 
R4E1FC$_{L}$ at $t-t_{b}\approx$ 25 ms, which explains
the strong GW emission of the latter model.}
             \label{fig20.eps}
\end{figure}
%%%%%%%%%%%%%%%%%%%%%%%%%%%%%%%%%%%%%%%%%%%%%%%%%%%%%%%%%%%%%%%%%%%%%%%

\begin{figure*}
   \centering
    \includegraphics[width=8.8cm,height=6cm]{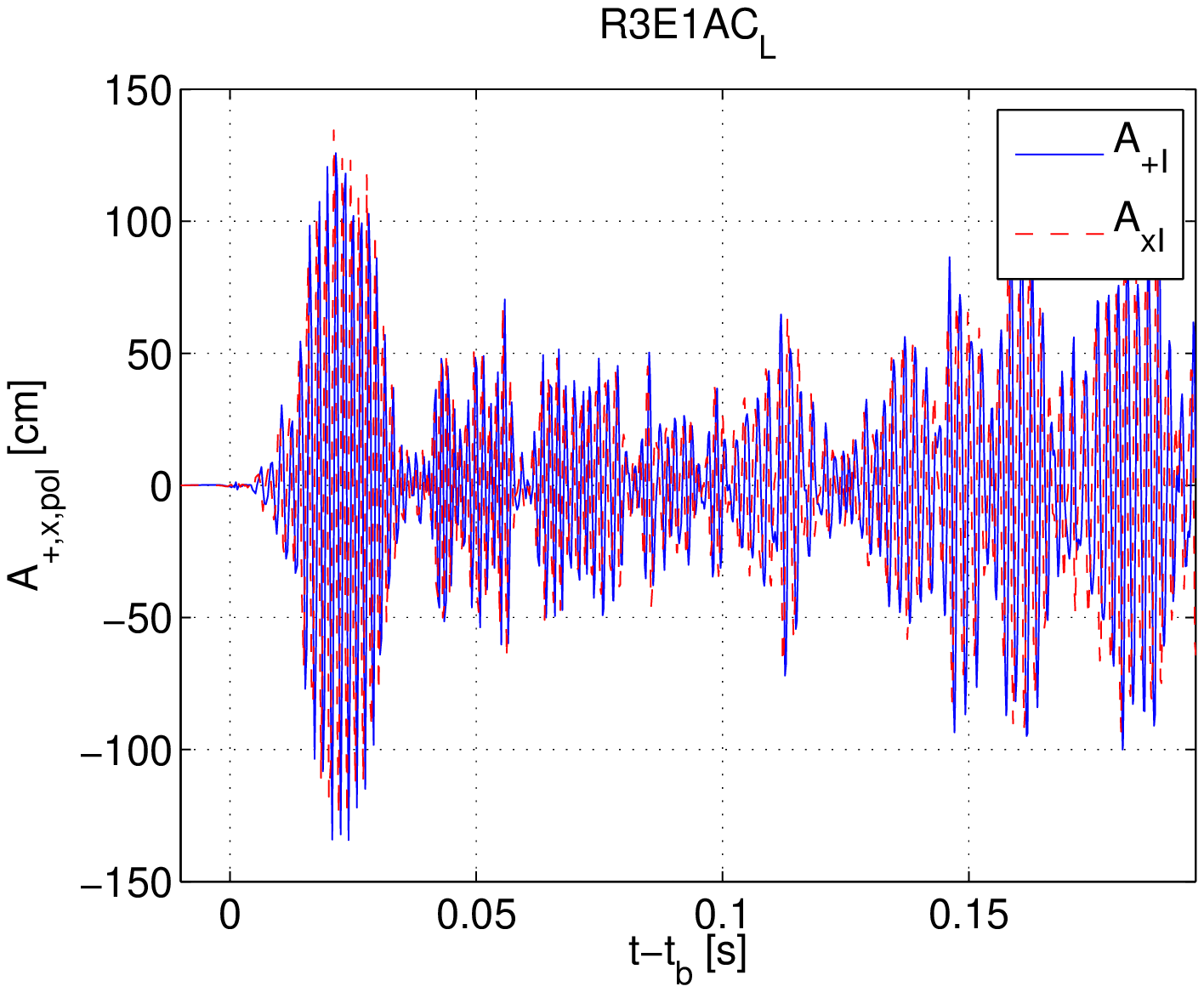}
    \includegraphics[width=8.8cm,height=6cm]{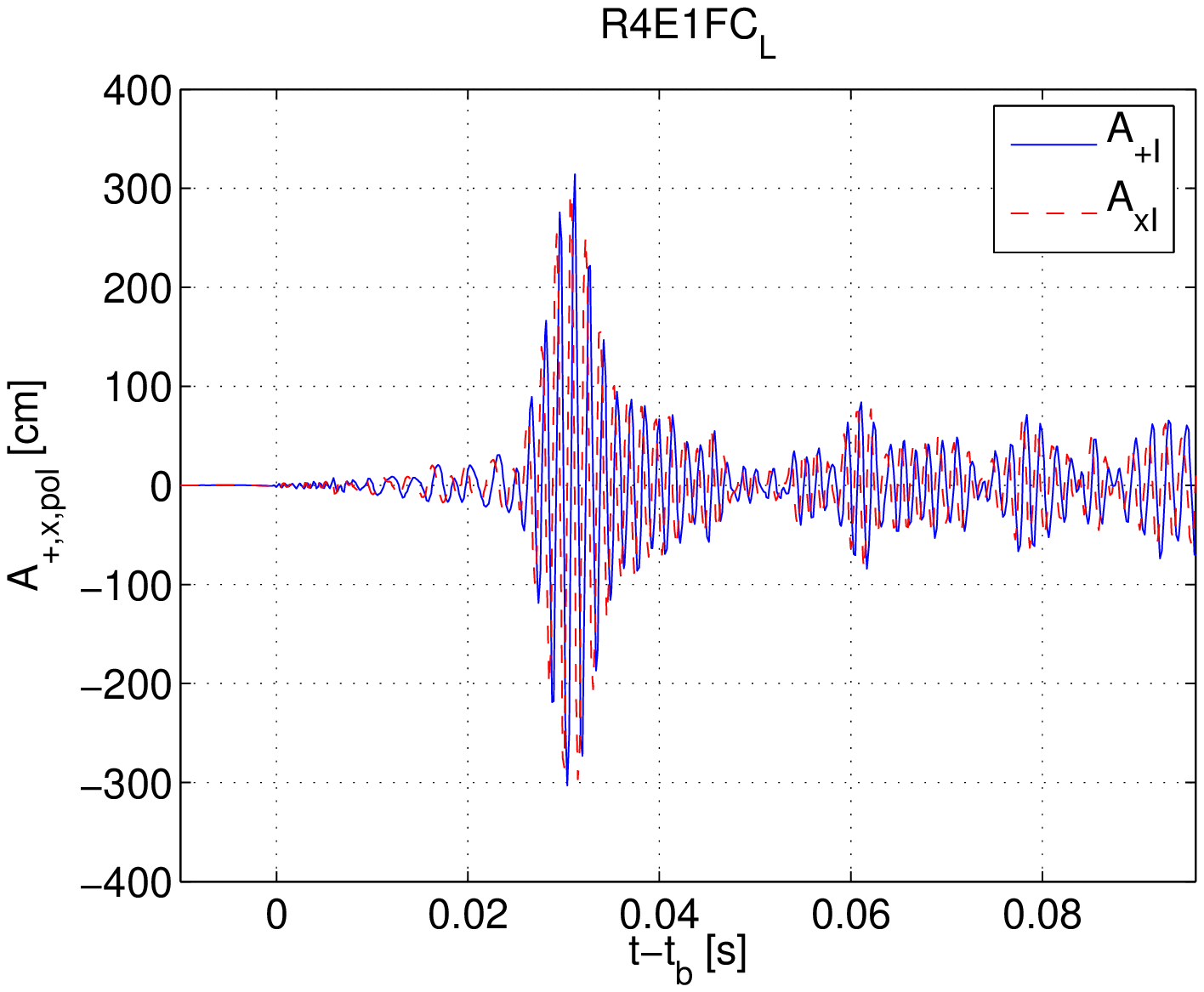}
    \includegraphics[width=8.8cm,height=6cm]{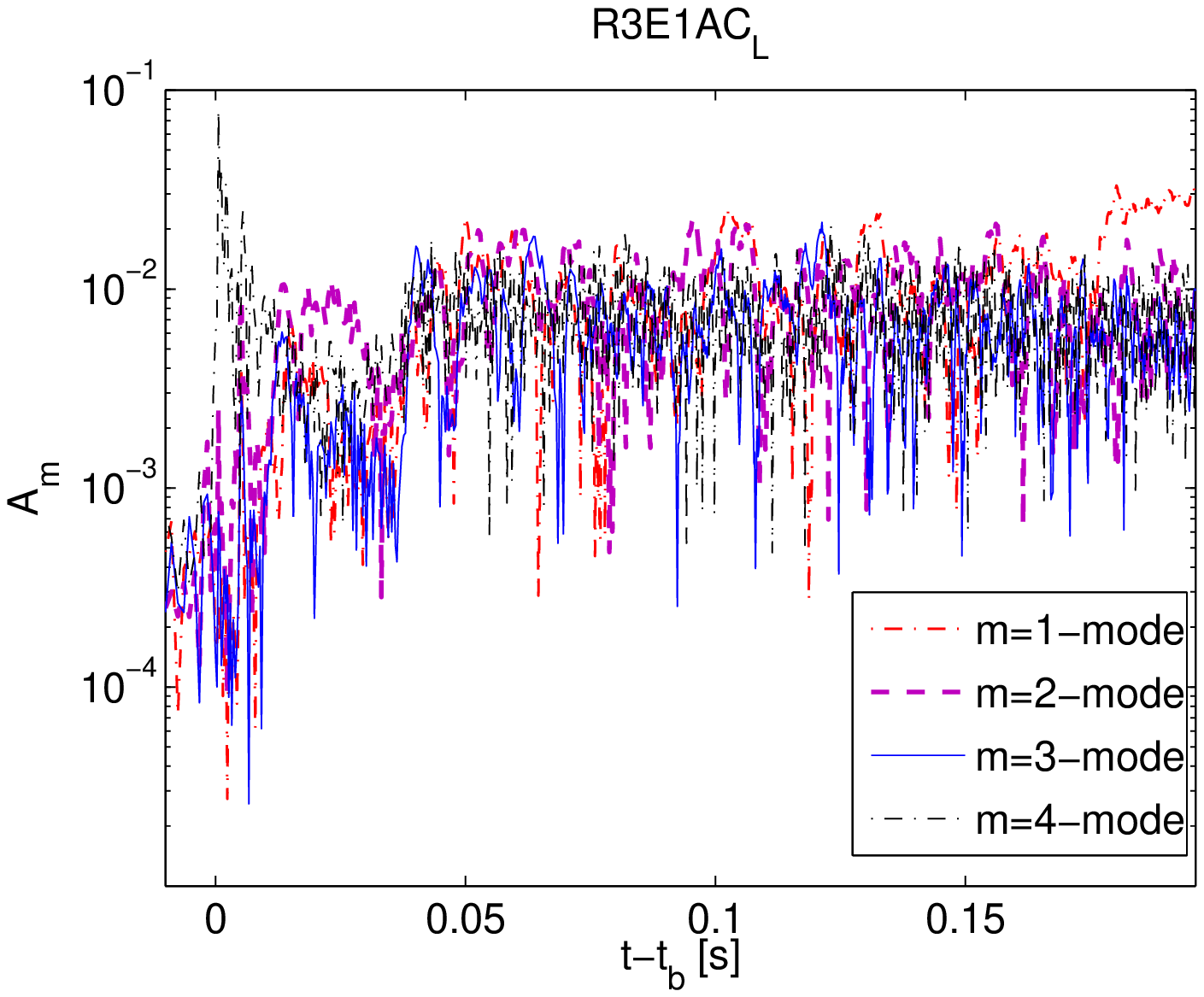}
    \includegraphics[width=8.8cm,height=6cm]{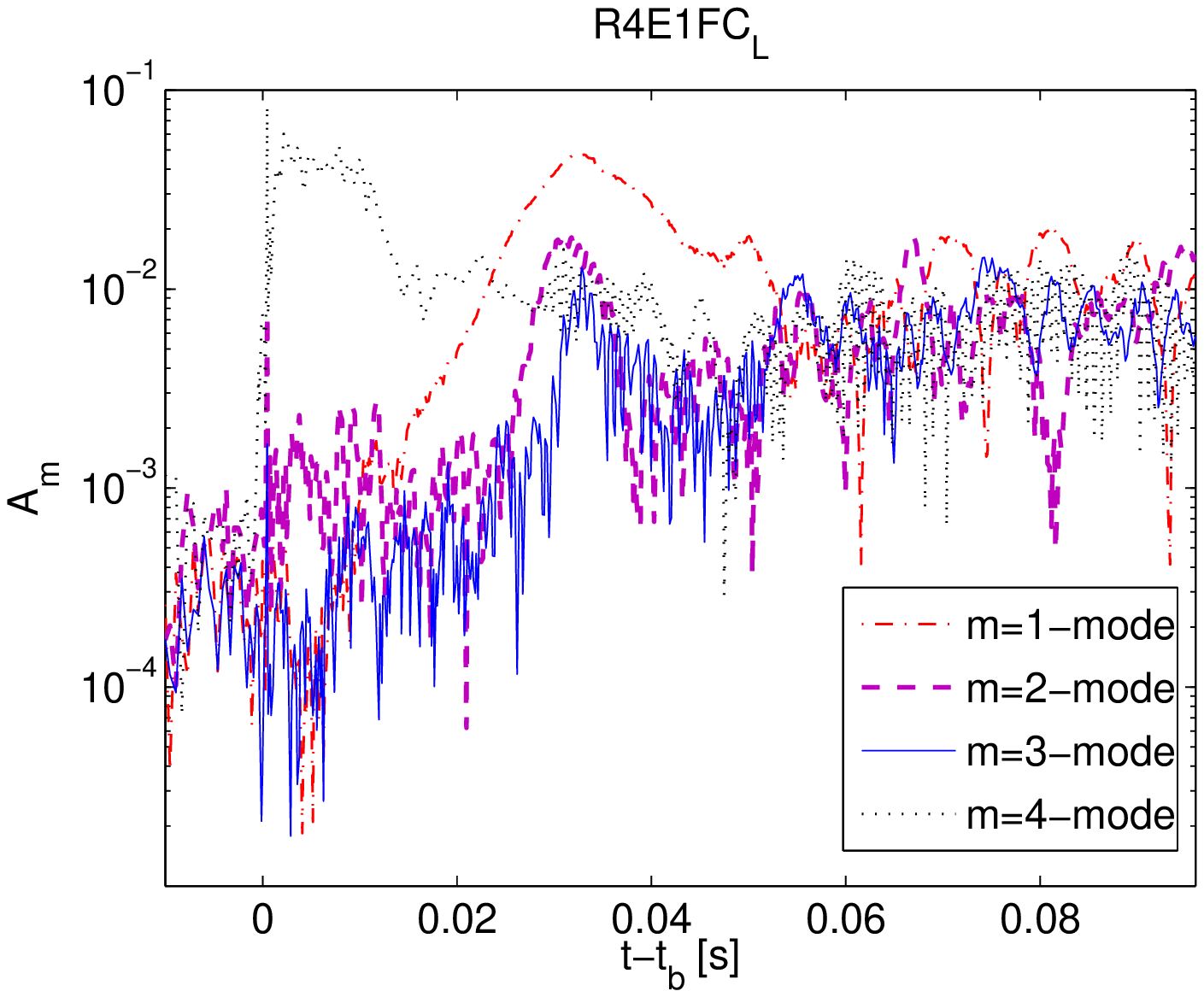}
        \caption{The upper left panel displays 
the emission of the A$_{+}$ and
the A$_{\times}$-amplitude along the pole 
for model R3E1AC$_{L}$. The upper right panel
displays the same for model R4E1FC$_{L}$. 
The lower two panels show for the same 
models as above
the normalised mode amplitudes 
$A_{m}$ for $m={1,2,3,4}$ extracted at a radius of
25km.}
             \label{fig21.eps}
\end{figure*}
%%%%%%%%%%%%%%%%%%%%%%%%%%%%%%%%%%%%%%%%%%%%%%%%%%%%%%%%%%%%%%%%%%%%%%%

The improved input physics of 
the two leakage models R3E1AC$_{L}$ and R4E1FC$_{L}$
allows us to address the question how the inclusion of 
deleptonisation in the postbounce phase quantitatively 
alters the GW signal of our 3D MHD models.
While the results show 
qualitative agreement with our earlier findings,  
they clearly deviate in quantitative terms.
As the most striking feature, these models show 
$5 - 10 \times$ bigger maximum GW amplitudes 
due to the nonaxisymmetric dynamics
compared to
their counterparts that neglect neutrino cooling, as one can see
in Table \ref{table:3} or 
when comparing Fig. \ref{fig17.eps}
with Fig. \ref{fig21.eps}.
This suggests that the treatment of postbounce neutrino cooling 
plays an important role when it comes to the quantitative 
forecast of GW signals from a low $\beta$ instability.
The neutrino cooling during the 
postbounce phase leads to a more 
condensed PNS with a shorter dynamical
timescale compared to the purely 
hydrodynamical treatment, as shown in Fig. \ref{fig20.eps}. 
This in turn is directly reflected in the dynamical evolution:
The shock wave stalls at considerably 
smaller radii and becomes more quickly unstable 
to azimuthal fluid modes (see Fig. \ref{fig16.eps}).
Since there is much more matter in the unstable 
region of these models, the unstable modes
grow faster and the triggered spiral density 
waves cause the emission of much more powerful
GW.

%%%%%%%%%%%%%%%%%%%%%%%%%%%%%%%%%%%%%%%%%%%%%%%%%%%%%%%%%%%%%%%%%
We close this subsection by pointing out that previous
core-collapse computations by 
\citet{2007CQGra..24..139O,2007PhRvL..98z1101O} and 
\citet{2008A&A...490..231S} show both qualitative and quantitative
agreement with the ones presented in the previous 
subsection as they incorporate
nearly identical micro-physics. However, they mismatch on a quantitative
scale with R3E1AC$_{L}$ \& R4E1FC$_{L}$ due to the absence of
the postbounce neutrino treatment. 
However, we point out that our leakage scheme overestimates
the compactification of the PNS due to neutrino
cooling. The `reality' for the strength of GW emission
therefore sould lay in between the results from the
pure MHD and the leakage treatment.
We also want to point out that another 
limitation which might affect the absolute values of the GW signature
from a $T/|W|$ dynamical instability is
the grid resolution.
Our choice of a uniform grid leads to more than 
sufficient resolution at the stalled
shock, but may underresolve 
the surface of the PNS.
For example, \citet{2007CoPhC.177..288C}
showed that resolution 
can have a significant effect on the
instability's developement and its GW signal.

%%%%%%%%%%%%%%%%%%%%%%%%%%%%%%%%%%%%%%%%%%%%%%%%%%%%%%%%%%%%%%%%%%%%%%%%%

\subsubsection{The influence of strong magnetic fields 
on the gravitational wave signature}
\label{section:Bfield}
%%%%%%%%%%%%%%%%%%%%%%%%%%%%%%%%%%%%%%%%%%%%%%%%%%%%%%%%%%%%%%%%%%%%%%%%%

 \begin{figure*}
   \centering
   \includegraphics[width=8.8cm,height=6cm]{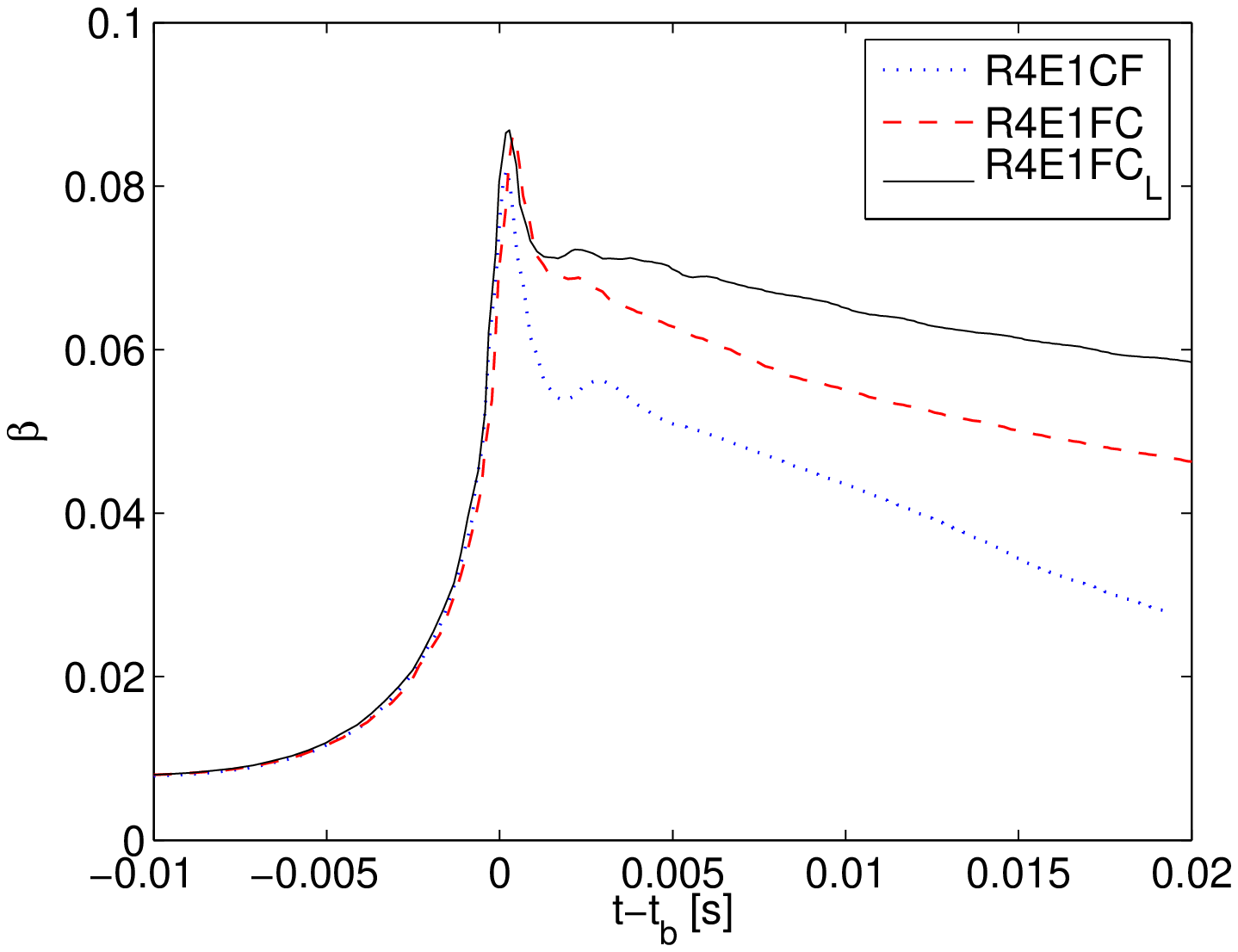} 
   \includegraphics[width=8.8cm,height=6cm]{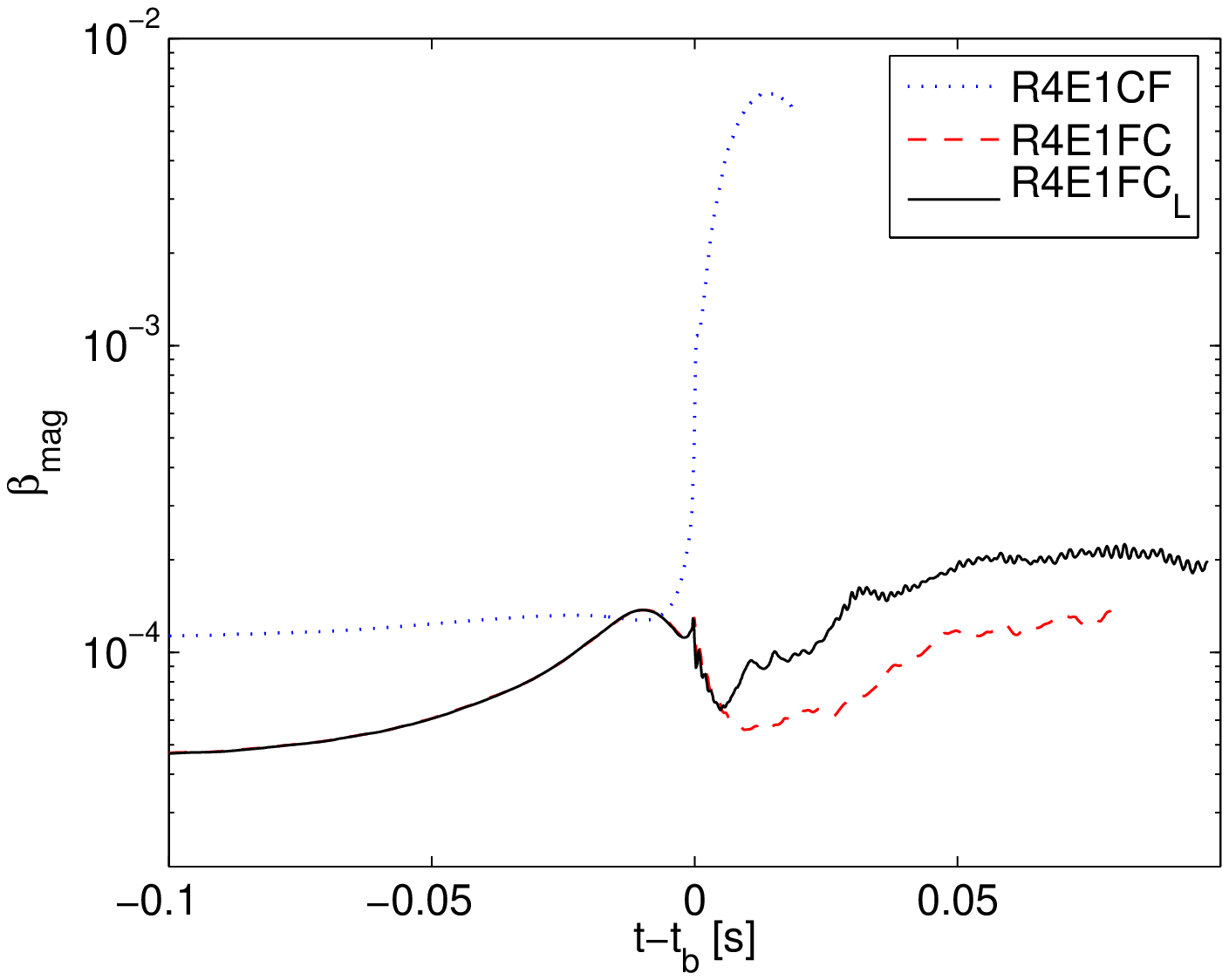} 
   \includegraphics[width=8.8cm,height=6cm]{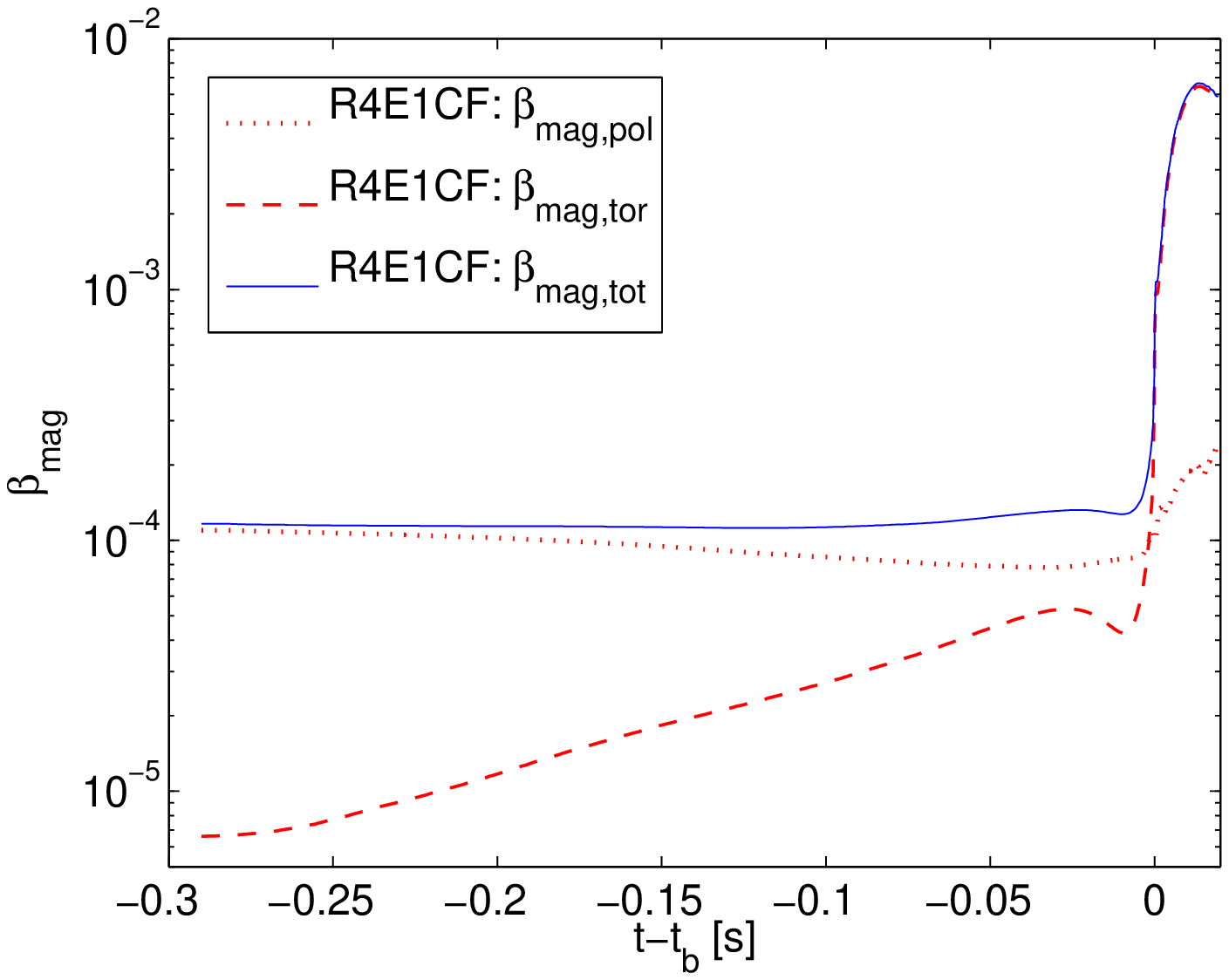} 
   \includegraphics[width=8.8cm,height=6cm]{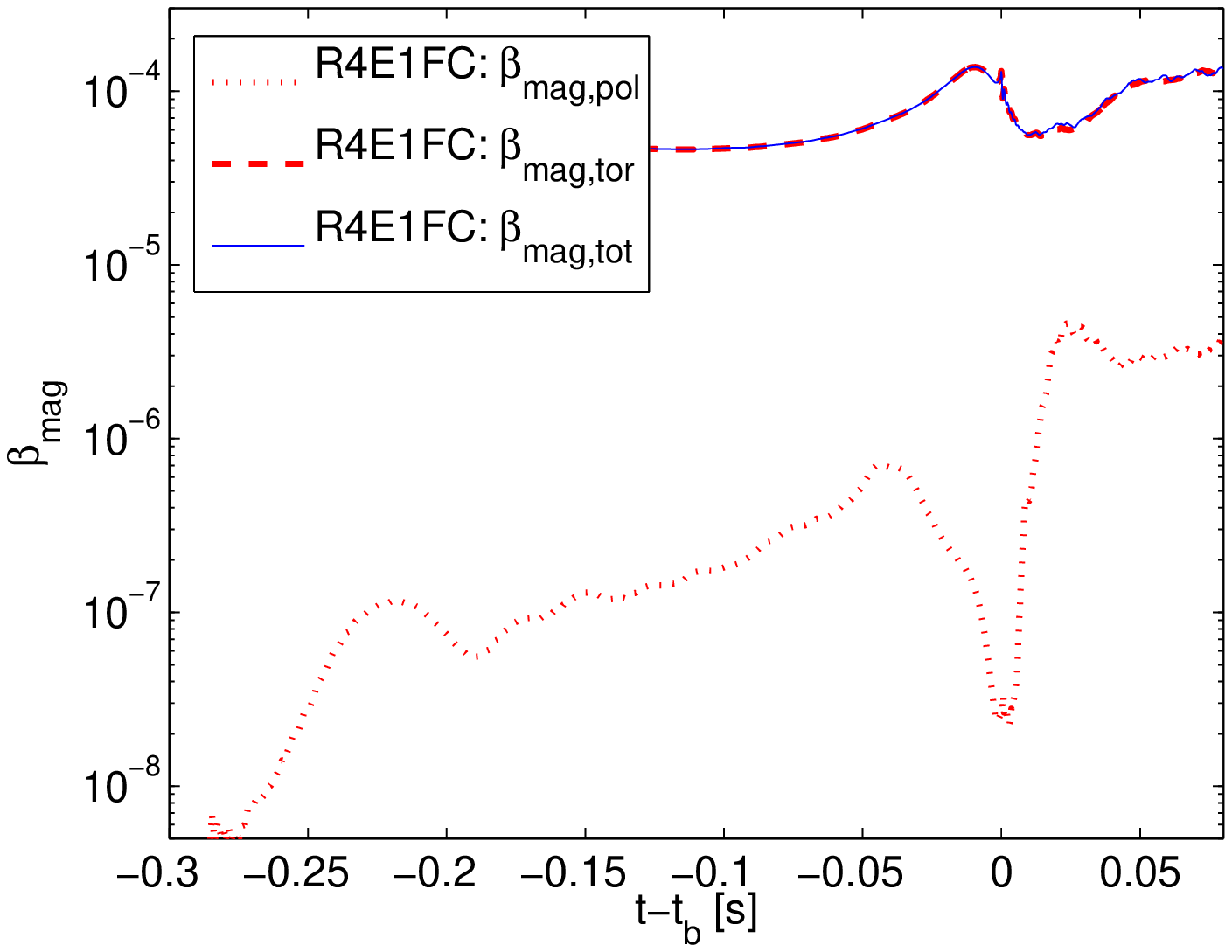} 
        \caption{The upper left panel 
shows the first 20ms of the time evolution of the rotational energy 
parameter $\beta=T/|W|$ of models R4E1CF, R4E1FC and R4E1FC$_{L}$.
The upper right panel displays the time evolution 
of the magnetic energy parameter $\beta_{mag}$ for 
the same models.
The small hump in the $\beta_{mag}$ profiles 
at $t-t_{b}=0$, best visible 
for R4E1FC and R4E1FC$_{L}$, stems from the 
fact that field compression
scales as $B\propto \rho^{2/3}$. 
When the core overshoots its
equilibrium position at bounce and re-expands, 
the density decreases temporarily, which
in turn is reflected in the energy density of 
the magnetic field.
The lower left panel shows the evolution of the 
magnetic parameter $\beta_{tor,mag}$,
$\beta_{pol,mag}$ and the total 
$\beta_{mag}$ for R4E1CF,
the lower right displays the same quantities 
for R4E1FC.}
             \label{fig22.eps}  
   \end{figure*}

%%%%%%%%%%%%%%%%%%%%%%%%%%%%%%%%%%%%%%%%%%%%%%%%%%%%%%%%%%%%%%%%%%%
  \begin{figure}
   \centering
    \includegraphics[width=8.8cm]{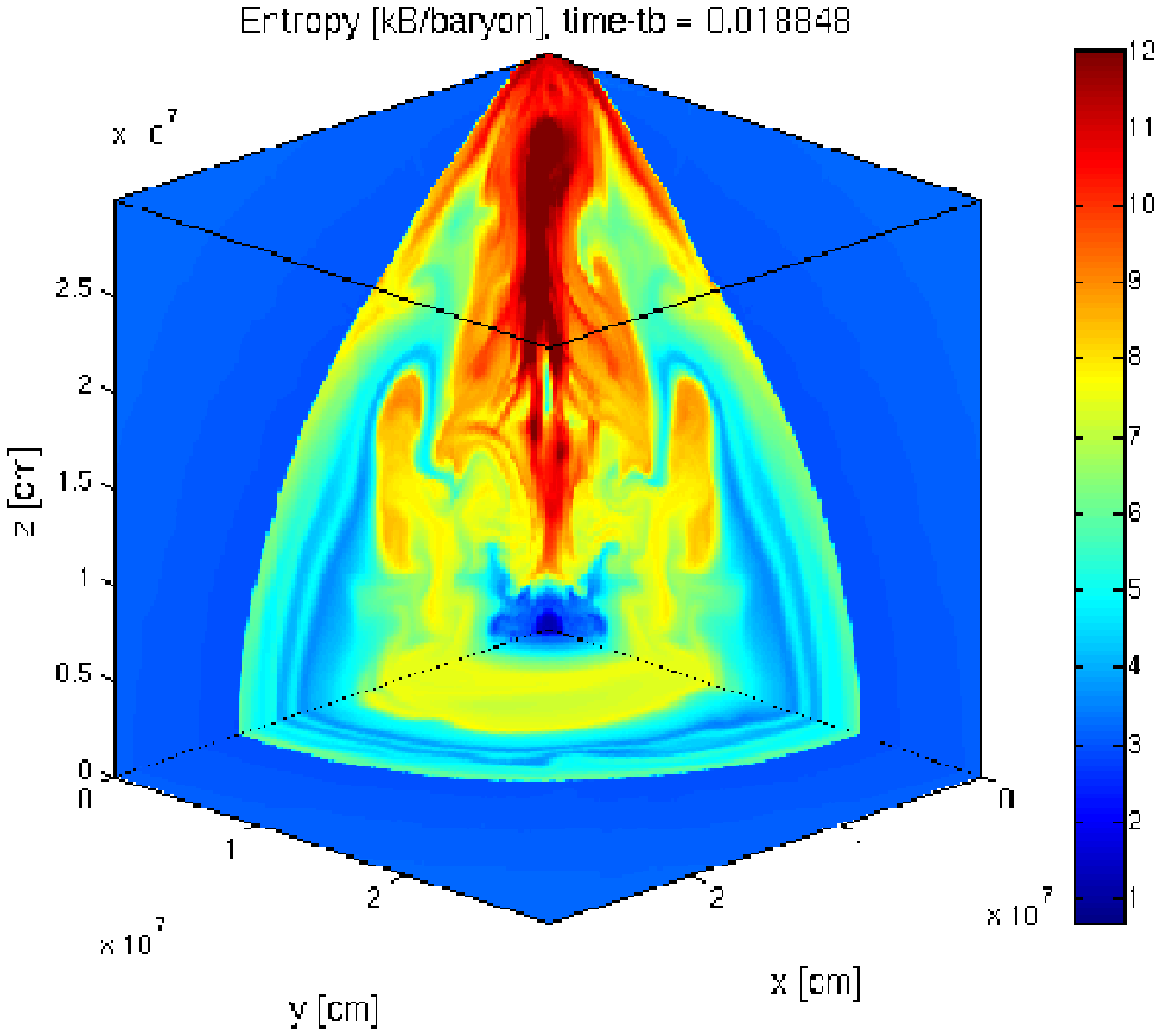}
        \caption{Snapshot of model R4E1CF's entropy 
distribution in the first octant at a representative instant 
of its evolution. The innermost 300$^3$km$^3$ are displayed.}
              \label{fig23.eps}
   \end{figure}

%%%%%%%%%%%%%%%%%%%%%%%%%%%%%%%%%%%%%%%%%%%%%%%%%%%%%%%%%%%%%%%%%%%

  \begin{figure}
   \centering
   \includegraphics[width=8.8cm]{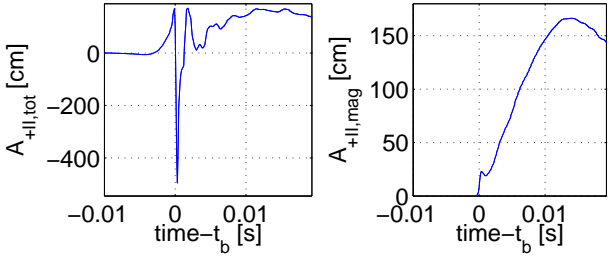} %BW
        \caption{Model R4E1CF's time evolution of 
the $total$ quadrupole amplitude A$_{\rm{+II,tot}}$ (left panel)
and its contributions 
from the magnetic field A$_{\rm{+II,mag}}$ (right panel).}
             \label{fig24.eps}  
   \end{figure}
%%%%%%%%%%%%%%%%%%%%%%%%%%%%%%%%%%%%%%%%%%%%%%%%%%%%%%%%%%%%%%%%%%%%%%%%

%%%%%%%%%%%%%%%%%%%%%%%%%%%%%%%%%%%%%%%%%%%%%%%%%%%%%%%%%%%%%%%%%%%%

In the previous three subsections we 
have shown that the core-collapse dynamics and thus
the GW signal of initially 
weakly or moderately strong 
magnetised stellar cores ($\lesssim10^{11}$G) 
is hardly affected
by magnetic fields.
However, in the case of strong initial fields 
($\sim10^{12}$G) things change dramatically.
The combined action of flux-freezing, 
field winding \citep{1976ApJ...204..869M} and 
also of the magneto-rotational 
instability MRI \citep{1998RvMP...70....1B}
generally may lead to growth of
magnetic fields 
by many orders of magnitude to values
where the magnetic pressure
reaches the order of magnitude of 
matter pressure.
This, in turn, triggers a
collimated, bipolar jet-explosion 
(see e.g. \citet{2007ApJ...664..416B},
and references therein) by 
converting magnetic energy into kinetic energy.
However, with our current grid setup we are unable
to resolve 
exponential growth triggered by the MRI
\citep{2006PhRvD..74d4030E};
the initial magnetic fields are solely 
amplified by compression during the 
infall phase and by magnetic winding.

%%%%%%%%%%%%%%%%%%%%%%%%%%%%%%%%%%%%%%%%%%%%%%%%%%%

For two reasons, magnetically-driven explosions are 
of interest for the prediction of GWs.
Firstly, the bipolar outflow of matter results in 
a `memory effect' \citep{Thorne1989} in the GW signature, as
observed e.g. by 
\citet{Obergaulinger2006} and \citet{2006PhRvD..74j4026S}.
As the out-stream of matter usually happens 
along the rotational axis, 
the amplitude $A_{\rm{+II}}$ will
grow over time, scaling as $A_{\rm{+II}}\propto 2mv_{z}^2$,
where the ejected mass $m$ increases constantly in the early stage
of a magneto-rotational supernova.
Secondly, the GW amplitude of these 
(not necessarily) realistic models
is also affected passively
by strong magnetic fields,  
as they rise during collapse and 
early postbounce phase to values as high
as $\sim10^{16-17}$G. 
The magnetic energy density
may provide sizable contributions to 
the overall amplitude, as estimated
approximately by the following formula 
(cf. \citet{Kotake2004} Eq. 22):  
\begin{equation}
\eta_{mag}= \frac{\frac{B_{c}^2}{8\pi}}
{\rho_{c}c^2}\sim 10\%\left(\frac{B_{c}}{\rm{several}
\times10^{17} \rm{G}}\right)^2\left
(\frac{\rho_{c}}{10^{13} \rm{gcm^{-3}}}\right)^{-1},
\end{equation}
%%%%%%%%%%%%%%%%%%%%%%%%%%%%
\noindent
where $B_{c}$ and $\rho_{c}$ stand for the central values
of the magnetic field and density.
If magnetic contributions become significant it is 
necessary to include them in Eq. \ref{equ:8}, 
which then turns into Eq. \ref{equ:bfieldgw}.

For the purpose of studying the effect of very strong magnetic 
fields on the GW signal
in 3D, we have carried out three runs: R4E1CF, R4E1FC and 
a leakage model R4E1FC$_{L}$. In order 
to overcome the technically challenging 
simulation of magnetic field growth through
small scale \citep{2009A&A...498..241O} or 
long-term processes,
we applied initial configurations
which may not be realised in nature, but 
can deliver the formation of jets in a
similar way as natural field growth would.

%%%%%%%%%%%%%%%%%%%%%%%%%%%%%%%%%%%%%%%%%%%%%
Model R4E1CF, which was set up with very strong initial 
poloidal fields (see Table \ref{table:1}), shows a typical
behaviour for a magneto-rotational core collapse followed
by a jet-like explosion. Field compression during 
the collapse phase strongly amplifies both 
the toroidal- and poloidal magnetic fields, 
since flux-freezing in stellar collapse
guarantees the B-field to scale as $B\sim \rho^{2/3}$.
Note that we evaluate \textit{toroidal} and 
\textit{poloidal} magnetic field components as:
\begin{eqnarray}
B_{tor}& = & \sqrt{B_{x}^2 + B_{y}^2} \\
\label{equ:btor}
B_{pol} &= &B_{z}. 
\label{equ:bpol}
\end{eqnarray}
Furthermore, the toroidal field 
is build up by tapping the magnetic energy contained
in the poloidal component of the field trough winding, 
whilst the poloidal field component grows only little 
during the postbounce phase by action of
meridional motions in the core as 
shown in the lower left panel of Fig. \ref{fig22.eps}.
In turn, the generated hoop stresses grow fast to reach high values 
near the polar region, and the ratio of magnetic to matter 
to fluid pressure
reaches $P_{mag}/P_{matter} \sim 1$.
A bipolar collimated jet explosion 
is launched, quickly leaving the computational 
domain (see Fig. \ref{fig23.eps}). 
 
Another remarkable feature is that, 
in contrast to weakly magnetised models,
magnetic breaking efficiently decelerates the 
inner core by redistributing angular momentum
\citep{1976ApJ...204..869M}.
This is shown in the upper left panel of 
Fig. \ref{fig22.eps}. Note that model R4E1FC 
slows down
at a higher rate than
R4E1FC$_{L}$ as its fluid 
can re-expand to larger radii due to the absence 
of neutrino cooling.
%%%%%%%%%%%%%%%%%%%%%%%%%%%%%%%%%%%%%%%%
The GW signal is displayed in Fig. \ref{fig24.eps}.
In order to distinguish the different contributions to the total
GW amplitude $h_{ij,tot}^{TT}$, we 
split the magnetic and fluid part in the following way:
\begin{equation}
 h_{ij,tot}^{TT} =  h_{ij,matter}^{TT} +  h_{ij,mag}^{TT}.
\label{equ:split}
\end{equation}

\noindent
Around core bounce, the structure of 
the GW signal is very similar to that of less magnetised
cores, exhibiting a clear type I signature.
However, shortly after bounce we observe a growing, 
slowly time-varying 
offset of $A_{\rm{+II}}$
relative to the horizontal axis compared to 
ring-down oscillations 
of the weakly-magnetised simulations. 
There are two reasons for this behaviour.
Firstly, the magnetic
contribution $A_{\rm{+II,mag}}$ to the total 
GW amplitude grows strongly as 
magnetic forces act on the core.
%%%%%%%%%%%%%%%%%%%%%%%%%%%%%%%%%%%%%%%%%%%%%%
\noindent
Secondly, as we have explained earlier in this subsection, 
matter-outflow along the z-axis also contributes to the 
signal.
This behaviour was already observed by 
\citet{Obergaulinger2006} in axisymmetric
simulations 
and its characteristical GW amplitude was named 
`\textit{type IV} signal'  
\citep{Obergaulinger2006}. 
The contribution of the magnetic amplitudes 
shrinks with the onset of the jet. The emerging matter gains its
kinetic energy by tapping the energy stored in the magnetic field, which 
causes a drop in $\beta_{mag}$ and hence also in $A_{\rm{+II,mag}}$, as
it can be seen in Figs. \ref{fig22.eps} and \ref{fig24.eps}
at about $t-t_{b}\approx10$ms.
The non-axisymmetric amplitudes are negligible 
compared to 
the axisymmetric part of the wave train
(see Tab. \ref{table:3})
and the 
result of some prompt convective
motions and complicated time-variations in the postbounce
magnetic field configuration.
However, even if a Galactic supernova was
optimally orientated for the detection
of a type IV signal, 
it is still
questionable whether we could 
distinguish it in early stages
from an ordinary type I signal.
The characteristic offset of this particular 
signal type 
was suggested e.g. 
by \citet{Obergaulinger2006} to be a valid 
measure of aspherity of the ongoing supernova
explosion. However, since the memory effect in the amplitude
appears on the long timescale of several times 10ms,
it would be out of the LIGO 
band.
However, the planned space-based DECIGO instrument
\citep{Kawamura}
could permit to track the low frequency contribution
of such a GW signal in the future. 

We point out that at the onset of jet 
formation around $t-t_{b}\approx 10$ms, 
the absolute value 
of the magnetic field in the polar region at the edge of the PNS
is $\sim10^{16}$G, which translates into a mildly relativistic 
fast Alfv\'en speed. Additionally, the 
velocity of the ejected matter accelerates 
up to radial velocities of $\sim0.1c$ 
when leaving the boundary of our computational domain. 
These two points challenge the Newtonian 
treatment of the dynamics in our scheme
and suggest to use a special or general
relativistic code for the 
further evolution of the jet dynamics.

%%%%%%%%%%%%%%%%%%%%%%%%%%%%%%%%%%%%%%%%%%%%%%%%%%%%%%%%%%%%%%%%%%%

  \begin{figure*}
   \centering
   \includegraphics[width=\textwidth]{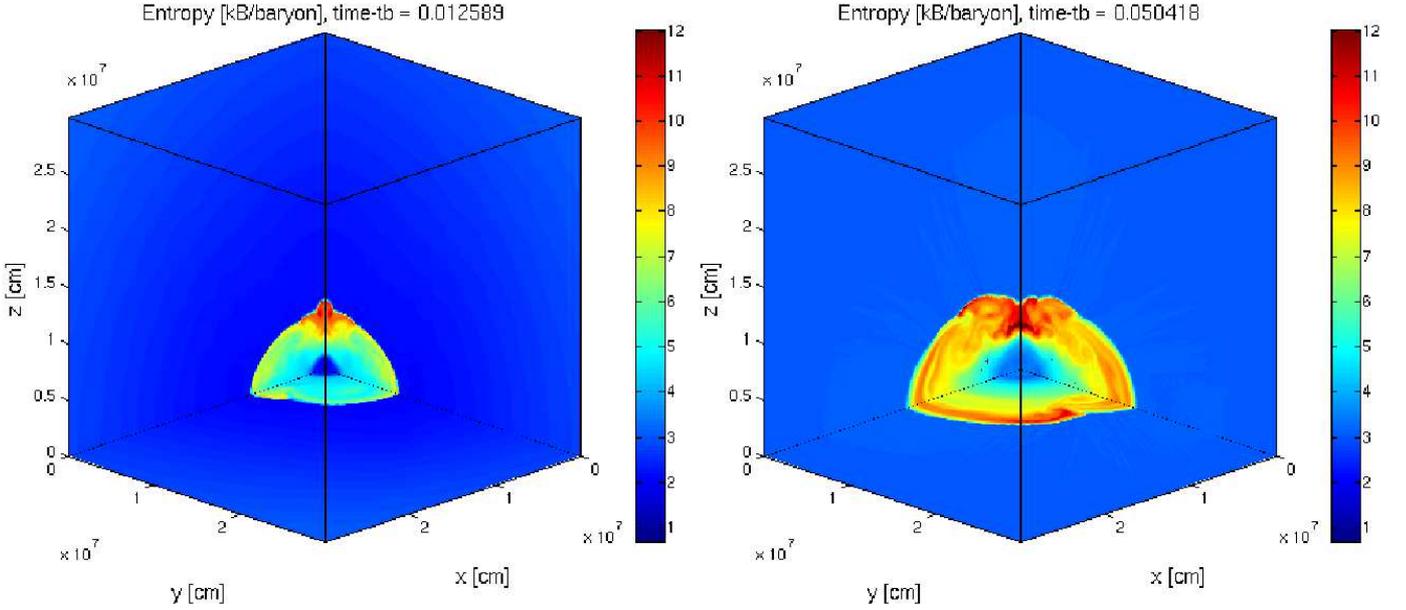}
        \caption{Snapshots of model R4E1FC$_{L}$'s entropy 
distribution in the first octant at two representative instants 
of its evolution. The innermost 300$^3$km$^3$ are displayed.}

              \label{fig25.eps}
   \end{figure*}
%%%%%%%%%%%%%%%%%%%%%%%%%%%%%%%%%%%%%%%%%%%%%%%%%%%%%%%%%%%%%%%%%%%

Models R4E1FC and R4E1FC$_{L}$ show a very different
dynamical outcome.
For these two simulations, we assumed 
the initial toroidal field to be
$10^3$ times the values of the poloidal field, 
as suggested from stellar evolution
calculations by \citet{Heger2005}.
During collapse the magnetic field 
components primarily grow due
to compression as one can see in the 
lower right panel of Fig. \ref{fig22.eps}.
Right after bounce, the strong toroidal 
fields in both models cause the onset of a jet, 
as one can see in the left panel of Fig. 
\ref{fig25.eps}.
However, surprisingly and in contrast to 
the previously discussed models, 
or simulations from
\citet{Kotake2004} where they 
applied similar initial conditions 
and obtained jet explosions in 2D, the wind-up of the
poloidal- into the toroidal field does not 
occur efficiently enough.
The spiral wave of the standing accretion shock 
instability (SASI, see e.g. \citet{Blondin2003}), 
which forms at the same time,
hinders and delays 
the growth of a jet, 
as displayed in the right panel 
of Fig. \ref{fig25.eps}. 
We interpret this phenomenon as follows: 
Matter can easily slip along, 
but not move perpendicular to magnetic field. 
In the x-y plane, the poloidal field's stress acts on 
matter in the opposite
direction of the fluid motion.
This is also reflected in the upper 
left panel of Fig. \ref{fig22.eps}.
It displays clearly that the magnetic forces are not capable
of slowing down the inner core in case of strong initial
toroidal fields as effectively as the 
poloidal ones that are anchored in the outer
stellar layers. 
Since models R4E1FC and 
R4E1FC$_{L}$ have relatively weak initial poloidal
magnetic fields, the fluid, which 
rotates around the z-axis, can develop 
nearly unhindered instabilities. 
The developing spiral waves then counteract
successfully the formation of a jet-like explosion
by turning matter aside the pole 
(see Fig. \ref{fig25.eps}), where the 
magnetic hoop stresses are strongest
(and in axisymmetry most probably would be able
to launch a jet at this stage of the simulation).
This corresponds also to our observation 
from the last subsection where we stated
that the low $T/|W|$ instability 
grows slower in the presence of 
dominant poloidal fields, because they 
cause stresses that act against the spiral instabilities.
However, further discussion of this
phenomenon is beyond the scope of this work
and will be investigated in a subsequent study.

%%%%%%%%%%%%%%%%%%%%%%%%%%%%%%%%%%%%%%%%%%%%%%%%%%%%%%%%%%%%%%%%%%%%%%%%

The resulting GW signals consequently show
a type I signal at core bounce,
subsequently followed by a low $\beta$
instability (see the right panels of \ref{fig21.eps}).
Although the GW contributions 
due to magnetic stresses in model
R4E1FC$_{L}$
show qualitatively the same 
features as the ones from model R4E1CF,
they are smaller and
dominated by the hydrodynamics part of the amplitude
(e.g., $A_{\rm{+II,mag}}\sim10$cm vs. $A_{\rm{+II,matter}}\sim$150cm 
at 30ms after bounce).

%%%%%%%%%%%%%%%%%%%%%%%%%%%%%%%%%%%%%%%%%%%%%%%%%%%%%%%%%%%%%%%%%%%%%%%%
\section{Summary and Conclusions}

Core-collapse supernovae are
a source of GWs 
that can only be modelled imperfectly 
due to large uncertainties in the initial 
conditions, input physics and the technically
challenging 3D neutrino transport. 
The parameter space of possible initial conditions
is huge since
many progenitor 
configurations are possible at the onset of collapse.
In this paper we tried to outline some
dependences of the 3D GW form
upon a variety of these conditions, since
for un-modelled burst analysis techiques,
partial information on wave forms is already useful, e.g. 
typical waveform features, approximate spectra and
the type of polarisation and burst duration.

With our model series containing 25 
three-dimensional MHD core-collapse 
supernova simulations, we have tried to probe 
the GW signature with respect to 
different nuclear equations of state, rotation
rates, poloidal and toroidal magnetic fields,
and a postbounce deleptonisation scheme.

Similar to the findings of e.g.
\citet{M2004,2008PhRvD..78f4056D,2009A&A...496..475M,
2009CQGra..26f3001O,2009arXiv0912.1455S},
our results show that all 
non- and slowly rotating 
models release GWs due to 
prompt convection within the first $\sim 30$ms 
postbounce that is caused by the presence of a 
negative radial entropy gradient.
Furthermore, we could show in simulations without deleptonisation
in the postbounce phase that the 
waveforms obtained from this early stage of
a supernova explosion contain indirect information 
about the 
underlying EoS.
Due to different radial locations
of the convectively unstable region and 
the amount of matter it contains, 
we were able to distinguish
the LS EoS from the one of Shen.
While the LS EoS leads to GW emission in a 
frequency band peaking 
between $\sim150-500$Hz, the spectrum of the models 
using the Shen EoS is 
restricted to roughly $\sim150-350$Hz.
However, LIGO's current sensitivity
makes it impossible to see the high-frequency component 
of the LS models. Thus, a distinction
between the two EoS is currently not possible.
Nevertheless, planned upgrades of the interferometers in the near future
should enable the discrimination between the prompt convection GW 
signal of the LS- and the Shen EoS.
We also found minor deviations in the GW characteristics 
for simulations which were carried out with different compressibility
versions of the LS EoS.
However, the differences in the frequency domain of the GW signal are 
negligibly small 
and thus not likely to be constrained by observation.

With the inclusion of a neutrino leakage scheme in a 
slowly rotating model for the postbounce phase, we could show
that the GWs emitted during the first $\sim 20$ms after bounce
are predominantly due to entropy driven 
`prompt' convection.
In purely hydrodynamical models, the GW emission ends with 
the decaying negative entropy gradient.
In more realistic models, a negative radial 
lepton gradient, caused by the neutronisation burst
and the subsequent deleptonisation, takes over as driving
force of the convective activity at the edge of the PNS.
The long-lasting GW emission associated 
with this dynamical feature
is roughly of the size of $\sim 1$cm and has a spectral
peak around $\sim 700-1200$Hz.

In our set of models, simulations 
with a precollapse core angular velocity
within the parameter range of 
$\Omega_{c,i}=\pi\ldots4\pi$rads$^{-1}$
undergo a rotational core collapse.
The models all exhibit 
a so-called type I GW
burst at core bounce.
As the most important outcome, 
our 3D MHD models could confirm
the recent findings of \citet{2008PhRvD..78f4056D}
that the purely axisymmetric (l=2, m=0) peak amplitude
scales about linearly with the rotation 
rate $\beta_{b}$ at core bounce 
($|h_{max}| \propto \beta_{b}$) for $\beta_{b} \lesssim 10\%$,
while the Fourier-transform of the bounce wave trains 
for most models in the indicated parameter range align 
around a spectral peak of $\sim 800-900$Hz.
However, for very fast initial rotation rates of 
$\Omega_{c,i}\gtrsim 3\pi$rads$^{-1}$,
centrifugal forces significantly decelerate collapse and core bounce.
The longer timescale and the weakened spin-up of the core 
due to the action of centrifugal forces 
leads generically 
to a decrease of the peak amplitude and a broadened spectral peak
at lower frequencies.
Furthermore, our results indicate that the particular 
choice of the nuclear EoS has little influence on the GW signal 
from rotational core bounce.
These findings are in good qualitative agreement with the ones 
\citet{2008PhRvD..78f4056D} derived from axisymmetric models.

Models with a rotation rate of $\beta_{b}\gtrsim 5$\%
at core bounce become subject to a so-called low $T/|W|$
instability of dominant $m=1$ or $m=2$ 
character within the first several tens
of ms after bounce, depending on
the individual model.
This nonaxisymmetric dynamical shear instability leads to 
prolonged narrow band GW emission at a frequency of twice the rotation
rate of the innermost part of the PNS that rotates as a solid body.
The fact that the effectively measured GW 
amplitude scales with the number of GW cycles $N$ as
$h_{eff}\propto h\sqrt{N}$ suggests that the detection of such
a signal is tremendously enhanced compared to e.g. the short-lived
GW bursts from core bounce, and would allow us to probe
the rotational state of the PNS over a long period.
However, we point out that such a mechanism only operates if the progenitor
is rotating much faster than predicted for most stars by stellar evolution
calculations \citep{Heger2005}.
We also find that centrifugal forces set a limit to the maximum frequency
of this periodic GW signal somewhere around $\sim 935$Hz as they suppress
the inward advection of angular momentum.
Besides that, we point out that GWs from a low $T/|W|$ instability
are highly degenerate with respect to initial 
rotation rate, EoS and magnetic fields.
Thus, it is very difficult to extract individual features 
of the input physics from the GW 
signal that can 
clearly be attributed to the initial conditions of a progenitor.
Rapidly rotating models that include the postbounce 
neutrino physics at a
qualitative level reproduce the previous findings. 
However, the GW signature from these more advanced models 
show huge quantitative deviations from the ones
that treat the postbounce phase purely hydrodynamically.
As the neutrino cooling during the postbounce phase leads to a more 
condensed PNS, the unstable regions contain 
considerably more mass, which then results in 5 to 10 times 
bigger GW amplitudes.
Motivated by these findings, 
we will continue to improve the postbounce neutrino physics
in future simulations by including neutrino heating
and the emission of $\mu/\tau$ neutrinos.

Our simulations show that the impact of 
magnetic fields on the overall supernova dynamics is 
generally small in cores
with relatively weak precollapse 
fields (B $< 10^{11}$G).
Nevertheless, if we impose very strong and 
probably unrealistic 
initial poloidal magnetic fields ($\sim 10^{12}$G),
the combined action of flux-freezing and field winding 
allows the toroidal field component to grow over many 
orders of magnitude to values where 
the magnetic pressure can trigger a
jet-like explosion along the poles.
The bipolar outflow of matter then causes a type IV gravitational signal.
However, if we assume a strong initial toroidal magnetic field,
the onset of a jet is effectively suppressed by the fast growing 
spiral waves of the SASI.
This finding stands in contradiction to 2D simulations, where similar
configurations led to `jet'-like explosions, as e.g. in \citet{Kotake2004}.
Thus, in simulations where the toroidal component of the magnetic field dominantes
over the poloidal one, the magnetic contributions to the GW signal are
dominated by the hydrodynamical part of the amplitude.
This effect is even stronger in simulations with a deleptonisation scheme 
in the postbounce phase.
%%%%%%%%%%%%%%%%%%%%%%%%%%%%%%%%%%%%%%%%%%%%%%%%%%%%%%%%%%%%%%%%%%%%%%%%

The major limitation of our code now is
in the monopole treatment of gravity, since it 
cannot account for spiral structures, which could be reflected
in GW. 
We are currently working on the improvement of this issue.
The IDSA includes at present only the dominant 
reactions relevant to the neutrino transport problem 
(see \cite{2009ApJ...698.1174L} for details).
Future upgrades will also include contributions from 
electron-neutrino scattering, which are indispensable
during the collapse phase. The inclusion of this reaction 
will also make the cumbersome switch of the neutrino 
parametrisation scheme to
the IDSA at bounce obsolete.
Finally, we work on the inclusion of $\mu$ and $\tau$
neutrinos, which are very important for the cooling of the PNS
to its final stage as neutron star.
%%%%%%%%%%%%%%%%%%%%%%%%%%%%%%%%%%%%%%%%%%%%%%%%%%%%%%%%%%%%%%%%%%%%%%%%

\begin{acknowledgements}
We thank C. D. Ott for carefully reading and commenting the manuscript.
Further acknowledgements go to F.-K. Thielemann 
from the University of Basel for his support,
John Biddiscombe and Sadaf Alam from the Swiss Supercomputing Centre CSCS
for the smooth and enjoyable collaboration, and
C. von Arx and E. O'Connor for helpful comments.
We also thank the referee
for his valuable suggestions to improve our manuscript.
This work would not have been possible without the 
support by the Swiss National 
Supercomputing Centre-CSCS under project ID 168.
We acknowledge support by the Swiss National Science Foundation under grant
No. 200020-122287 and PP0022-106627.
Moreover, this work was supported by CompStar, 
a Research Networking Programme of the European Science Foundation.
\end{acknowledgements}

%\bibliographystyle{/home/simon/Documents/Bibitex/apj}   % Setzt style File fuer Bibliographie (Moeglich sind prsty, apj, h-elsevier2, h-physrev3)
%\bibliography{/home/simon/Documents/Bibitex/References}

\end{document}